\documentclass[aps,showpacs,twocolumn,twoside,amsmath,amssymb,superscriptaddress,prc]{revtex4-2}
\usepackage{multirow}
\usepackage{tikz}
\usepackage{epstopdf}
\usepackage[percent]{overpic}  % insert LaTex text over figures
\usepackage{scrextend}  % to use \footref - refer to the same footnote multiple times.
\usepackage{xcolor,xspace}
\usepackage[ulem=normalem]{changes}
\usepackage{hyperref}
\hypersetup{colorlinks=True,urlcolor=blue,linkcolor=blue,citecolor=blue,filecolor=black}

\colorlet{Changes@Color}{red}  % changes in red color
\usepackage{amsmath}
\usepackage{dcolumn,color,footnote,bm,braket}
\usepackage{url,longtable,tabularx}

\usepackage{fancybox}
\usetikzlibrary{matrix}

\newcommand\+{\dagger}

\newcommand{\omin}{\beta_{3}^{\mathrm{min}}}

\begin{document}

\title{
Quadrupole-octupole coupling and the evolution of collectivity 
in neutron-deficient Xe, Ba, Ce, and Nd isotopes
}

\author{K.~Nomura}
\email{knomura@phy.hr}
\affiliation{Department of Physics, Faculty of Science, 
University of Zagreb, HR-10000 Zagreb, Croatia}

\author{R.~Rodr\'iguez-Guzm\'an}
\affiliation{Physics Department, Kuwait University, 13060 Kuwait, Kuwait}

\author{L.~M.~Robledo}
\affiliation{Departamento de F\'\i sica Te\'orica and CIAFF, Universidad
Aut\'onoma de Madrid, E-28049 Madrid, Spain}

\affiliation{Center for Computational Simulation,
Universidad Polit\'ecnica de Madrid,
Campus de Montegancedo, Bohadilla del Monte, E-28660-Madrid, Spain
}

\date{\today}

\begin{abstract}
The evolution of quadrupole and octupole collectivity 
in neutron-deficient Xe, Ba, Ce, 
and Nd nuclei near the ``octupole magic'' neutron number $N=56$ is investigated
within the mapped $sdf$-IBM framework. Microscopic 
input is obtained via quadrupole and octupole constrained
Hartree-Fock-Bogoliubov 
calculations, based on the parametrization D1M of the Gogny
energy density functional. Octupole-deformed mean-field ground 
states are predicted for Ba and Ce isotopes near $N=56$. 
Excitation energies of positive- and negative-parity 
states as well as  electric transition rates  
are computed with wave functions resulting from the diagonalization 
of the mapped IBM Hamiltonian. The 
parameters of the Hamiltonian are determined via the  mapping of 
the mean-field potential energy surfaces 
onto the expectation value of the Hamiltonian 
in the condensate state of the $s$, $d$, and $f$ bosons. 
Enhanced octupolarity is predicted for Xe, Ba, and Ce isotopes 
near $N=56$. The shape/phase transition from 
octupole-deformed to strongly quadrupole-deformed near $N=60$
is analyzed in detail. 
\end{abstract}

\maketitle

\section{Introduction}

Octupole deformation emerges 
in specific regions of the nuclear chart, 
that correspond to 
``magic''
proton $Z$ and/or neutron $N$ numbers 
34, 56, 88 and 134 around which, 
octupole shapes are stabilized 
as a consequence of the coupling between 
states with opposite parity that 
differ in the angular momentum quantum numbers $j$ and $l$ by 
$\Delta{j}=\Delta{l}=3\,\hbar$.
The search for permanent octupole 
deformation represents a central topic in modern 
nuclear structure 
physics \cite{butler1996,butler2016,butler2020b}. 
Experiments using radioactive-ion beams have found 
evidence for  static octupole deformation 
in light actinides with $N\approx134$ 
\cite{gaffney2013,butler2020a,chishti2020} 
and in  neutron-rich nuclei 
with $N\approx88$ and $Z\approx56$. Typical
fingerprints of  octupole deformations 
are large electric octupole ($E3$) 
transition rates and low-lying negative-parity 
($\pi=-1$)
states forming an approximate alternating-parity 
doublet with the positive-parity ($\pi=+1$)
ground-state band \cite{bucher2016,bucher2017}.
 On the other hand, there is limited 
experimental information 
\cite{rugari1993,FAHLANDER1994,smith1998,degraaf1998,paul2000,rzacaurban2000,SMITH2001,DEANGELIS2002,capponi2016,gregor2017} on octupole 
correlations in 
neutron-deficient nuclei with $N\approx Z\approx56$ 
and $N\approx Z\approx34$
as well as in the case of 
neutron-rich nuclei with $N\approx56$ 
and $Z\approx34$. Note that  $N\approx Z$ 
nuclei are close to the proton drip line, and are 
currently not accessible  experimentally.

From a theoretical point of view, octupole related
properties 
have been investigated using a variety of approaches, 
such as macroscopic-microscopic models 
\cite{naza1984b,leander1985,moeller2008}, self-consistent 
mean-field (SCMF) and beyond-mean-field approaches 
\cite{MARCOS1983,BONCHE1986,BONCHE1991,heenen1994,ROBLEDO1987,ROBLEDO1988,EGIDO1990,EGIDO1991,EGIDO1992,GARROTE1998,GARROTE1999,long2004,robledo2010,robledo2011,erler2012,robledo2012,rayner2012,robledo2013,robledo2015,bernard2016,agbemava2016,agbemava2017,xu2017,xia2017,ebata2017,rayner2020,cao2020,rayner2020oct,rayner2021,nomura2021qoch}, 
interacting boson models (IBM)
\cite{engel1985,engel1987,kusnezov1988,yoshinaga1993,zamfir2001,zamfir2003,nomura2013oct,nomura2014,nomura2015,nomura2020oct,vallejos2021,nomura2021oct-u,nomura2021oct-ba}, 
geometrical collective models 
\cite{bonatsos2005,lenis2006,bizzeti2013}, 
and cluster models \cite{shneidman2002,shneidman2003}. 
Octupole correlations have also been studied 
within the framework of the symmetry-projected
Generator Coordinate Method (GCM) 
\cite{RS,bernard2016,robledo2019,rayner2020oct,rayner2021}. 
However, those
symmetry-conserving GCM calculations 
are quite time consuming and alternative 
schemes, such as the use of a collective 
Hamiltonian obtained via the Gaussian 
overlap approximation (GOA), have also been 
considered \cite{xia2017,nomura2021qoch}.  

Most of the theoretical studies already mentioned 
have concentrated on nuclei with $(N,Z)\approx(134,88)$ 
and $(88,56)$. On the other hand, octupole correlations 
are much less studied in  
lighter nuclei with  
$(N,Z)\approx(56,56)$, $(34,34)$, and $(56,34)$.
\cite{naza1984b,SKALSKI1990,heenen1994,cao2020,nomura2021qoch}. 
Exception made of Ref.~\cite{nomura2021qoch}, calculations 
for those nuclei have been carried out at the mean-field level 
or have 
been restricted to specific spectroscopic properties. Thus, considering 
the renewed experimental interest in  
octupole correlations, it is timely to carry out 
systematic reflection-asymmetric spectroscopic calculations in 
those  regions of the nuclear chart so far not sufficiently studied.

In this work, we investigate the low-energy collective 
quadrupole and octupole excitations in  neutron-deficient 
Xe, Ba, Ce, and Nd nuclei with $N\approx Z$. Special attention is 
paid to the onset of octupole deformation 
and to whether the octupole 
``magic number'' 56 is robust in the case of 
$N\approx Z$ nuclei. We employ the SCMF-to-IBM mapping procedure 
\cite{nomura2008}. Within this approach, constrained 
Hartree-Fock-Bogoliubov (HFB) calculations, based on the Gogny-D1M \cite{D1M}
energy density functional (EDF), are performed to obtain
the mean-field potential energy surfaces 
(denoted hereafter as SCMF-PESs) as functions 
of the axially-symmetric quadrupole $\beta_{2}$ and 
octupole $\beta_{3}$ deformations.
Spectroscopic properties are computed 
via the diagonalization of the IBM Hamiltonian, with the 
strength parameters determined by mapping the SCMF-PES 
onto the expectation value of the 
Hamiltonian in the condensate state of the monopole $s$ 
(with spin and parity $0^{+}$), quadrupole 
$d$ ($2^{+}$), and octupole $f$ ($3^{-}$) bosons. At variance
with the conventional IBM fit, the parameters of the 
model IBM Hamiltonian are completely determined 
from microscopic EDF calculations, which enables us to 
access those nuclei where experimental data are not available.

The mapping procedure, hereafter referred to as mapped 
$sdf$-IBM, was first employed to describe 
octupole shape/phase transitions 
in reflection-asymmetric light actinides 
and rare-earth nuclei \cite{nomura2013oct,nomura2014} 
based on the relativistic DD-PC1 \cite{DDPC1}
EDF as  microscopic input. A similar approach
has also been applied to study the low-energy spectroscopy 
of Gd and Sm nuclei using 
microscopic input from 
Gogny-D1M EDF calculations
\cite{nomura2015}. More recently, the mapped $sdf$-IBM, 
in combination with the Gogny-D1M EDF, has been successfully 
employed in
systematic studies on the evolution of the octupole collectivity 
in the Ra, Th, U, Pu, Cm, and Cf 
isotopic chains \cite{nomura2020oct,nomura2021oct-u} and 
in  neutron-rich Xe, Ba, Ce, and Nd nuclei 
\cite{nomura2021oct-ba}. 
Within this context, it is  reasonable to extend the 
mapped $sdf$-IBM calculations, based on Gogny-D1M 
microscopic input, to describe the low-lying states in $N\approx Z\approx56$ 
nuclei, where octupole correlations are expected to play an essential role.

The paper is organized as follows. The theoretical procedure is 
outlined in Sec.~\ref{sec:method}. Both the Gogny-D1M SCMF- and mapped IBM-PESs
are discussed in Sec.~\ref{sec:pes}. The results 
obtained for the spectroscopic properties
of the studied nuclei are presented in Sec.~\ref{sec:spec}. In this
section, attention is paid to
low-energy 
excitation spectra and electric transition probabilities. Alternating-parity doublets 
and signatures of 
the octupole shape/phase transitions are discussed in Sec.~\ref{sec:qpt}.
Finally, Section~\ref{sec:summary} is devoted to the concluding remarks.

\section{Theoretical method \label{sec:method}}

To obtain the SCMF-PES, the HFB equation is 
solved, with constrains on the axially symmetric 
quadrupole
$\hat{Q}_{20}$ and octupole $\hat{Q}_{30}$ operators
\cite{rayner2012,rayner2020oct}. 
The mean value 
$\langle \Phi_\mathrm{HFB} |\hat{Q}_{\lambda0}| \Phi_\mathrm{HFB} \rangle \equiv Q_{\lambda0}$
defines the deformation 
$\beta_{\lambda}$ 
($\lambda=2$ for quadrupole and $\lambda=3$ for octupole), 
through the relation  
$\beta_{\lambda}=\sqrt{4 \pi (2\lambda +1)}Q_{\lambda 0}/(3 R_{0}^{\lambda} A)$, with $R_0=1.2 A^{1/3}$ fm. 
The constrained  Gogny-D1M calculations 
provide a set of HFB states 
$\{\ket{\Phi_\mathrm{HFB}(\beta_{2},\beta_{3})}\}$, 
labeled by their static quadrupole 
$\beta_{2}$ and octupole $\beta_{3}$ deformations. The 
corresponding SCMF energies $E_\mathrm{HFB}(\beta_{2},\beta_{3})$ define the 
SCMF-PESs. 
Note that, since the interaction is reflection-symmetry invariant, the HFB energies satisfy the property 
$E_\mathrm{HFB}(\beta_{2},\beta_{3}) = E_\mathrm{HFB}(\beta_{2},-\beta_{3})$, and therefore
only positive $\beta_{3}$ values are considered in the plots and the discussion. The Gogny-D1M  SCMF-PES 
is subsequently 
mapped onto the $sdf$-IBM Hamiltonian via the  
procedure briefly described below.

%-----------------------------------------------------------------------
%
% 	Gogny PESs
%
%-----------------------------------------------------------------------
\begin{figure*}[htb!]
\begin{center}
\includegraphics[width=\linewidth]{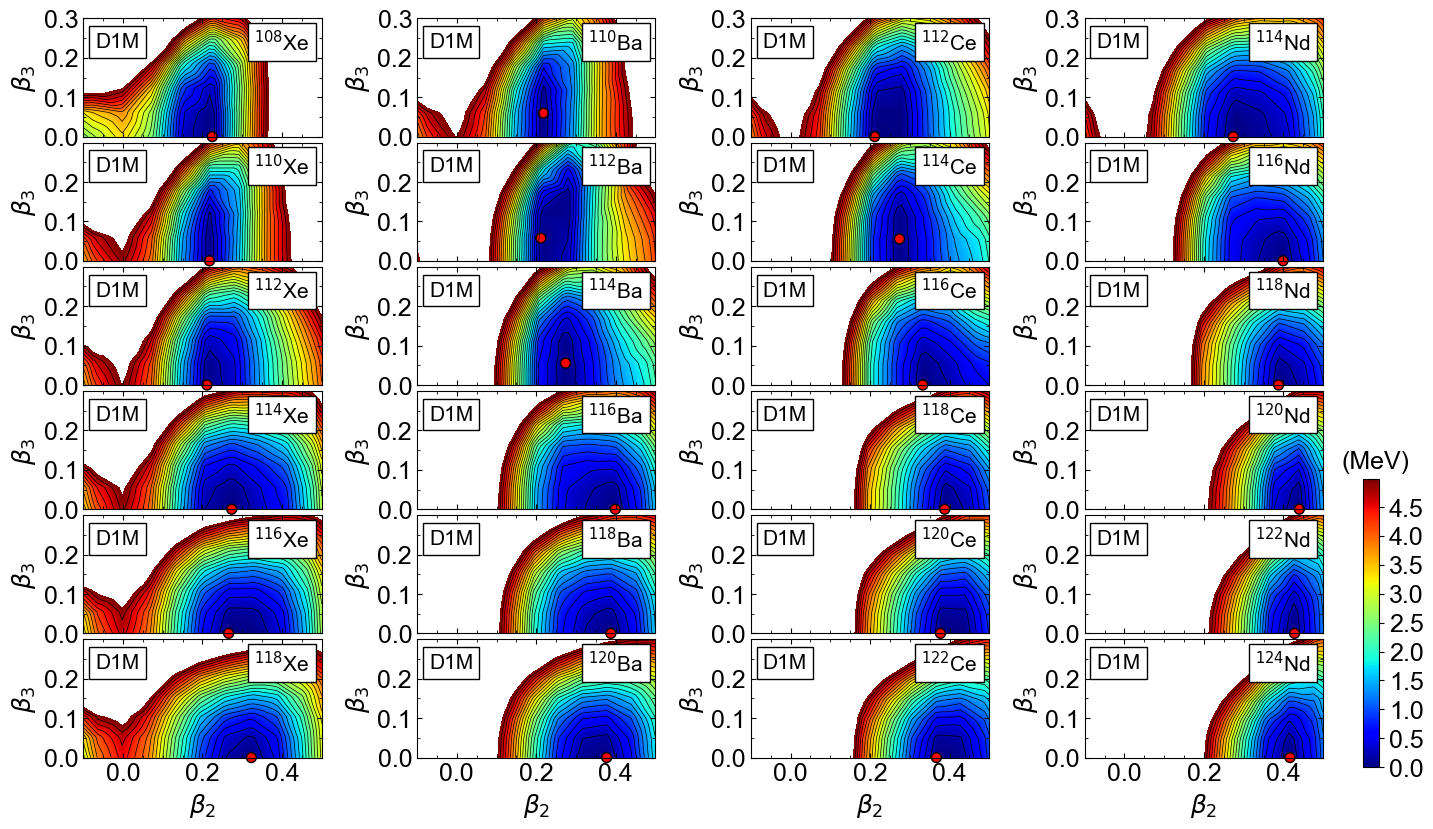}
\caption{SCMF-PESs, as functions of the 
quadrupole $\beta_{2}$ and octupole $\beta_{3}$ deformations, for 
$^{108-118}$Xe, $^{110-120}$Ba, $^{112-122}$Ce, 
and $^{114-124}$Nd. 
The color code indicates the total HFB energies (in MeV) 
plotted up to 5 MeV with respect to the global minimum. The 
energy
 difference between neighboring contours is 0.2 MeV. For each
 nucleus, the global minimum is
 indicated by a red solid circle. Results have been obtained with the 
 Gogny-D1M EDF.} 
\label{fig:pesdft}
\end{center}
\end{figure*}
%-----------------------------------------------------------------------
%
% 	Mapped PESs
%
%-----------------------------------------------------------------------
\begin{figure*}[htb!]
\begin{center}
\includegraphics[width=\linewidth]{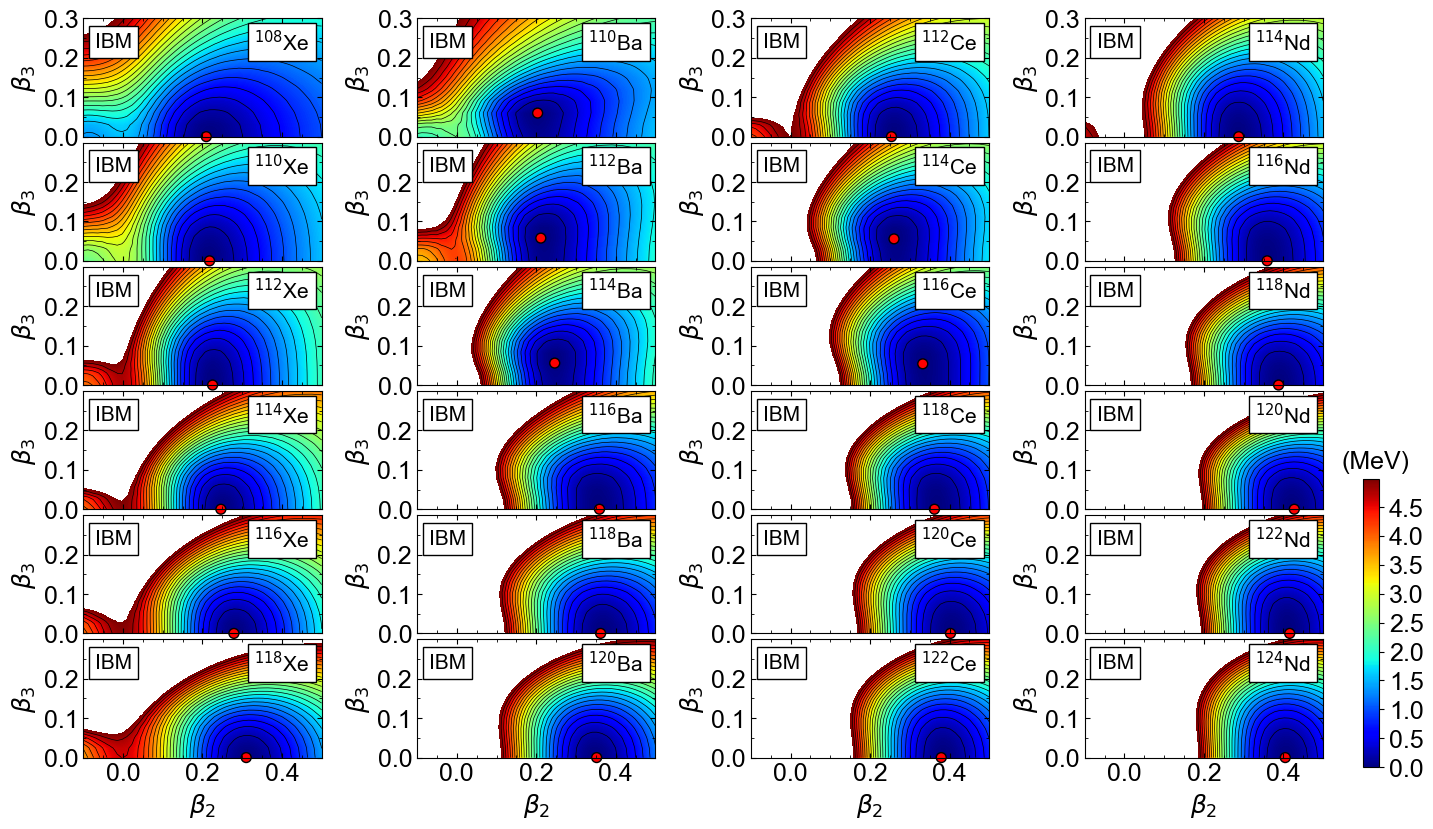}
\caption{The same as in Fig.~\ref{fig:pesdft}, 
but for the mapped IBM-PESs.} 
\label{fig:pesibm}
\end{center}
\end{figure*}

From a microscopic point of view, IBM bosons 
represent collective pairs of valence nucleons 
\cite{OAIT,OAI,IBM}. In principle, one should consider both 
neutron and proton bosons, which correspond to 
neutron-neutron 
and proton-proton pairs, respectively, within 
the framework of the proton-neutron IBM (IBM-2) \cite{OAI}.
However,  in order to keep 
our approach as simple as possible, we have employed 
the simpler IBM-1 framework, which does not make 
a distinction between proton and neutron boson 
degrees of freedom.

Within the standard IBM framework, neutron-proton 
pairs are not included. In medium-heavy and heavy nuclei, 
to which the IBM has 
been mainly applied, neutrons and protons occupy different 
major oscillator shells and, therefore, the contribution of the 
neutron-proton coupling is expected to be negligible. On the other
hand, for $N\approx Z$ nuclei
in addition to 
the neutron-neutron and proton-proton pairs  
the neutron-proton pairs should be considered, since, 
in this case, both protons 
and neutrons can occupy similar orbits. However, this 
would require the introduction of
additional isospin degrees of freedom 
in the IBM \cite{ELLIOT1980,ELLIOT1981}. This is probably the 
reason why the IBM has rarely been applied to $N\approx Z$ nuclei. 
For the same reason, the $sdf$-IBM framework employed in the present 
study does not consider the neutron-proton pairs.

The total number of bosons $n=n_{s}+n_{d}+n_{f}$ 
is equal to half the number of valence nucleons within the neutron 
and proton $N,Z=50-82$ major shells, and is conserved for a given 
nucleus. 
In the majority of the $sdf$-IBM phenomenology, certain truncation 
of the boson model space has been considered, that is, 
the maximum number of $f$ bosons $n_{f}^{\mathrm{max}}$ is 
limited to $n_{f}^{\mathrm{max}}=1$ or $n_{f}^{\mathrm{max}}=3$. 
In the present study, we do not make such an assumption, 
but allow  $n_{f}$ to vary between zero and $n$.

The $sdf$-IBM Hamiltonian employed in the present 
study has the form \cite{nomura2020oct,nomura2021oct-u}:
\begin{align}
\label{eq:ham}
\hat H= 
\epsilon_d\hat n_{d} + \epsilon_{f}\hat{n}_{f} 
+ \kappa_{2}\hat{Q}_{2}\cdot\hat{Q}_{2} + \rho\hat{L}\cdot\hat{L} 
+ \kappa_{3}\hat{Q}_{3}\cdot\hat{Q}_{3}. 
\end{align}
The first (second) term on the right-hand side 
represents the number operator for
the $d$ ($f$) bosons with $\epsilon_{d}$ ($\epsilon_{f}$), 
standing for the single
$d$ ($f$) boson energy relative to the $s$ boson one. 
The third, fourth  and fifth terms represent the quadrupole-quadrupole
interaction, the rotational term, and the octupole-octupole 
interaction, respectively. 
The quadrupole $\hat{Q}_{2}$, angular momentum $\hat{L}$, 
and octupole $\hat{Q}_{3}$
operators read
\begin{subequations}
 \begin{align}
\label{eq:q2}
& \hat Q_{2}=s^{\dagger}\tilde d+d^{\dagger}\tilde s+\chi_{d}(d^{\dagger}\tilde
  d)^{(2)}+\chi_{f}(f^{\dagger}\tilde f)^{(2)} \\
\label{eq:l}
& \hat{L}=
\sqrt{10}(d^{\dagger}\tilde{d})^{(1)}
+\sqrt{28}(f^\+\tilde{f})^{(1)} \\
\label{eq:q3}
&\hat{Q}_{3}=
s^{\dagger}\tilde{f}+f^{\dagger}\tilde{s}
+\chi_{3}(d^{\dagger}\tilde{f}
+f^{\dagger}\tilde{d})^{(3)}. 
\end{align}
\end{subequations}
Note that the term proportional to
$(d^{\+}\tilde{d})^{(1)}\cdot (f^{\+}\tilde{f})^{(1)}$ in the
$\hat{L}\cdot\hat{L}$ term has been neglected \cite{nomura2020oct}. 
Exception made of $\rho$, all the parameters 
of the $sdf$-IBM Hamiltonian 
are determined, for each nucleus, by mapping 
the SCMF-PES onto the 
corresponding IBM-PES \cite{nomura2015,nomura2020oct}.
This requires the approximate equality 
$E_\mathrm{HFB}(\beta_2,\beta_3)\approx E_\mathrm{IBM}(\beta_2,\beta_3)$ 
to be satisfied in the neighborhood of the global minimum. 
The IBM-PES is defined as 
the expectation value of the $sdf$-IBM Hamiltonian 
in the boson condensate state \cite{ginocchio1980} wave function
$\ket{n,\beta_{2},\beta_{3}}$, i.e., 
$E_\mathrm{IBM}(\beta_2,\beta_3)=
\bra{n,\beta_{2},\beta_{3}}\hat{H}\ket{n,\beta_{2},\beta_{3}}$, 
where
\begin{subequations}
 \label{eq:coherent}
 \begin{align}
|n,\beta_{2},\beta_{3}\rangle=
(n!)^{-1/2}(b_{c}^{\+})^{n}
\ket{0}
 \end{align}
with
 \begin{align}
b_{c}^{\+}=(1+{\bar\beta_{2}}^{2}+{\bar\beta_{3}}^{2})^{-1/2}
(s^\+ +\bar\beta_{2}d_0^\+ + \bar\beta_{3}f_{0}^{\+}). 
\end{align}
\end{subequations}
The ket $\ket{0}$ denotes the boson vacuum, or inert core. 
The doubly-magic nucleus $^{100}$Sn is taken 
as the inert core in the present study, hence 
$n=(A-100)/2$ for a nucleus with mass $A$. 
The amplitudes $\bar\beta_{2}$ and $\bar\beta_{3}$ entering the 
definition of the boson condensate wave function are proportional 
to the $\beta_{2}$ and $\beta_{3}$ deformations of the
fermionic space, $\bar\beta_{2}=C_2\beta_2$ and
$\bar\beta_{3}=C_3\beta_3$ 
\cite{ginocchio1980,nomura2014,nomura2015}, 
with dimensionless proportionality 
constants  $C_2$ and $C_3$. 
Their values are also determined by the
mapping procedure, 
so that the location of the global minimum in the SCMF-PES  
is reproduced. 
Finally, the parameter $\rho$ is fixed  
\cite{nomura2011rot} by equating the cranking 
moment of inertia 
obtained in the intrinsic frame of the IBM 
\cite{schaaser1986} at the global minimum 
to the corresponding 
Thouless-Valatin (TV) value \cite{TV} 
computed with the Gogny-HFB cranking method.  
For a more detailed description of the whole procedure 
the reader is referred to 
Ref.~\cite{nomura2020oct}. 
For the numerical diagonalization of the 
mapped Hamiltonian $\hat{H}$ (\ref{eq:ham}), 
we use the computer code \textsc{arbmodel} \cite{arbmodel}. 

%-----------------------------------------------------------------------
%
% 	PARAMETERS
%
%-----------------------------------------------------------------------
\begin{figure}[htb!]
\begin{center}
\includegraphics[width=\linewidth]{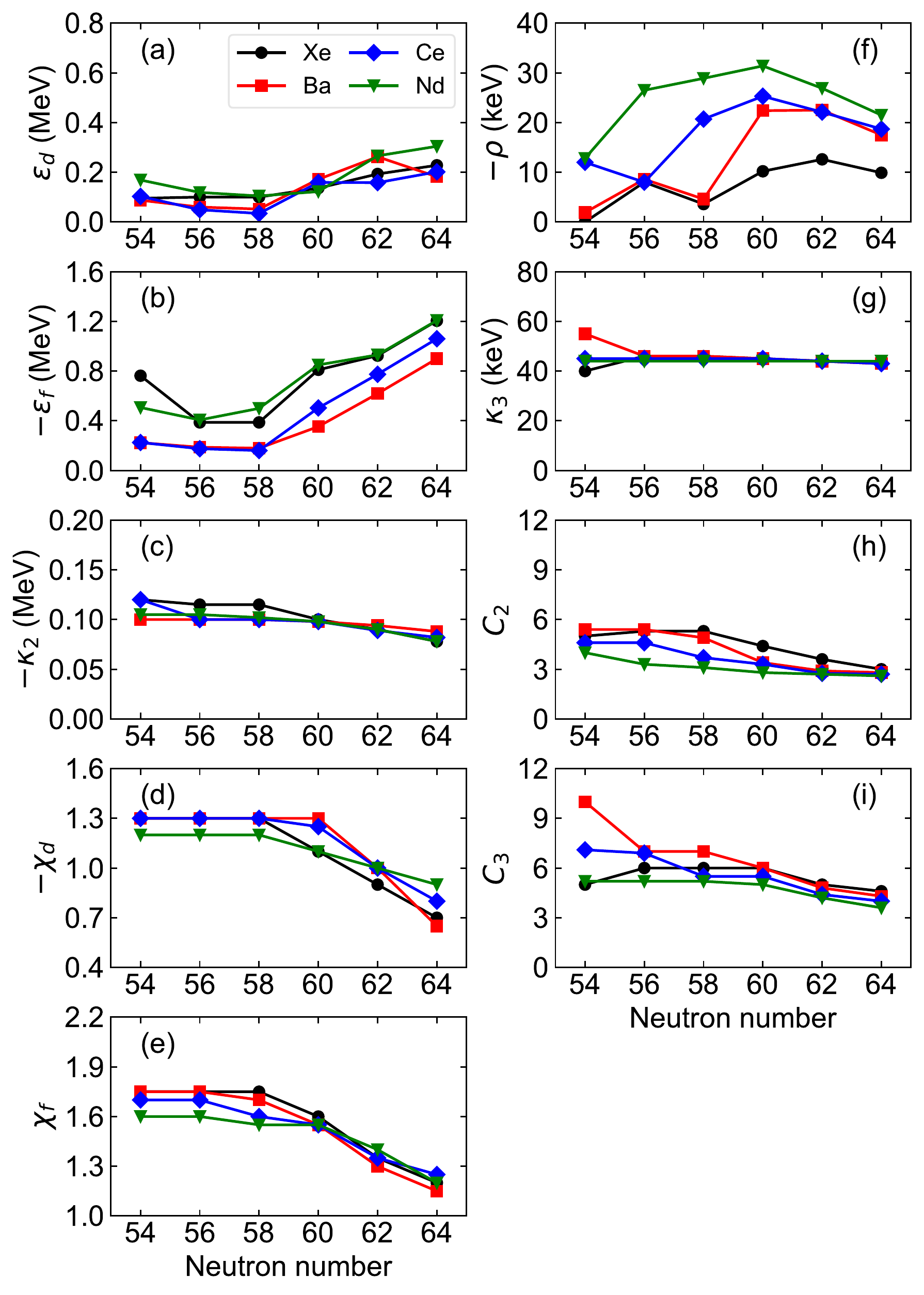}
\caption{Strength parameters 
(a) $\epsilon_d$, (b) $\epsilon_f$, (c) $\kappa_2$, 
(d) $\chi_{d}$, (e) $\chi_{f}$, (f) $\rho$, and 
(g) $\kappa_{3}$, of the $sdf$-IBM 
Hamiltonian (\ref{eq:ham}), and the coefficients 
(h) $C_{2}$ and (i) $C_{3}$. 
for the studied isotopic chains. For more details, see
the main text.
}
\label{fig:parameter}
\end{center}
\end{figure}

%-----------------------------------------------------------------------
%\section{Results and discussions\label{sec:results}}
%-----------------------------------------------------------------------

%-----------------------------------------------------------------------
\section{Potential energy surfaces\label{sec:pes}}
%-----------------------------------------------------------------------

The Gogny-D1M SCMF-PESs obtained for the 
studied Xe, Ba, Ce, and Nd nuclei 
are depicted in Fig.~\ref{fig:pesdft}.
A shallow 
octupole-deformed  
minimum is observed for the 
$N \approx 56$ nuclei $^{110}$Ba, $^{112}$Ba, 
$^{114}$Ba, and $^{114}$Ce. On the other
hand, the ground states of all the considered
Xe and Nd nuclei are reflection-symmetric.
Nevertheless, especially for $^{108}$Xe, $^{110}$Xe and 
$^{112}$Xe the SCMF-PESs are rather soft along the
 $\beta_{3}$ direction. 
For  $N\gg56$, the octupole minimum 
disappears in all the 
isotopic chains while well quadrupole-deformed 
ground states emerge. The SCMF-PESs 
resulting from the Gogny-HFB calculations 
are qualitatively similar to the ones 
obtained from the constrained 
relativistic mean-field calculations 
based 
on the DD-PC1 EDF \cite{nomura2021qoch}. The major
difference is that the latter predict $\omin=0$ for $^{110}$Ba.

The  mapped IBM-PESs are plotted in 
Fig.~\ref{fig:pesibm}. They reproduce the overall 
features of the SCMF-PESs, such as the location of the 
global minimum  and the 
softness along the  $\beta_{3}$-direction. 
In comparison to the SCMF-PESs, the IBM ones
are flat especially in  regions 
corresponding to large $\beta_{2}$ and $\beta_{3}$ 
values that are far from the global minimum. 
This is a common feature within the IBM framework 
arising from the fact that the IBM consists of valence 
nucleon pairs in one major shell, while the SCMF model 
involves all nucleon degrees of freedom. 
For a detailed account of this problem, the reader is 
referred to Ref.~\cite{nomura2021oct-u}.

The parameters for the Hamiltonian (\ref{eq:ham}), 
determined by the mapping procedure, 
are shown in Fig.~\ref{fig:parameter}. 
Each parameter exhibits a weak dependence on the 
neutron number $N$ and, in some cases, is almost constant. 
For most of the parameters, there is no striking 
difference in their values and  $N$-dependence 
for the considered isotopic chains. 
For the sake of simplicity, the parameters 
$\chi_{f}$ and $\chi_{3}$ 
are assumed to have the same magnitude 
$\chi_{3}=-\chi_{f}$, 
and only the $\chi_{f}$ value is plotted 
in panel (e).

%-----------------------------------------------------------------------
%
%	EXCITATION SPECTRA
%
%-----------------------------------------------------------------------

\begin{figure}[htb!]
\begin{center}
\includegraphics[width=\linewidth]{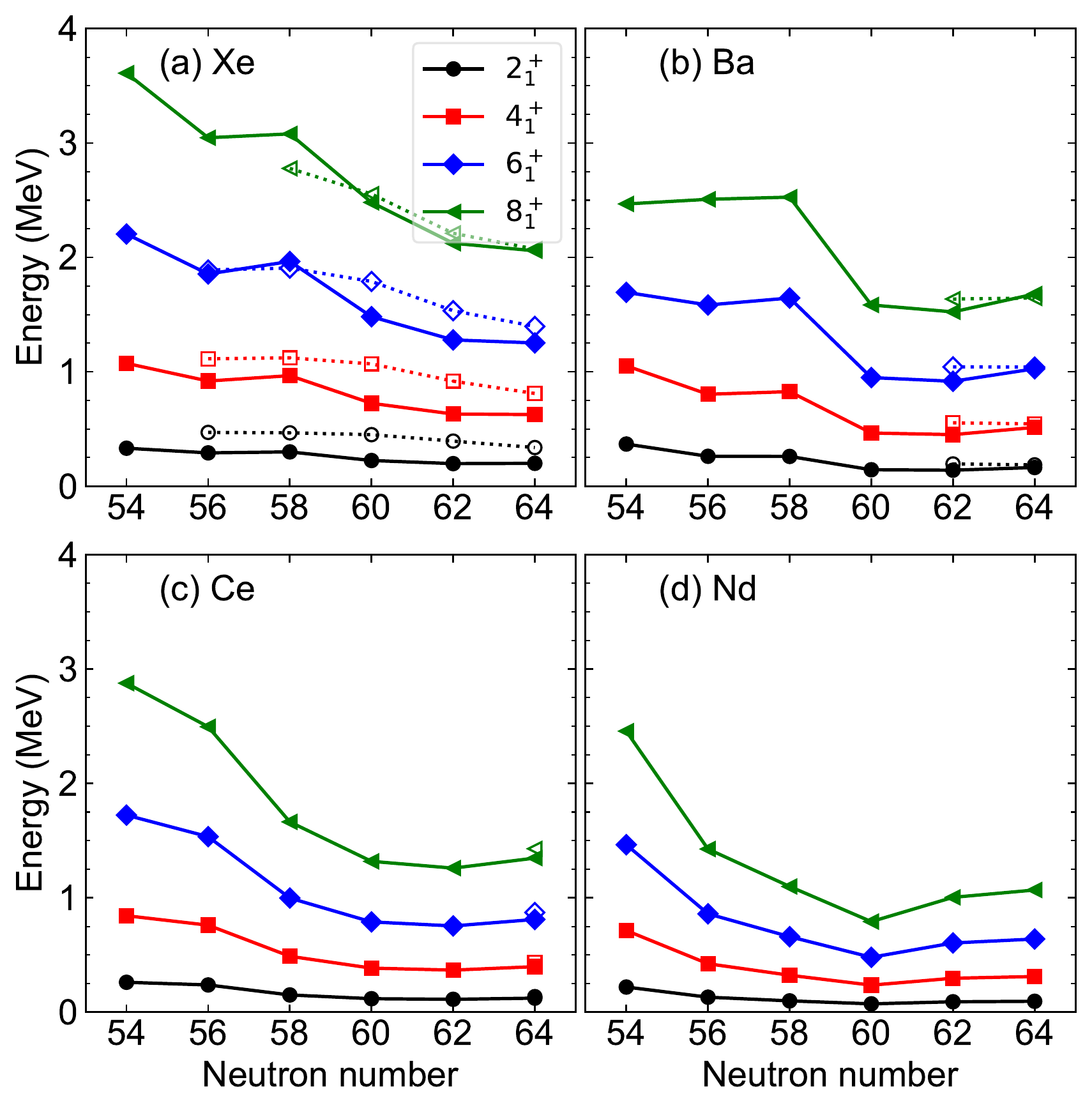}
\caption{The low-energy excitation spectra of 
positive-parity ($\pi=+1$) even-spin yrast states  obtained 
for $^{108-118}$Xe, $^{110-120}$Ba, $^{112-122}$Ce, 
and $^{114-124}$Nd are compared with the available
experimental data \cite{data}. 
Theoretical (experimental) values are represented by 
filled (open) symbols connected by 
solid (dotted) lines. 
} 
\label{fig:level-pos}
\end{center}
\end{figure}
%-----------------------------------------------------------------------
%
%	EXCITATION SPECTRA
%
%-----------------------------------------------------------------------
\begin{figure}[htb!]
\begin{center}
\includegraphics[width=\linewidth]{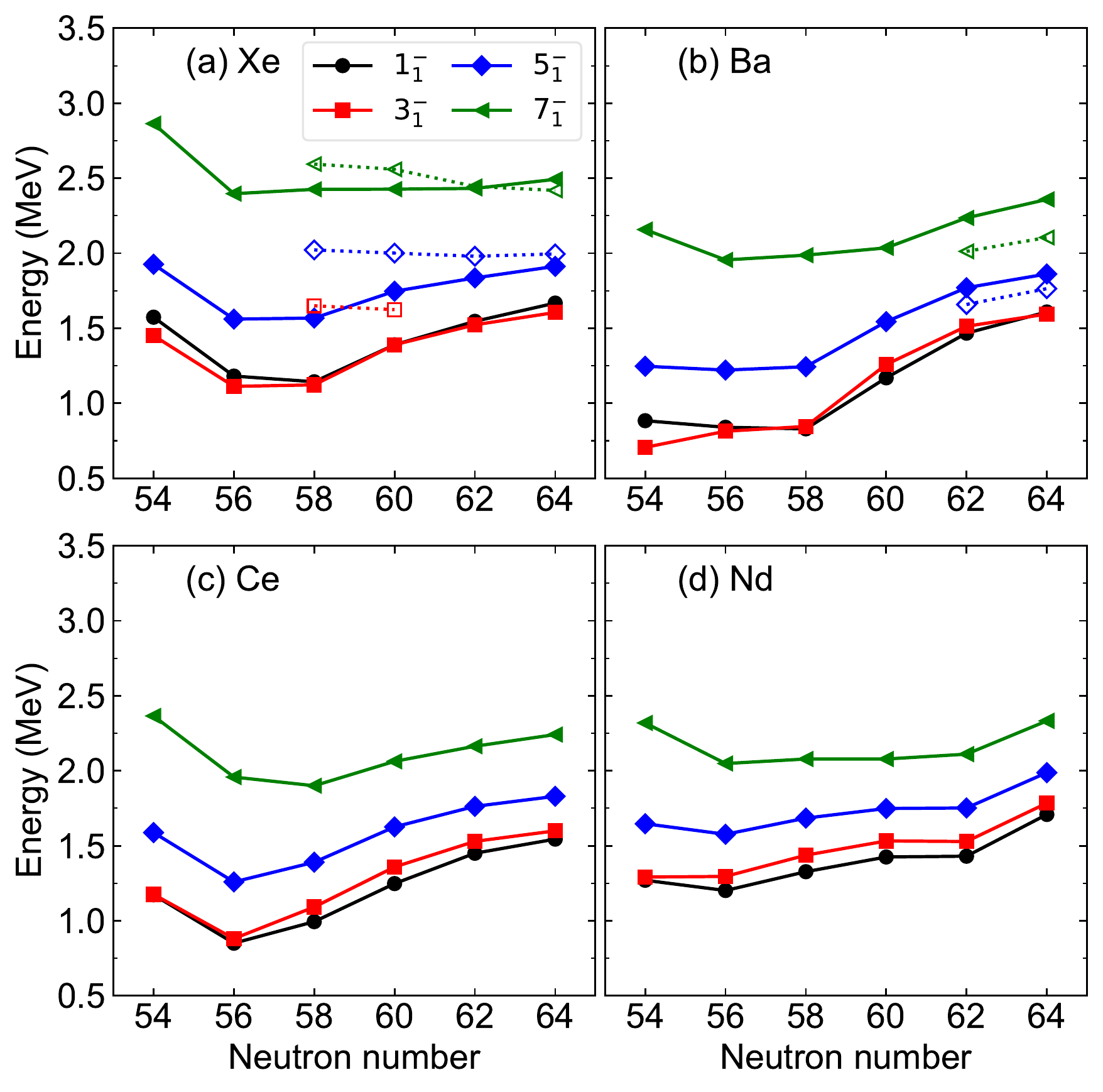}
\caption{The same as in  Fig.~\ref{fig:level-pos}, 
but for the negative-parity ($\pi=-1$) odd-spin states.} 
\label{fig:level-neg}
\end{center}
\end{figure}
%-----------------------------------------------------------------------
%
%	EXCITATION SPECTRA
%
%-----------------------------------------------------------------------
\begin{figure}[htb!]
\begin{center}
\includegraphics[width=\linewidth]{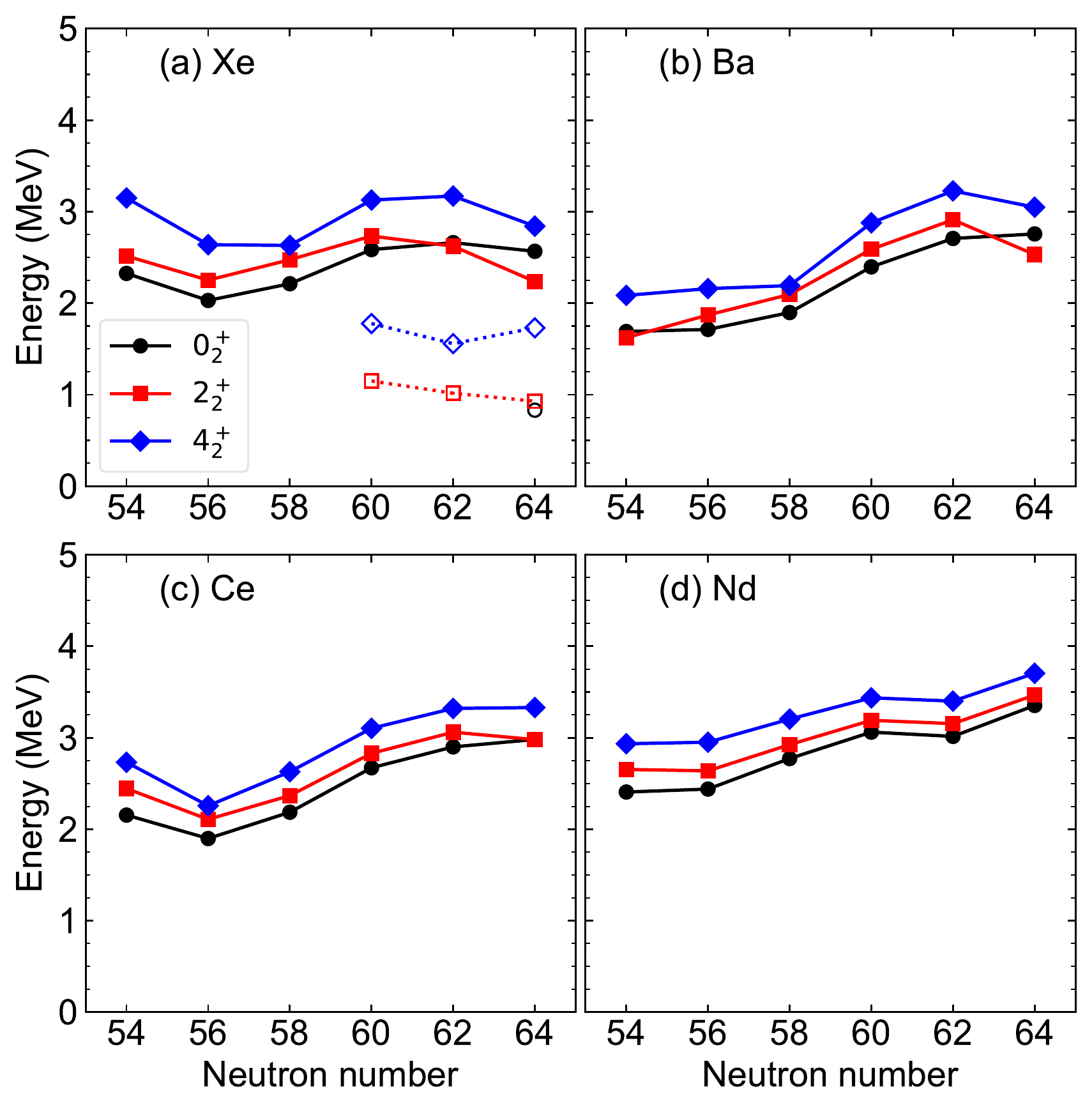}
\caption{The same as in  Fig.~\ref{fig:level-pos}, 
but for the quasi-$\beta$ band states.}
\label{fig:level-beta}
\end{center}
\end{figure}
%-----------------------------------------------------------------------
%
%	EXCITATION SPECTRA
%
%-----------------------------------------------------------------------
\begin{figure}[htb!]
\begin{center}
\includegraphics[width=\linewidth]{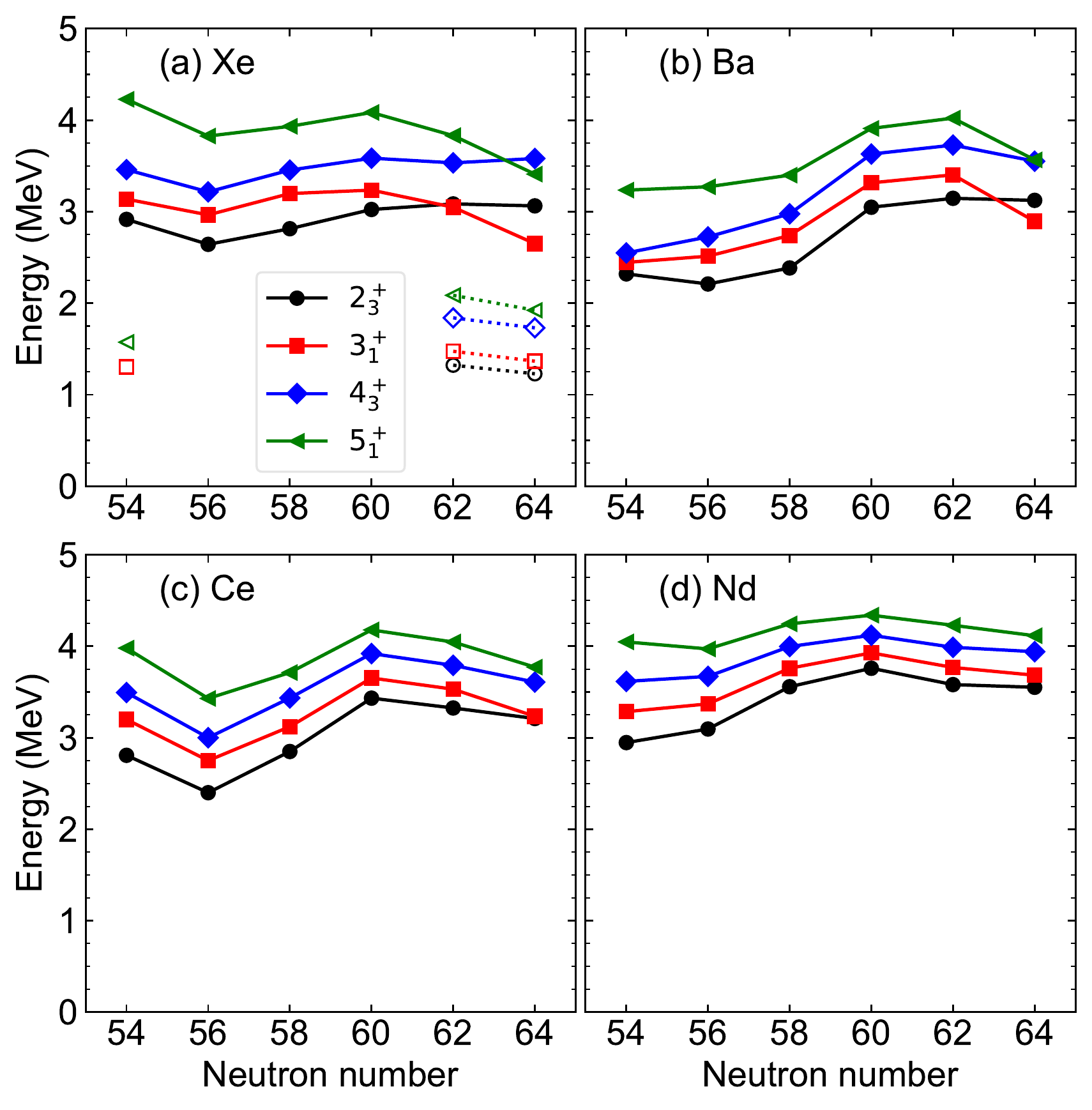}
\caption{The same as in Fig.~\ref{fig:level-pos}, 
but for the quasi-$\gamma$ band states.}
\label{fig:level-gamma}
\end{center}
\end{figure}
%-----------------------------------------------------------------------
%
%	F-BOSON CONTENT
%
%-----------------------------------------------------------------------
\begin{figure}[htb!]
\begin{center}
\includegraphics[width=\linewidth]{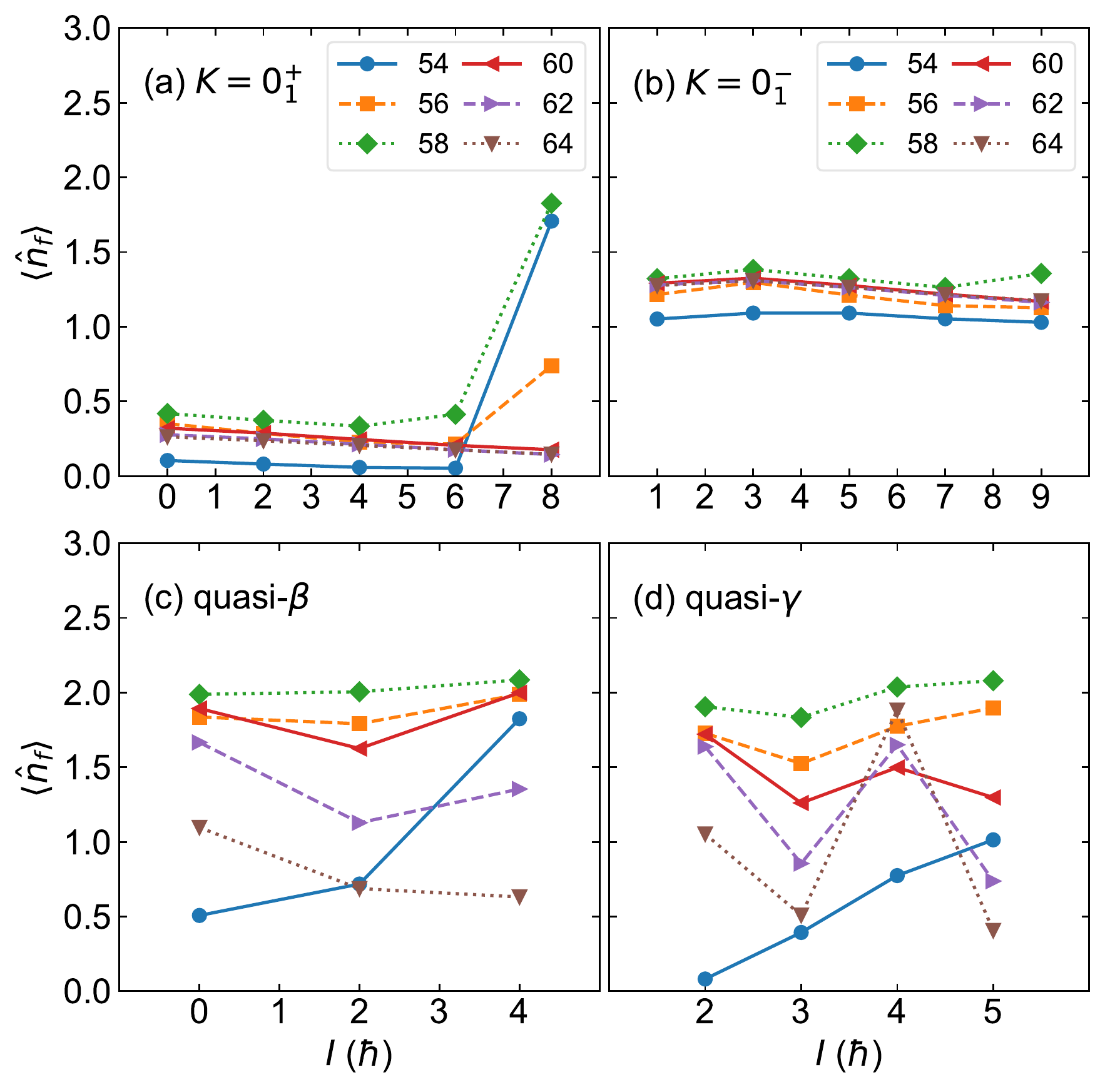}
\caption{The expectation values 
of the $f$-boson number operator 
$\braket{\hat{n_{f}}}$ in the IBM wave functions
corresponding to the $K=0^{+}_{1}$(a) and 
 $K=0^{-}_{1}$ (b) states, the  quasi-$\beta$ (c) and  
quasi-$\gamma$ (d) bands of Xe isotopes.} 
\label{fig:nf}
\end{center}
\end{figure}

%-----------------------------------------------------------------------
\section{Results for spectroscopic properties\label{sec:spec}}
%-----------------------------------------------------------------------

\subsection{Excitation energies for yrast states}

The low-energy excitation 
spectra corresponding to the 
$\pi=+1$ even-spin yrast states in 
$^{108-118}$Xe, $^{110-120}$Ba, $^{112-122}$Ce, 
and $^{114-124}$Nd
are 
plotted in Fig.~\ref{fig:level-pos}. As can be seen
from the figure, the $\pi=+1$ excitation energies decrease 
with increasing neutron number.
The calculated $\pi=+1$ ground-state ($K=0^{+}_{1}$) band 
for those Xe nuclei with $N\geqslant60$ appear to be 
more compressed than the experimental ones. 
The  $\pi=+1$  spectra for  Ba and Ce nuclei 
with $N\geqslant62$ agree reasonably well with the 
experiment. For Xe and Ba isotopes, the predicted 
excitation energies exhibit a pronounced 
decrease from $N=58$ to 60, 
suggesting the onset of a pronounced quadrupole 
collectivity. This correlates well with the 
features observed in the corresponding 
Gogny-D1M SCMF-PESs in Fig.~\ref{fig:pesdft}.
Note, that the SCMF-PESs become rather 
 soft along the $\beta_{2}$-direction 
for $N\geqslant60$.

The spectra corresponding to the $\pi=-1$ odd-spin yrast states
are depicted in Fig.~\ref{fig:level-neg}. The predicted 
excitation energies 
exhibit 
a weak dependence with neutron number, reaching 
a minimal value around $N=56$. 
This tendency reflects that for that neutron number the 
SCMF-PESs exhibit an octupole-deformed 
minimum 
or are notably soft along the $\beta_{3}$-direction (see Fig.~\ref{fig:pesdft}). 
The mapped $sdf$-IBM predicts $\pi=-1$ levels  lower in 
energy than the experimental ones. Similar results
have been obtained in Ref.~\cite{nomura2021qoch} 
for the odd-spin $\pi=-1$ band centered 
around $N=56$. However, the $\pi=-1$  excitation
energies obtained in those calculations 
are higher than the 
present results.

\subsection{Excitation energies for non-yrast states}

The excitation energies of the non-yrast states $0^{+}_{2}$, 
$2^{+}_{2}$, and $4^{+}_{2}$ are shown in 
Fig.~\ref{fig:level-beta}. In most cases,  
these states 
are the lowest-spin members of the quasi-$\beta$ band, 
interconnected by strong $E2$ transitions. For 
$^{116}$Xe, $^{118}$Xe, $^{120}$Ba 
and $^{122}$Ce, the predicted quasi-$\beta$ band is comprised of the 
$0^{+}_{2}$, $2^{+}_{3}$, and $4^{+}_{3}$ states. This explains 
the inversion of the $0^{+}_{2}$ and $2^{+}_{2}$ 
levels in  Fig.~\ref{fig:level-beta} 
for these particular nuclei.
The predicted quasi-$\beta$  states 
exhibit a weak parabolic dependence 
as functions of $N$, with a minimum around $N=56$.
The excitation energy of the band-head state 
$0^{+}_{2}$ is systematically high (above 2 MeV excitation 
from the ground state), overestimating the experimental values for 
$^{114}$Xe, $^{116}$Xe, and $^{118}$Xe  
by a factor of two. 
The mapped IBM procedure often yields excitation energies of 
non-yrast states higher than the experimental ones.
The discrepancy suggest that some of the values obtained for the
Hamiltonian 
parameters might not be reasonable. 
Specifically, for the quadrupole-quadrupole boson interaction strength 
we have obtained 
$\kappa_{2} \approx -0.1$ MeV (see Fig.~\ref{fig:parameter}(c)), 
while 
purely phenomenological IBM calculations 
(see, for example, \cite{muellergatermann2020} for $^{118}$Xe) 
usually employ a value for this parameter that is an order of magnitude 
smaller. 
%
%For most nuclei with $N\leqslant60$, the 
%parameter $\chi_{d}$ has a value 
%close to the SU(3) limit of the IBM 
%($\chi_\textnormal{SU(3)}=-1.32$) \cite{IBM}, and starts 
%to decrease in magnide for $N>60$ 
%(see Fig.~\ref{fig:parameter}(e)). 
%The derived $\chi_{d}$ values are consistent with the average 
%values of the $\chi_{\nu}$ and $\chi_{\pi}$ parameters for the 
%neutron and proton boson quadrupole operators, respectively, 
%employed in the earlier 
%phenomenological IBM-2 calculation in the same mass region 
%of Ref.~\cite{puddu1980}. 
%The more recent phenomenological IBM fit of Ref.~\cite{muellergatermann2020} 
%for $^{118}$Xe finds the value $\chi_{d}=-0.69$, 
%which is in agreement with our value for the same nucleus. 
%
In the mapped $sdf$-IBM framework, the large magnitude of the 
derived $\kappa_{2}$ parameter often leads to 
non-yrast $\pi=+1$ bands lying quite high in energy with respect to the 
ground-state band \cite{nomura2021oct-ba}. 
The calculations \cite{puddu1980,muellergatermann2020} 
with the parameters fitted to experimental data, 
on the other hand, 
generally reproduced the observed quasi-$\beta$ 
and quasi-$\gamma$ bands quite nicely. 
The values of the derived IBM parameters, however, reflect the topology 
of the SCMF-PES. Many of the available EDFs 
yield SCMF-PESs with a steep valley along the $\beta_{2}$-direction 
around the global minimum.  This often requires to choose 
IBM parameters quite 
different from those in phenomenological studies.

The predicted excitation energies for the 
$2^{+}_{3}$, 
$3^{+}_{1}$, $4^{+}_{3}$, and $5^{+}_{1}$ states are 
shown in Fig.~\ref{fig:level-gamma}. Those states 
are  considered as members of the quasi-$\gamma$ band, 
exception made of $^{116}$Xe, $^{118}$Xe, $^{120}$Ba, 
and $^{122}$Ce for which 
the $2^{+}_{2}$ and $4^{+}_{2}$ are the states 
to be assigned as the even-$I$ members of the $\gamma$ band.
As with the quasi-$\beta$ band, the predicted 
energies display a parabolic trend centered around $N=56$, 
whereas the band-head 
energy is too high with respect to the 
yrast band.

\subsection{$f$-boson content of the bands}

We have analyzed the relevance of the octupole degree of freedom 
in the predicted bands. In particular,  
changes in the $f$-boson 
content in the bands with the neutron number can 
be considered as  signatures of shape/phase transitions involving
octupolarity. The expectation value of 
the $f$-boson number operator $\braket{\hat{n}_{f}}$
for states in the ground-state ($K=0^{+}_{1}$), 
lowest $\pi=-1$ ($K=0^{-}_{1}$),
quasi-$\beta$, 
and quasi-$\gamma$ bands is plotted in Fig.~\ref{fig:nf}. Results
are shown for Xe isotopes as illustrative examples. 

As seen from Fig.~\ref{fig:nf}(a), for
$I\leqslant6^{+}$ the   
members 
of the ground-state band  
are dominated by the positive-parity $s$ and $d$ bosons, 
while the contribution from the negative-parity $f$ boson 
is minor ($\braket{\hat{n}_{f}}<1$). In the case 
of transitional nuclei with 
$N\leqslant58$, the $f$-boson 
components start to dominate the higher-spin states 
with $I\geqslant8^{+}$. 
The odd-$I$ states in the $K=0^{-}_{1}$ band for all the Xe 
nuclei have expectation values 
which are typically within the range $1<\braket{\hat{n}_{f}}<1.5$ and therefore can be interpreted as being made of one $f$ boson 
coupled to the $sd$ bosons space.

In Fig.~\ref{fig:nf}(c), the structure of the
quasi-$\beta$ band, which includes the $0^{+}_{2}$, $2^{+}_{2}$, 
and $4^{+}_{2}$ states, substantially differs
from one nucleus to another. For the transitional nuclei 
with $56\leqslant N\leqslant60$, 
the quasi-$\beta$ band is considered to be of double-octupole-phonon 
nature with $\braket{\hat{n}_{f}}\approx2$. On the other hand, 
for well-quadrupole deformed nuclei with $N=62$ and $N=64$, 
the expectation value $\braket{\hat{n}_{f}}$ decreases
and the contribution from the $f$ boson becomes less important.
An irregularity is observed for $N=54$. However, in this case, the 
 number of bosons is only  $n=4$ and 
 the IBM description can be  
expected to be worse than for nuclei with a larger number of bosons. 
Similar observations can be made for the quasi-$\gamma$ band 
(Fig.~\ref{fig:nf}(d)), which is comprised of the $2^{+}_{3}$, 
$3^{+}_{1}$, $4^{+}_{3}$, and $5^{+}_{1}$ states.

%-----------------------------------------------------------------------
%
%	Level schemes
%
%-----------------------------------------------------------------------
\begin{figure}[htb!]
\begin{center}
\includegraphics[width=\linewidth]{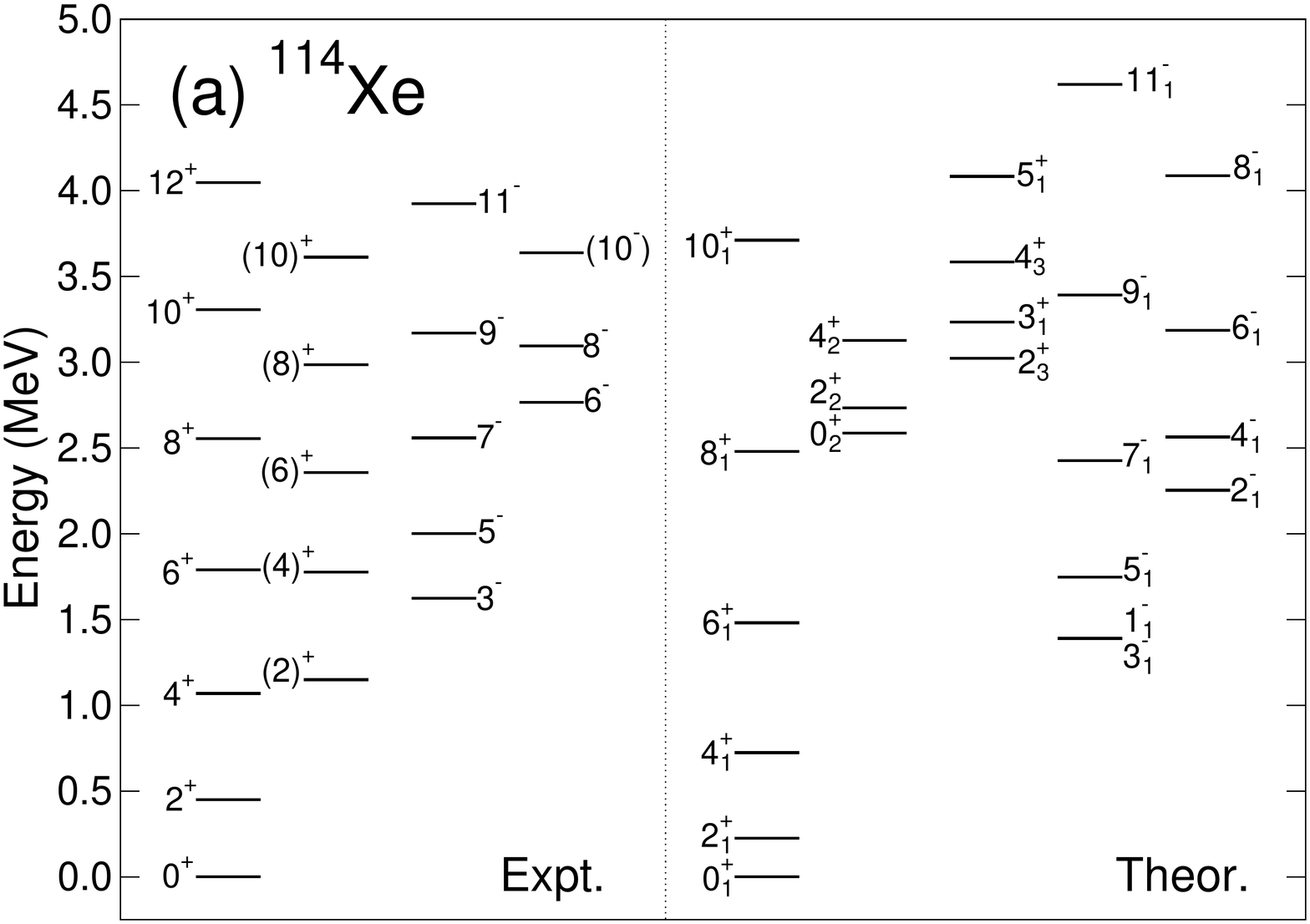}\\
\includegraphics[width=\linewidth]{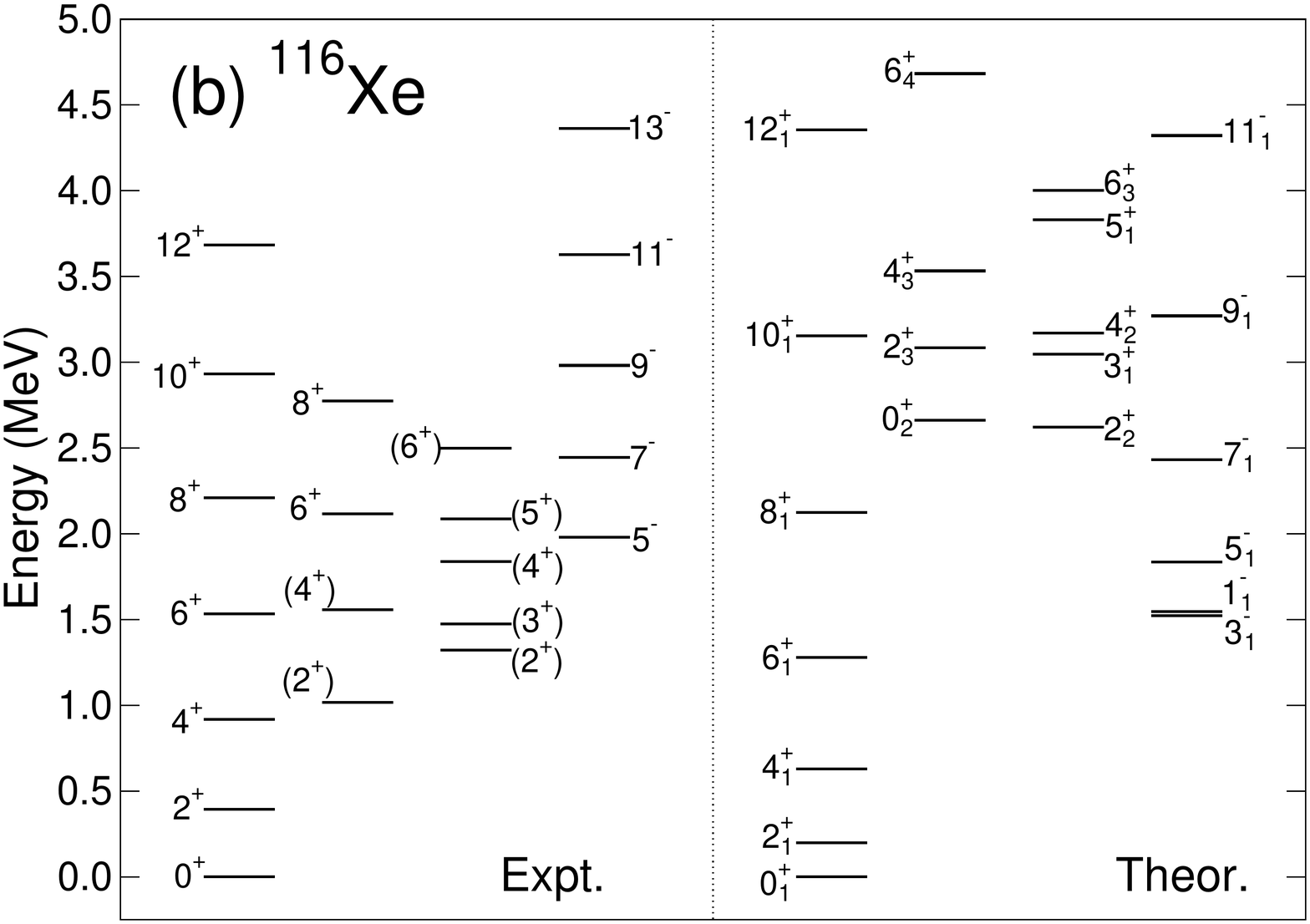}\\
\includegraphics[width=\linewidth]{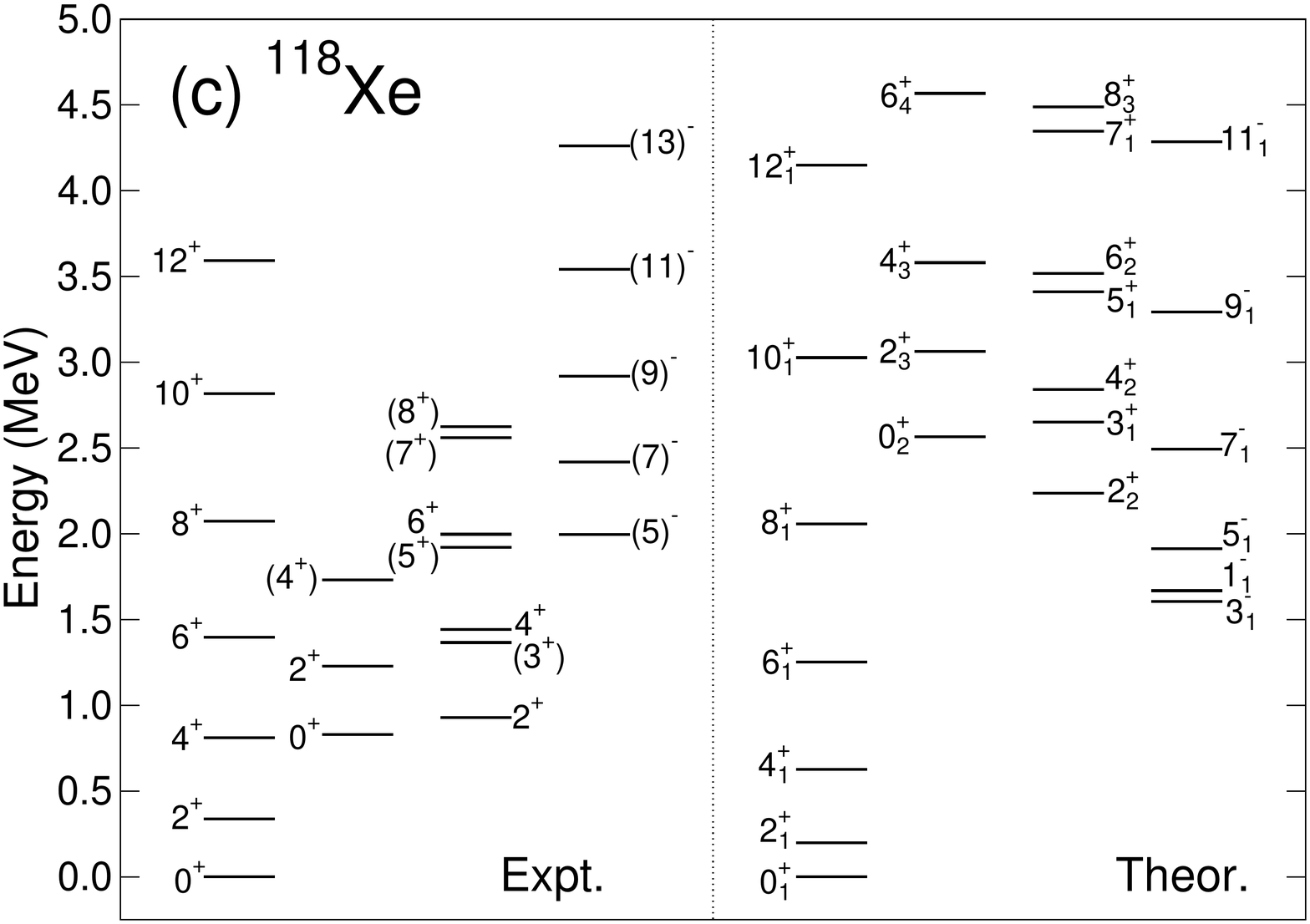}
\caption{Experimental and predicted band structures 
in $^{114}$Xe, $^{116}$Xe, and $^{118}$Xe. }
\label{fig:band}
\end{center}
\end{figure}

\subsection{Band structure of individual nuclei}

Experimental information is available for
$^{114}$Xe, $^{116}$Xe, and $^{118}$Xe nuclei. This  
allows to assess the quality of the model description. 
The  partial 
level schemes predicted for $^{114}$Xe, $^{116}$Xe, and $^{118}$Xe
are compared in Fig.~\ref{fig:band} with the available 
experimental
data \cite{data}. 

The nucleus $^{114}$Xe represents a transitional 
system between the nearly spherical and 
the strongly quadrupole deformed shapes (see, Fig.~\ref{fig:pesdft}).
As can be seen from Fig.~\ref{fig:band}(a) 
the mapped $sdf$-IBM calculations 
reproduce the lowest $\pi=\pm1$ ($K=0^{\pm}_{1}$) bands 
reasonably well. While the observed $\pi=+1$ spectra look 
harmonic, the theoretical ones 
exhibit rotational 
SU(3) features: the ground-state band resembles a rotational band with a 
moment of inertia larger than the experimental one;
both the quasi-$\beta$ and quasi-$\gamma$ bands are 
high in energy with respect to the ground-state band;
the moments of inertia for the ground-state, quasi-$\beta$, 
and quasi-$\gamma$ bands are almost equal to each other. 
Both empirically and theoretically, the $\pi=-1$ 
even-$I$ states are found above 2 MeV excitation 
from the ground state.

Within the Gogny-HFB framework $^{116}$Xe 
displays  a large quadrupole deformation 
(see Fig.~\ref{fig:pesdft}). 
The experimental band structure of $^{116}$Xe is 
similar to the band structure of  $^{114}$Xe. 
The mapped $sdf$-IBM reproduces the ground-state and lowest $\pi=-1$ 
($K=0^{-}_{1}$) bands, though they are 
stretched compared to the 
experimental bands. The band-heads of the 
quasi-$\beta$ and quasi-$\gamma$ bands have an 
excitation energy around 2.5 MeV.
In comparison with $^{114}$Xe, the predicted quasi-$\beta$ band 
looks more irregular, as seen, for example, from the large energy gap between 
the $4^{+}_{3}$ and $6^{+}_{4}$ levels. 
This is a consequence of the strong level repulsion between low-spin 
states due to a considerable amount of shape mixing. 
From the experimental point of view, the band built on 
the $2^{+}_{3}$ state is tentatively 
assigned to be quasi-$\gamma$ band.
Such a band exhibits 
the staggering pattern  
$(2^{+}_{\gamma},3^{+}_{\gamma}),(4^{+}_{\gamma},5^{+}_{\gamma}),\ldots$
and resembles the level structure predicted in the rigid-triaxial rotor 
model \cite{Davydov58}. In contrast, the quasi-$\gamma$ band 
obtained in the present calculation shows the 
staggering pattern $2^{+}_{\gamma}, (3^{+}_{\gamma},4^{+}_{\gamma}),
(5^{+}_{\gamma},6^{+}_{\gamma}),\ldots$  
characteristic of the $\gamma$-unstable-rotor picture \cite{gsoft}.

Figure~\ref{fig:band}(c) displays the results 
obtained for $^{118}$Xe. The
$K=0^{\pm}_{1}$ bands predicted within the mapped 
$sdf$-IBM approach agree well with the experimental data.
The $0^{+}_{2}$ state and the band 
built on it are known experimentally, with a 
band-head 
energy below 1 MeV. 
The quasi-$\gamma$ band with the staggering pattern, 
$2^{+}_{\gamma}, (3^{+}_{\gamma},4^{+}_{\gamma}),
(5^{+}_{\gamma},6^{+}_{\gamma}),\ldots$ is also known 
experimentally . As can be seen, the quasi-$\beta$ and quasi-$\gamma$ bands
are much higher than their experimental counterparts. 
Nevertheless, overall features of the bands, such as the moments of 
inertia and energy splitting 
members of the bands, agree well with the experiment.
Note, that the band-head energy of the predicted quasi-$\beta$ 
band is slightly lower than that of the quasi-$\gamma$ band. 

% herehere

At this point it is worth to make a few remarks on some of the
features of the predicted spectra shown 
in Fig.~\ref{fig:band}. 
First, the moments of inertia 
for the predicted ground-state bands at low spin  
are systematically larger than the experimental ones, 
with too low $2^{+}_{1}$ excitation energies 
and  $E(4^{+}_{1})/E(2^{+}_{1})$ ratios close to the 
rotor limit value $3.33$. The responsible for this
behavior is the TV moment of inertia obtained in the 
cranking calculation with the Gogny force. The cranking rotational band 
corresponds to  
a very good rotor and the moment of inertia is roughly a factor of two larger than the
experimental data. As the moment of inertia is inversely proportional to
the square of the pairing gap, the disagreement is probably a consequence of the missing
proton-neutron pairing present in $N\approx Z$ nuclei and not considered in
neither the cranking calculation nor the $sdf$-IBM Hamiltonian.
Second, for the three Xe nuclei
one observes almost degenerated
$1^{-}_{1}$ and $3^{-}_{1}$ energy levels corresponding to the $K=0^{-}_{1}$ band  
with the $3^{-}_{1}$ slightly below the $1^{-}_{1}$, which is at variance with 
the empirical trend of negative parity rotational bands. 
Such irregularity may suggest that  
there is strong configuration mixing in the low-spin 
$\pi=-1$ states. 
%in which considerable amount of the 
%$f$-boson components can contribute to the corresponding 
%wave functions (see, ). 
%The fact that the $1^{-}_{1}$ state always appears 
%above the $3^{-}_{1}$ state 
It could also reflect the lack of  
the dipole $p$ 
boson degree of freedom with spin and parity $1^{-}$ 
in our model space. Its inclusion could allow a 
more accurate description of the low-spin part of the 
$0^{-}_{1}$ band. 

%-----------------------------------------------------------------------
%
%	B(E1,2,3) systematics
%
%-----------------------------------------------------------------------
\begin{figure}[htb!]
\begin{center}
\includegraphics[width=.75\linewidth]{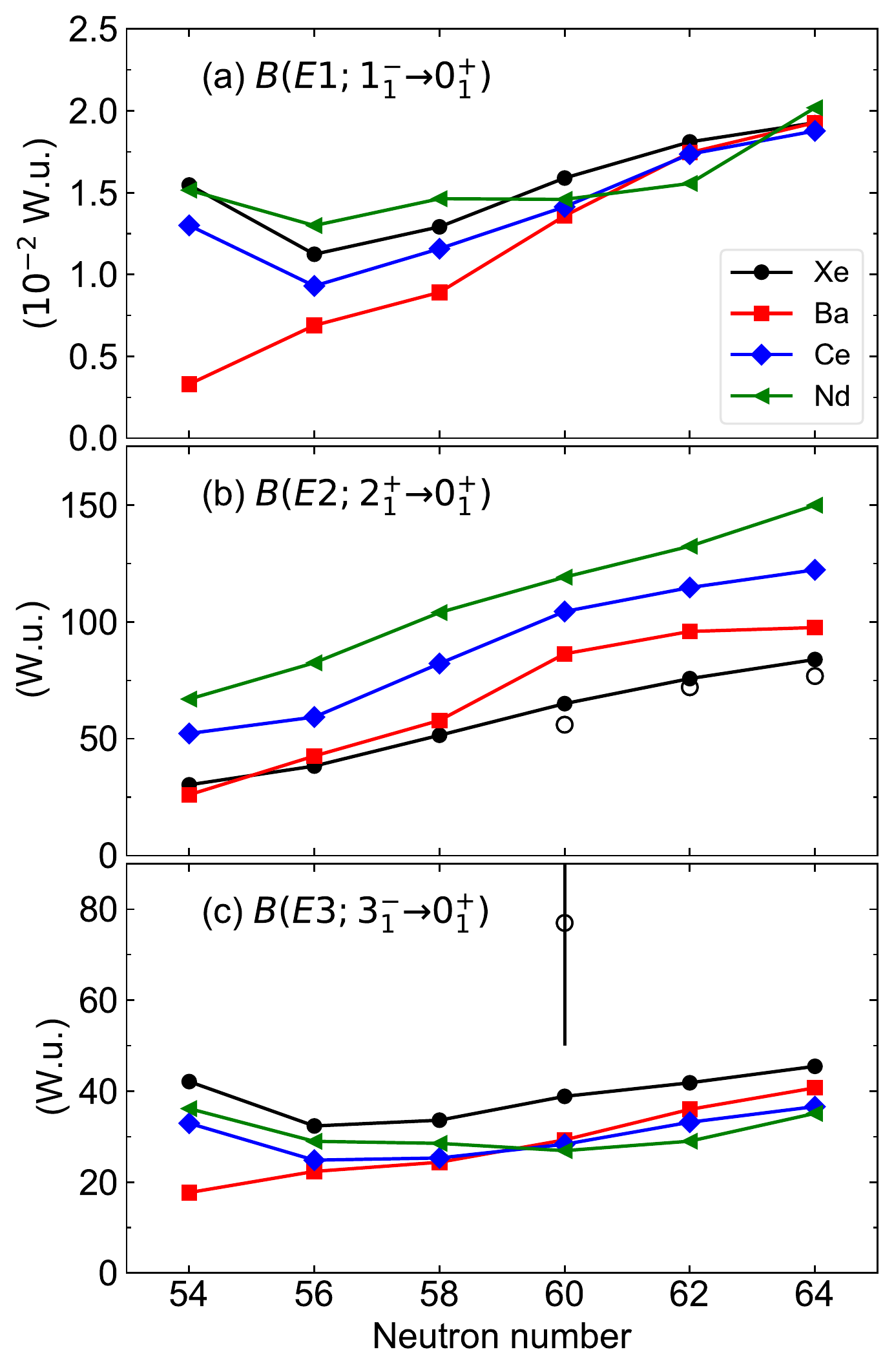}
\caption{
The calculated $B(E1;1^{-}_{1}\to 0^{+}_{1})$ (a), 
$B(E2;2^{+}_{1}\to 0^{+}_{1})$ (c), and 
$B(E3;3^{-}_{1}\to 0^{+}_{1})$ (e) 
reduced transition probabilities in Weisskopf units (W.u.), 
represented by the solid symbols connected by lines. 
Experimental data for the Xe nuclei 
are taken from \cite{data,degraaf1998,DEANGELIS2002,muellergatermann2020}, 
and are represented by the open symbols. 
Note that the error bars for the experimental 
$B(E2)$ rates of $^{114,116,118}$Xe are not 
shown, as they are 
smaller than the marker size.}
\label{fig:trans}
\end{center}
\end{figure}

%-----------------------------------------------------------------------
\subsection{Electric transition rates\label{sec:trans}}
%-----------------------------------------------------------------------

The electric dipole ($E1$), quadrupole ($E2$), and octupole $(E3)$ 
transition probabilities are computed  using the corresponding 
operators defined as   
$\hat{T}(E1)=e_{1}(d^{\+}\tilde{f}+f^{\+}\tilde{d})^{(1)}$, 
$\hat{T}(E2)=e_{2}\hat{Q}_{2}$, 
and $\hat{T}(E3)=e_{3}\hat{Q}_{3}$, respectively. 
The operators $\hat{Q}_{2}$ and
$\hat{Q}_{3}$ have the same forms and parameters 
as those  in Eqs.~(\ref{eq:q2}) and (\ref{eq:q3}). 
The boson effective $E1$ and $E3$ charges $e_{1}=0.01$ 
$e\,$b$^{1/2}$ and $e_{3}=0.12$ 
$e\,$b$^{3/2}$ are adopted from our previous study 
on the neutron-rich Ba region \cite{nomura2014,nomura2021oct-ba}. 
The $E2$ boson charge $e_{2}=0.107$ $e\,$b is 
fixed so that the experimental 
$B(E2;2^{+}_{1}\to 0^{+}_{1})$ transition rates 
for  Xe nuclei \cite{data,degraaf1998,muellergatermann2020} 
are reproduced reasonably well. 
The computed $B(E1;1^{-}_{1}\to 0^{+}_{1})$, 
$B(E2;2^{+}_{1}\to 0^{+}_{1})$, and $B(E3;3^{-}_{1}\to 0^{+}_{1})$ 
transition probabilities are shown in Fig.~\ref{fig:trans}. 
The available experimental data are the  
$B(E2)$ transition rates for $^{114}$Xe 
\cite{degraaf1998}, $^{116}$Xe \cite{degraaf1998}, 
and $^{118}$Xe \cite{muellergatermann2020}, 
and the $B(E3)$ value for $^{114}$Xe \cite{DEANGELIS2002}.

The $B(E1)$ transition probabilities obtained for 
Xe, Ce, and Nd  nuclei are plotted in Fig.~\ref{fig:trans}(a). They
exhibit a parabolic dependence on $N$ with a minimum 
around $N=56$. This trend  correlates with the 
systematic of the $E(1^{-}_{1})$ energy (see Fig.~\ref{fig:level-neg}). 
A similar trend was obtained for Xe isotopes in Ref.~\cite{nomura2021qoch}. 
However, one should keep in mind that 
the $E1$ properties may have a strong component 
determined by noncollective (single-particle) degrees 
of freedom, which are by construction not included 
in the configuration space of the $sdf$-IBM. 
Thus the mapped IBM framework, in its current 
version, does not provide an accurate description of the 
$E1$ transitions. The $B(E2)$ rates in 
Fig.~\ref{fig:trans}(b)
increase monotonously with $N$, 
which confirms the increasing quadrupole collectivity. 
The $2^{+}_{1}$ excitation energy also becomes 
lower, as one approaches the middle of the 
neutron major shell $N=66$ (see Fig.~\ref{fig:level-pos}). 

In the case of nuclei with pronounced octupole
deformation effects, the  $B(E3;3^{-}_{1}\to0^{+}_{1})$ transition
probabilities are expected to be large. As can be see
from Fig.~\ref{fig:trans}(c), the predicted $B(E3)$ values
do not show this pattern.
To take into account the empirical isotopic dependence of 
the $B(E3;3^{-}_{1}\to0^{+}_{1})$ rate, 
in earlier studies \cite{nomura2021oct-u,nomura2021oct-ba}  
we assumed the boson effective $E3$ charge to have a certain 
boson-number dependence. In the present study, however, we 
have used the 
constant $E3$ charge  $e_{3}=0.12$ $e\,$b$^{2/3}$, 
mainly due to the lack of $B(E3)$ data.
The present calculations underestimate the large experimental 
$B(E3;3^{-}_{1}\to0^{+}_{1})$ value for $^{114}$Xe ($77\pm27$ W.u.) 
\cite{DEANGELIS2002}. The data, however, also has a large error bar.

%-----------------------------------------------------------------------
%       B(EL) Table
%-----------------------------------------------------------------------
\begin{table}[!htb]
\begin{center}
\caption{\label{tab:em}
The  $B(E1)$, $B(E2)$, and $B(E3)$ transition probabilities 
(in W.u.) obtained for $^{112}$Xe, $^{114}$Xe, $^{116}$Xe and $^{118}$Xe
are compared with experimental data
from 
Refs.~\cite{data,sears1998,degraaf1998,SMITH2001,DEANGELIS2002,muellergatermann2020}.
}
 \begin{ruledtabular}
 \begin{tabular}{lccccc}
 & $E\lambda$  & $I_{i}$ & $I_{f}$ & Expt. & IBM \\ 
\hline
$^{112}$Xe
& $E1$ 
  & $5^-_1$ & $4^+_1$ & $(1.0\pm0.3)\times10^{-4}$ & $5.4\times10^{-3}$ \\ % PLB 523 (2001) 13
& & $7^-_1$ & $6^+_1$ & $(6\pm2)\times10^{-5}$ & $7.4\times10^{-3}$ \\ % PLB 523 (2001) 13
\hline
$^{114}$Xe
& $E1$ 
  & $3^-_1$ & $4^+_1$ & $(7.4\pm2.1)\times10^{-5}$ & $2.4\times10^{-5}$ \\ % PLB535 (2002) 93
& &         & $2^+_1$ & $(2.0\pm0.5)\times10^{-5}$ & $3.6\times10^{-3}$ \\ % PLB535 (2002) 93
& & $5^-_1$ & $6^+_1$ & $(1.6\pm0.2)\times10^{-4}$ & $2.0\times10^{-6}$ \\ % PLB535 (2002) 93
& $E2$ 
  & $2^+_1$ & $0^+_1$ & 56$\pm4$ & 65 \\ % PRC58 (1998) 164
& & $4^+_1$ & $2^+_1$ & 56$\pm3$ & 90 \\ % PRC58 (1998) 164
& & $6^+_1$ & $4^+_1$ & 43$\pm6$ & 93 \\ % PRC58 (1998) 164
& & $5^-_1$ & $3^-_1$ & 94$\pm11$ & 58 \\ % PRC58 (1998) 164
& & $7^-_1$ & $5^-_1$ & 69$\pm15$ & 63 \\ % PRC58 (1998) 164
& $E3$ 
  & $3^{-}_1$ & $0^+_1$ & 77$\pm27$ & 39 \\ % PLB535 (2002) 93
& & $5^{-}_1$ & $2^+_1$ & 68$\pm17$ & 59 \\ % PLB535 (2002) 93
\hline
$^{116}$Xe
& $E1$ 
  & $7^-_1$ & $6^+_1$ & $(1.4\pm0.6)\times10^{-4}$ & $8.2\times10^{-3}$ \\ % ENSDF
& & $9^-_1$ & $8^+_1$ & $(9.4\pm1.2)\times10^{-5}$ & $1.1\times10^{-2}$ \\ % ENSDF
& $E2$ 
  & $2^+_1$ & $0^+_1$ & 72$\pm3$ & 76 \\ % PRC58 (1998) 164
& & $4^+_1$ & $2^+_1$ & 127$\pm5$ & 106 \\ % PRC58 (1998) 164
& & $6^+_1$ & $4^+_1$ & 113$\pm10$ & 112 \\ % PRC58 (1998) 164
& & $8^+_1$ & $6^+_1$ & 100$\pm12$ & 108 \\ % PRC58 (1998) 164
& & $10^+_1$ & $8^+_1$ & 113$\pm21$ & 97 \\ % PRC58 (1998) 164
& & $7^-_1$ & $5^-_1$ & 82$\pm44$ & 72 \\ % PRC58 (1998) 164
& & $9^-_1$ & $7^-_1$ & 90$\pm15$ & 79 \\ % PRC58 (1998) 164
& & $11^-_1$ & $9^-_1$ & 86$\pm10$ & 79 \\ % PRC58 (1998) 164
\hline
$^{118}$Xe
& $E1$ 
  & $7^-_1$ & $6^+_1$ & $(2.3\pm0.1)\times10^{-4}$ & $8.2\times10^{-3}$\\ % PRC57 (1998) 2991
& & $9^-_1$ & $8^+_1$ & $(2.2\pm0.1)\times10^{-4}$ & $1.1\times10^{-2}$ \\ % PRC57 (1998) 2991
& & $11^-_1$ & $10^+_1$ & $(1.2\pm0.3)\times10^{-4}$ & $1.3\times10^{-2}$ \\ % PRC57 (1998) 2991
& $E2$ 
  & $2^+_1$ & $0^+_1$ & 76.8$\pm1.1$ & 84 \\ % PRC102 (2020) 064318
& & $4^+_1$ & $2^+_1$ & 118$\pm2$ & 119 \\ % PRC102 (2020) 064318
& & $6^+_1$ & $4^+_1$ & 156$^{+7}_{-6}$ & 127 \\ % PRC102 (2020) 064318
& & $8^+_1$ & $6^+_1$ & 143$^{+17}_{-13}$ & 125 \\ % PRC102 (2020) 064318
 \end{tabular}
 \end{ruledtabular}
\end{center} 
\end{table}

% ------------------------------------------- hereherehere

Table~\ref{tab:em} lists the $B(E\lambda)$ values 
for $^{112}$Xe, $^{114}$Xe, $^{116}$Xe, and $^{118}$Xe, 
for which experimental data are available 
\cite{data,sears1998,degraaf1998,DEANGELIS2002,muellergatermann2020}. 
For the $B(E2)$ rates, only the data for inband transitions 
in the lowest-energy $\pi=\pm1$ bands are known. The
calculations account reasonably well for the $B(E2;I\to{I-2})$ 
values in $^{116,118}$Xe. For $^{114}$Xe, the mapped IBM 
overestimates these inband $E2$ transitions, which  
suggests a  much stronger quadrupole collectivity than expected 
experimentally. 
For completeness, some $B(E1)$ rates are also included
in the table.

%-----------------------------------------------------------------------
%
%	Energy displacement
%
%-----------------------------------------------------------------------
\begin{figure}[htb!]
\begin{center}
\includegraphics[width=\linewidth]{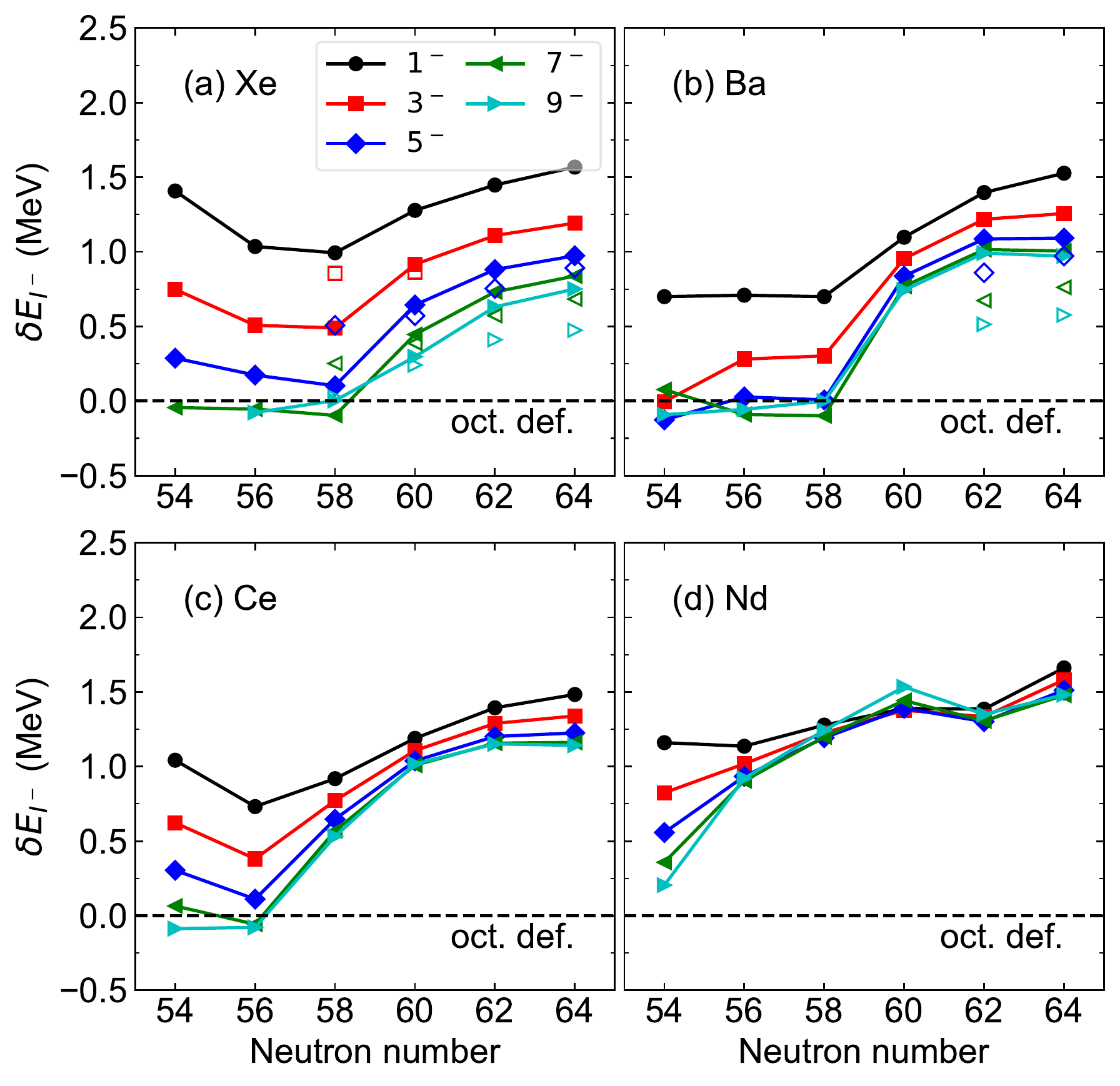}
\caption{The energy displacement $\delta{E_{I^{-}}}$ 
(\ref{eq:de}) is plotted as a function of the neutron number. 
Theoretical values are connected by lines. Experimental 
 values \cite{data}
for the $I^\pi=3^{-}$, 
 $5^{-}$, $7^{-}$, and $9^{-}$ yrast 
states are represented by open
 squares,  diamonds, and left- 
and right-pointing triangles, respectively. 
A broken horizontal line in each panel stands for 
the limit of stable octupole deformation 
$\delta{E_{I^{-}}}=0$.}
\label{fig:de}
\end{center}
\end{figure}
%-----------------------------------------------------------------------
%
%	Alternating parity
%
%-----------------------------------------------------------------------
\begin{figure}[htb!]
\begin{center}
\includegraphics[width=\linewidth]{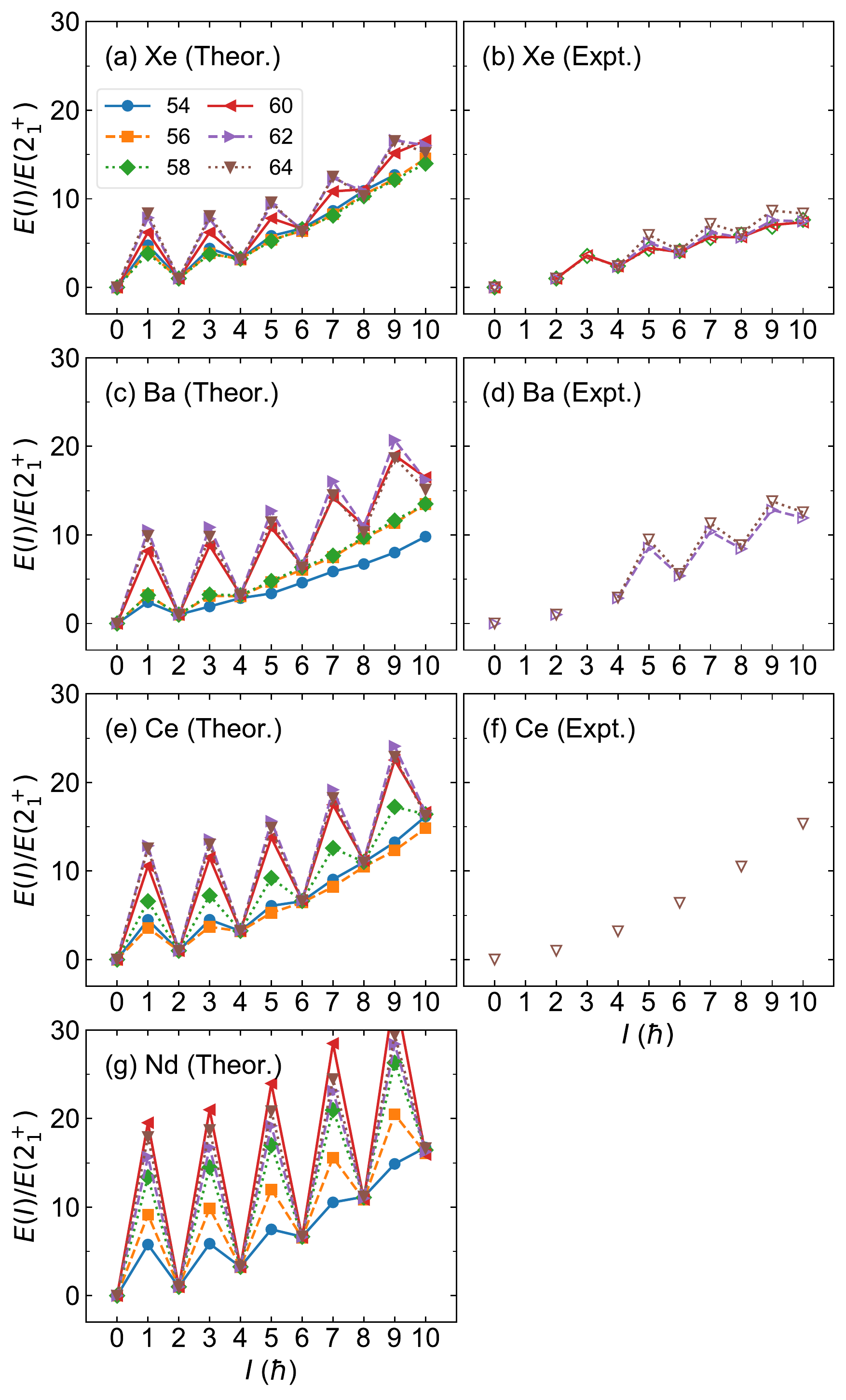}
\caption{The energy ratio 
$E(I^{\pi})/E(2^{+}_{1})$ is plotted as a function of 
spin $I^{\pi}$, with $\pi=+1$ for even-$I$ 
and $\pi=-1$ for odd-$I$ yrast states. 
The available experimental data for the Xe, Ba, and Ce isotopes are 
taken from \cite{data}.}
\label{fig:alt}
\end{center}
\end{figure}
%-----------------------------------------------------------------------
%
%	Effective betas and fluctuations
%
%-----------------------------------------------------------------------
\begin{figure}[htb!]
\begin{center}
\includegraphics[width=\linewidth]{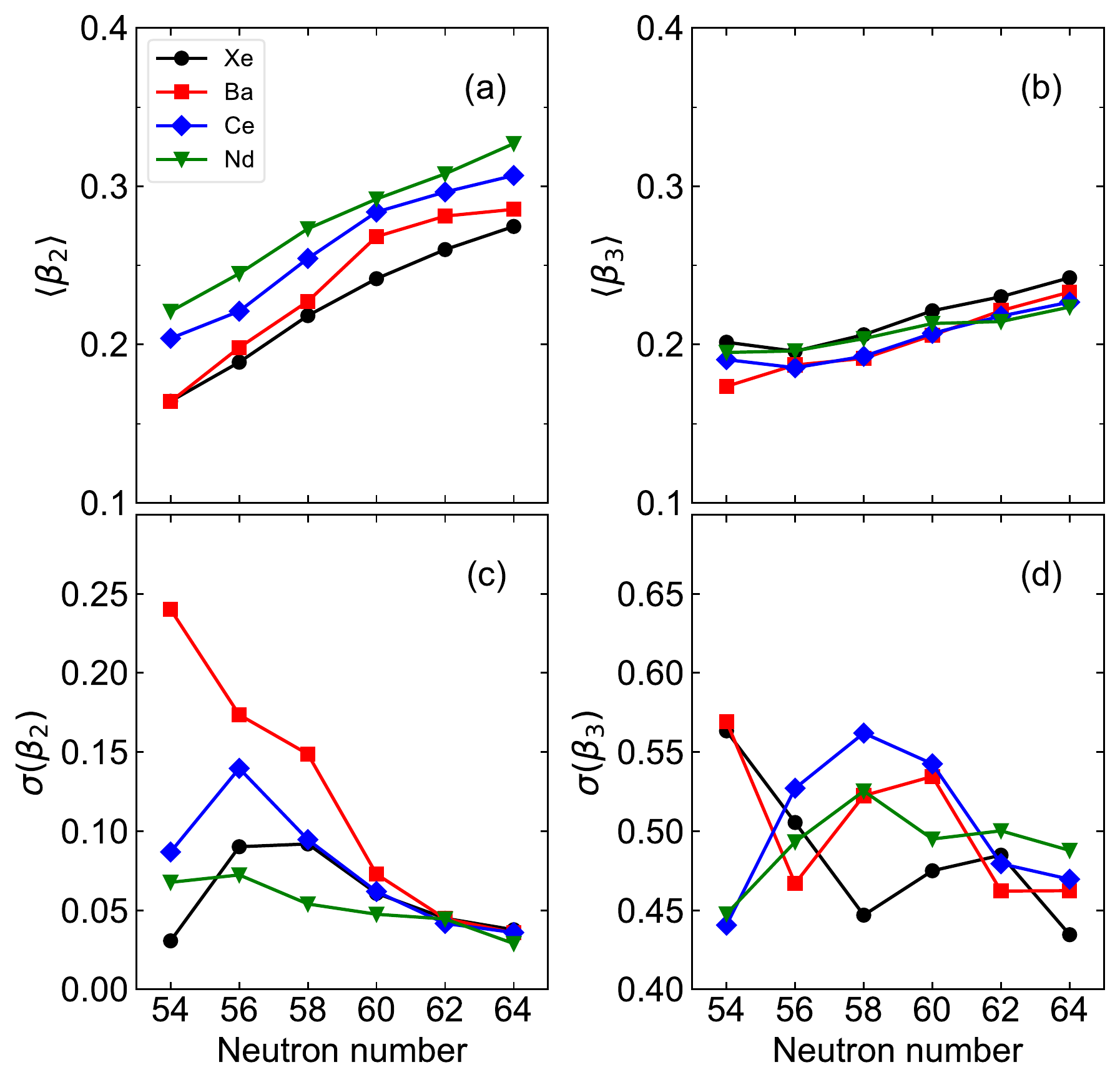}
\caption{Effective  quadrupole (a) and  octupole (b)
deformations $\braket{\beta_{\lambda}}$
 and 
fluctuations $\sigma(\beta_{\lambda})$ in  the $\beta_{2}$ (c) 
and  $\beta_{3}$ (d) deformations. For more details, see the main text.}
\label{fig:bet23}
\end{center}
\end{figure}

%-----------------------------------------------------------------------
\section{Signatures of octupole shape phase transition\label{sec:qpt}}
%-----------------------------------------------------------------------

%-----------------------------------------------------------------------
\subsection{Possible alternating-parity band structure\label{sec:alt}}
%-----------------------------------------------------------------------

In order to distinguish whether the members 
of $\pi=-1$ rotational bands are octupole-deformed 
or octupole vibrational states, it is convenient to 
analyze the energy displacement, defined by 
\begin{align}
\label{eq:de}
\delta{E_{I^{-}}} = E_{I^{-}} - \frac{E_{(I+1)^{+}} + E_{(I-1)^{+}}}{2}, 
\end{align}
where $E_{I^{-}}$ and $E_{(I\pm1)^{+}}$ 
represent the excitation energies of the 
$\pi=-1$ odd-spin and $\pi=+1$ even-spin yrast states,
respectively. 
If the two lowest bands with opposite parity 
share an octupole deformed band-head they form an 
alternating-parity doublet and the quantity 
$\delta{E_{I^{-}}}$ should be equal to zero. 
The deviation from the limit $\delta{E_{I^{-}}}=0$, 
implies that the states generating the $\pi=\pm1$ 
bands are different in nature, and therefore 
the $\pi=-1$ state has an octupole 
vibrational character.

The $\delta{E_{I^{-}}}$  values are displayed 
in Fig.~\ref{fig:de}. This quantity is close to zero for Xe, Ba, and Ce nuclei with 
$54\leqslant N\leqslant58$, especially for 
higher-spin states.
One observes a pronounced  increase
of  $\delta{E_{I^{-}}}$
from $N=58$ to 60 in the Ba and Xe isotopes and from 
$N=56$ to 58 in the Ce isotopes. 
The deviation from the limit 
($\delta{E_{I^{-}}}=0$) becomes more significant for 
larger neutron numbers. Note, that for 
 Xe and Ba isotopes, the  $\delta{E_{I^{-}}}$ 
values agree well with the 
experimental ones \cite{data}.

We have also examined the 
energy ratio $E(I^{\pi})/E(2^{+}_{1})$, with $\pi=+1$ 
for even-$I$ and $\pi=-1$ for odd-$I$ yrast states.  
For the ideal alternating-parity rotational 
band the ratio would depend quadratically on the spin $I$. 
If the $\pi=\pm1$ yrast bands are decoupled, as in the case of 
octupole vibrational states, the ratio 
is expected to show an odd-even-spin staggering. As can be 
seen from Fig.~\ref{fig:alt}, the energy ratios for the  Xe, Ba and Ce 
isotopes with $N<60$ increase quadratically 
with $I$. On the other hand, a pronounced odd-even-spin 
staggering occurs for $N\geqslant60$. The 
staggering is much 
more pronounced for heavier-$Z$ isotopes. 
These results confirm that 
octupole correlations are enhanced around $N=56$ and $Z=56$.

\subsection{Quadrupole and octupole shape invariants}

We consider shape invariants \cite{cline1986,werner2000} 
computed using the $E2$ and $E3$ matrix elements
as another signature of shape/phase transitions. 
The relevant shape invariants are defined as
\begin{align}
\label{eq:qinvar2}
q_{2}^{(\lambda)}=\sum_{i}
&(-1)^{I}
\braket{0^+_1     \| \hat{T}(E\lambda) \| I^{\pi}_i}
\braket{I^{\pi}_i \| \hat{T}(E\lambda) \| 0^+_1}\\ 
\label{eq:qinvar4}
q_{4}^{(\lambda)}=\sum_{i,j,k}
&\braket{0^+_1     \| \hat{T}(E\lambda) \| I^{\pi}_i}
\braket{I^{\pi}_i \| \hat{T}(E\lambda) \| 0^+_j}
\nonumber \\
&\times
\braket{0^+_j     \| \hat{T}(E\lambda) \| I^{\pi}_k}
\braket{I^{\pi}_k \| \hat{T}(E\lambda) \| 0^+_1}
\end{align}
where $\ket{0^{+}_{1}}$ is the $sdf$-IBM ground state, 
$\braket{\|(\cdots)\|}$ represents the reduced $E2$ ($E3$) 
matrix element and $I^{\pi}=2^{+}$ ($3^{-}$) 
for $\lambda=2$ ($\lambda=3$). 
The sums in Eqs.~(\ref{eq:qinvar2}) and (\ref{eq:qinvar4}) include up
to ten lowest $0^+$, $2^+$, and $3^-$ states. 
The effective quadrupole 
and octupole deformations
\begin{align}
\label{eq:beta}
\braket{\beta_{\lambda}} =\frac{4\pi}{3eZR_{0}^{\lambda}}
\sqrt{q_{2}^{(\lambda)}}
\end{align}
as well as the fluctuations \cite{werner2000}
\begin{align}
\label{eq:fluc}
 \sigma(\beta_{\lambda})=q_{4}^{(\lambda)}/(q_{2}^{(\lambda)})^{2}-1,
\end{align}
which measure the softness along the $\beta_{\lambda}$-directions 
are shown in  Fig.~\ref{fig:bet23}.

The steady increase of $\braket{\beta_{2}}$ in 
Fig.~\ref{fig:bet23}(a), corroborates
the increasing quadrupole collectivity along 
the considered isotopic chains. In contrast, the 
$\braket{\beta_{3}}$ values, in 
Fig.~\ref{fig:bet23}(b), change  much less with $N$.
The fluctuations $\sigma(\beta_{2})$, 
in Fig.~\ref{fig:bet23}(c), appear to reach a maximum 
around $N=56$, 
%except for the $^{110}$Ba nucleus, 
and show a notable decrease from $N=56$ toward $N=60$. 
Note, that the corresponding SCMF-PESs become 
more rigid along the $\beta_{2}$-direction from $N=56$ on, 
and a more distinct prolate minimum appears 
(see Fig.~\ref{fig:pesdft}). 
This result is also consistent with the behavior of the 
predicted $\pi=+1$ energy spectra, 
which suggest the onset of strongly 
quadrupole deformed shapes at $N\approx60$ 
(see Fig.~\ref{fig:level-pos}). 
As seen in Fig.~\ref{fig:bet23}(d), 
the fluctuations $\sigma(\beta_{3})$ are systematically 
larger in magnitude than $\sigma(\beta_{2})$. 
The $\sigma(\beta_{3})$ values 
also exhibit a significant variation for $54\leqslant N\leqslant62$.
The behavior of the $\sigma(\beta_{3})$ fluctuations reflect 
a considerable degree of octupole mixing
and an enhanced octupole collectivity.

% ----------------------------------------------------------------------
\section{Summary\label{sec:summary}}
% ----------------------------------------------------------------------

The quadrupole-octupole coupling 
in the low-lying states of  neutron-deficient 
Xe, Ba, Ce, and Nd nuclei has been studied within the mapped 
$sdf$-IBM framework. The strength parameters for the $sdf$-IBM Hamiltonian 
have been 
obtained via the mapping of the (microscopic) axially-symmetric 
$(\beta_{2},\beta_{3})$-PESs, obtained from 
constrained Gogny-D1M HFB calculations, 
onto the expectation value of the IBM Hamiltonian in the condensate 
state of the $s$, $d$, and $f$ bosons. Excitation spectra and 
electric transition probabilities have been obtained by the 
diagonalization of the mapped Hamiltonian.

The Gogny-D1M SCMF-PESs for nuclei near the 
neutron octupole ``magic number'' $N=56$ 
are notably soft along the $\beta_{3}$-direction. 
An octupole-deformed mean-field ground state
has been obtained for $^{110}$Ba, $^{112}$Ba, $^{114}$Ba 
and $^{114}$Ce. Beyond the HFB level, the systematic
of the properties of the positive parity states points towards an increased 
quadrupole collectivity with increasing neutron number. A notable 
change is found in Xe and Ba isotopes with $N=58$ and 60.
The negative parity yrast 
states exhibit a parabolic behavior as functions 
of $N$, with a minimum around $N=56$. Moreover, 
the predicted $\pi=\pm1$ yrast bands 
form an approximate alternating-parity doublet 
for most of the Xe isotopes as well as Ba 
and Ce nuclei in the vicinity of $N=56$. 
Another signature of octupole correlations 
can be associated with the large 
fluctuations of the effective $\beta_{3}$ deformation 
around $N=56$.

We have further assessed the predictive power of the mapped $sdf$-IBM 
to describe spectroscopic properties in the 
$N\approx Z$ mass region. The excitation energies of the 
$\pi=\pm1$ 
yrast bands agree reasonably well with the 
available experimental data for Xe and Ba nuclei.
However, non-yrast $\pi=+1$ bands have been predicted 
much 
higher in energy than the experimental ones. This 
indicates that certain extensions of the model are 
required to improve the description of non-yrast bands
in regions of the nuclear chart where octupolarity
plays a role. 
A reasonable approach to address 
this problem is to identify whether the deficiency in the 
model description of the 
non-yrast $\pi=+1$ bands is due to the deficiencies of the chosen 
EDF for the mass region under study, or 
that the employed $sdf$-IBM Hamiltonian lacks important degrees of freedom, 
or a combination of the two. 
Work along these lines is in progress
and will be reported elsewhere.

\begin{acknowledgments}
This work has been supported by the Tenure Track Pilot Programme of the 
Croatian Science Foundation and the \'Ecole Polytechnique F\'ed\'erale 
de Lausanne, and the Project TTP-2018-07-3554 Exotic Nuclear Structure 
and Dynamics, with funds of the Croatian-Swiss Research Programme. The  
work of LMR was supported by Spanish Ministry of Economy and 
Competitiveness (MINECO) Grant No. PGC2018-094583-B-I00. 
\end{acknowledgments}

\bibliography{refs}

\begin{thebibliography}{93}%
\makeatletter
\providecommand \@ifxundefined [1]{%
 \@ifx{#1\undefined}
}%
\providecommand \@ifnum [1]{%
 \ifnum #1\expandafter \@firstoftwo
 \else \expandafter \@secondoftwo
 \fi
}%
\providecommand \@ifx [1]{%
 \ifx #1\expandafter \@firstoftwo
 \else \expandafter \@secondoftwo
 \fi
}%
\providecommand \natexlab [1]{#1}%
\providecommand \enquote  [1]{``#1''}%
\providecommand \bibnamefont  [1]{#1}%
\providecommand \bibfnamefont [1]{#1}%
\providecommand \citenamefont [1]{#1}%
\providecommand \href@noop [0]{\@secondoftwo}%
\providecommand \href [0]{\begingroup \@sanitize@url \@href}%
\providecommand \@href[1]{\@@startlink{#1}\@@href}%
\providecommand \@@href[1]{\endgroup#1\@@endlink}%
\providecommand \@sanitize@url [0]{\catcode `\\12\catcode `\$12\catcode
  `\&12\catcode `\#12\catcode `\^12\catcode `\_12\catcode `\%12\relax}%
\providecommand \@@startlink[1]{}%
\providecommand \@@endlink[0]{}%
\providecommand \url  [0]{\begingroup\@sanitize@url \@url }%
\providecommand \@url [1]{\endgroup\@href {#1}{\urlprefix }}%
\providecommand \urlprefix  [0]{URL }%
\providecommand \Eprint [0]{\href }%
\providecommand \doibase [0]{https://doi.org/}%
\providecommand \selectlanguage [0]{\@gobble}%
\providecommand \bibinfo  [0]{\@secondoftwo}%
\providecommand \bibfield  [0]{\@secondoftwo}%
\providecommand \translation [1]{[#1]}%
\providecommand \BibitemOpen [0]{}%
\providecommand \bibitemStop [0]{}%
\providecommand \bibitemNoStop [0]{.\EOS\space}%
\providecommand \EOS [0]{\spacefactor3000\relax}%
\providecommand \BibitemShut  [1]{\csname bibitem#1\endcsname}%
\let\auto@bib@innerbib\@empty
%</preamble>
\bibitem [{\citenamefont {Butler}\ and\ \citenamefont
  {Nazarewicz}(1996)}]{butler1996}%
  \BibitemOpen
  \bibfield  {author} {\bibinfo {author} {\bibfnamefont {P.~A.}\ \bibnamefont
  {Butler}}\ and\ \bibinfo {author} {\bibfnamefont {W.}~\bibnamefont
  {Nazarewicz}},\ }\href {https://doi.org/10.1103/RevModPhys.68.349} {\bibfield
   {journal} {\bibinfo  {journal} {Rev. Mod. Phys.}\ }\textbf {\bibinfo
  {volume} {68}},\ \bibinfo {pages} {349} (\bibinfo {year} {1996})}\BibitemShut
  {NoStop}%
\bibitem [{\citenamefont {Butler}(2016)}]{butler2016}%
  \BibitemOpen
  \bibfield  {author} {\bibinfo {author} {\bibfnamefont {P.~A.}\ \bibnamefont
  {Butler}},\ }\href {https://doi.org/10.1088/0954-3899/43/7/073002} {\bibfield
   {journal} {\bibinfo  {journal} {J. Phys. G: Nucl. Part. Phys.}\ }\textbf
  {\bibinfo {volume} {43}},\ \bibinfo {pages} {073002} (\bibinfo {year}
  {2016})}\BibitemShut {NoStop}%
\bibitem [{\citenamefont {Butler}(2020)}]{butler2020b}%
  \BibitemOpen
  \bibfield  {author} {\bibinfo {author} {\bibfnamefont {P.~A.}\ \bibnamefont
  {Butler}},\ }\href {https://doi.org/10.1098/rspa.2020.0202} {\bibfield
  {journal} {\bibinfo  {journal} {Proc. R. Soc. A}\ }\textbf {\bibinfo {volume}
  {476}},\ \bibinfo {pages} {20200202} (\bibinfo {year} {2020})}\BibitemShut
  {NoStop}%
\bibitem [{\citenamefont {Gaffney}\ \emph {et~al.}(2013)\citenamefont
  {Gaffney}, \citenamefont {Butler}, \citenamefont {Scheck}, \citenamefont
  {Hayes}, \citenamefont {Wenander}, \citenamefont {Albers}, \citenamefont
  {Bastin}, \citenamefont {Bauer}, \citenamefont {Blazhev}, \citenamefont
  {B\"onig}, \citenamefont {Bree}, \citenamefont {Cederk\"all}, \citenamefont
  {Chupp}, \citenamefont {Cline}, \citenamefont {Cocolios}, \citenamefont
  {Davinson}, \citenamefont {Witte}, \citenamefont {Diriken}, \citenamefont
  {Grahn}, \citenamefont {Herzan}, \citenamefont {Huyse}, \citenamefont
  {Jenkins}, \citenamefont {Joss}, \citenamefont {Kesteloot}, \citenamefont
  {Konki}, \citenamefont {Kowalczyk}, \citenamefont {Kröll}, \citenamefont
  {Kwan}, \citenamefont {Lutter}, \citenamefont {Moschner}, \citenamefont
  {Napiorkowski}, \citenamefont {Pakarinen}, \citenamefont {Pfeiffer},
  \citenamefont {Radeck}, \citenamefont {Reiter}, \citenamefont {Reynders},
  \citenamefont {Rigby}, \citenamefont {Robledo}, \citenamefont {Rudigier},
  \citenamefont {Sambi}, \citenamefont {Seidlitz}, \citenamefont {Siebeck},
  \citenamefont {Stora}, \citenamefont {Thoele}, \citenamefont {Duppen},
  \citenamefont {Vermeulen}, \citenamefont {von Schmid}, \citenamefont
  {Voulot}, \citenamefont {Warr}, \citenamefont {Wimmer}, \citenamefont
  {Wrzosek-Lipska}, \citenamefont {Wu},\ and\ \citenamefont
  {Zielinska}}]{gaffney2013}%
  \BibitemOpen
  \bibfield  {author} {\bibinfo {author} {\bibfnamefont {L.~P.}\ \bibnamefont
  {Gaffney}}, \bibinfo {author} {\bibfnamefont {P.~A.}\ \bibnamefont {Butler}},
  \bibinfo {author} {\bibfnamefont {M.}~\bibnamefont {Scheck}}, \bibinfo
  {author} {\bibfnamefont {A.~B.}\ \bibnamefont {Hayes}}, \bibinfo {author}
  {\bibfnamefont {F.}~\bibnamefont {Wenander}}, \bibinfo {author}
  {\bibfnamefont {M.}~\bibnamefont {Albers}}, \bibinfo {author} {\bibfnamefont
  {B.}~\bibnamefont {Bastin}}, \bibinfo {author} {\bibfnamefont
  {C.}~\bibnamefont {Bauer}}, \bibinfo {author} {\bibfnamefont
  {A.}~\bibnamefont {Blazhev}}, \bibinfo {author} {\bibfnamefont
  {S.}~\bibnamefont {B\"onig}}, \bibinfo {author} {\bibfnamefont
  {N.}~\bibnamefont {Bree}}, \bibinfo {author} {\bibfnamefont {J.}~\bibnamefont
  {Cederk\"all}}, \bibinfo {author} {\bibfnamefont {T.}~\bibnamefont {Chupp}},
  \bibinfo {author} {\bibfnamefont {D.}~\bibnamefont {Cline}}, \bibinfo
  {author} {\bibfnamefont {T.~E.}\ \bibnamefont {Cocolios}}, \bibinfo {author}
  {\bibfnamefont {T.}~\bibnamefont {Davinson}}, \bibinfo {author}
  {\bibfnamefont {H.~D.}\ \bibnamefont {Witte}}, \bibinfo {author}
  {\bibfnamefont {J.}~\bibnamefont {Diriken}}, \bibinfo {author} {\bibfnamefont
  {T.}~\bibnamefont {Grahn}}, \bibinfo {author} {\bibfnamefont
  {A.}~\bibnamefont {Herzan}}, \bibinfo {author} {\bibfnamefont
  {M.}~\bibnamefont {Huyse}}, \bibinfo {author} {\bibfnamefont {D.~G.}\
  \bibnamefont {Jenkins}}, \bibinfo {author} {\bibfnamefont {D.~T.}\
  \bibnamefont {Joss}}, \bibinfo {author} {\bibfnamefont {N.}~\bibnamefont
  {Kesteloot}}, \bibinfo {author} {\bibfnamefont {J.}~\bibnamefont {Konki}},
  \bibinfo {author} {\bibfnamefont {M.}~\bibnamefont {Kowalczyk}}, \bibinfo
  {author} {\bibfnamefont {T.}~\bibnamefont {Kröll}}, \bibinfo {author}
  {\bibfnamefont {E.}~\bibnamefont {Kwan}}, \bibinfo {author} {\bibfnamefont
  {R.}~\bibnamefont {Lutter}}, \bibinfo {author} {\bibfnamefont
  {K.}~\bibnamefont {Moschner}}, \bibinfo {author} {\bibfnamefont
  {P.}~\bibnamefont {Napiorkowski}}, \bibinfo {author} {\bibfnamefont
  {J.}~\bibnamefont {Pakarinen}}, \bibinfo {author} {\bibfnamefont
  {M.}~\bibnamefont {Pfeiffer}}, \bibinfo {author} {\bibfnamefont
  {D.}~\bibnamefont {Radeck}}, \bibinfo {author} {\bibfnamefont
  {P.}~\bibnamefont {Reiter}}, \bibinfo {author} {\bibfnamefont
  {K.}~\bibnamefont {Reynders}}, \bibinfo {author} {\bibfnamefont {S.~V.}\
  \bibnamefont {Rigby}}, \bibinfo {author} {\bibfnamefont {L.~M.}\ \bibnamefont
  {Robledo}}, \bibinfo {author} {\bibfnamefont {M.}~\bibnamefont {Rudigier}},
  \bibinfo {author} {\bibfnamefont {S.}~\bibnamefont {Sambi}}, \bibinfo
  {author} {\bibfnamefont {M.}~\bibnamefont {Seidlitz}}, \bibinfo {author}
  {\bibfnamefont {B.}~\bibnamefont {Siebeck}}, \bibinfo {author} {\bibfnamefont
  {T.}~\bibnamefont {Stora}}, \bibinfo {author} {\bibfnamefont
  {P.}~\bibnamefont {Thoele}}, \bibinfo {author} {\bibfnamefont {P.~V.}\
  \bibnamefont {Duppen}}, \bibinfo {author} {\bibfnamefont {M.~J.}\
  \bibnamefont {Vermeulen}}, \bibinfo {author} {\bibfnamefont {M.}~\bibnamefont
  {von Schmid}}, \bibinfo {author} {\bibfnamefont {D.}~\bibnamefont {Voulot}},
  \bibinfo {author} {\bibfnamefont {N.}~\bibnamefont {Warr}}, \bibinfo {author}
  {\bibfnamefont {K.}~\bibnamefont {Wimmer}}, \bibinfo {author} {\bibfnamefont
  {K.}~\bibnamefont {Wrzosek-Lipska}}, \bibinfo {author} {\bibfnamefont
  {C.~Y.}\ \bibnamefont {Wu}},\ and\ \bibinfo {author} {\bibfnamefont
  {M.}~\bibnamefont {Zielinska}},\ }\href {https://doi.org/10.1038/nature12073}
  {\bibfield  {journal} {\bibinfo  {journal} {Nature (London)}\ }\textbf
  {\bibinfo {volume} {497}},\ \bibinfo {pages} {199} (\bibinfo {year}
  {2013})}\BibitemShut {NoStop}%
\bibitem [{\citenamefont {Butler}\ \emph {et~al.}(2020)\citenamefont {Butler},
  \citenamefont {Gaffney}, \citenamefont {Spagnoletti}, \citenamefont
  {Abrahams}, \citenamefont {Bowry}, \citenamefont {Cederk\"all}, \citenamefont
  {de~Angelis}, \citenamefont {De~Witte}, \citenamefont {Garrett},
  \citenamefont {Goldkuhle}, \citenamefont {Henrich}, \citenamefont {Illana},
  \citenamefont {Johnston}, \citenamefont {Joss}, \citenamefont {Keatings},
  \citenamefont {Kelly}, \citenamefont {Komorowska}, \citenamefont {Konki},
  \citenamefont {Kr\"oll}, \citenamefont {Lozano}, \citenamefont {Nara~Singh},
  \citenamefont {O'Donnell}, \citenamefont {Ojala}, \citenamefont {Page},
  \citenamefont {Pedersen}, \citenamefont {Raison}, \citenamefont {Reiter},
  \citenamefont {Rodriguez}, \citenamefont {Rosiak}, \citenamefont {Rothe},
  \citenamefont {Scheck}, \citenamefont {Seidlitz}, \citenamefont {Shneidman},
  \citenamefont {Siebeck}, \citenamefont {Sinclair}, \citenamefont {Smith},
  \citenamefont {Stryjczyk}, \citenamefont {Van~Duppen}, \citenamefont
  {Vinals}, \citenamefont {Virtanen}, \citenamefont {Warr}, \citenamefont
  {Wrzosek-Lipska},\ and\ \citenamefont {Zieli\ifmmode~\acute{n}\else
  \'{n}\fi{}ska}}]{butler2020a}%
  \BibitemOpen
  \bibfield  {author} {\bibinfo {author} {\bibfnamefont {P.~A.}\ \bibnamefont
  {Butler}}, \bibinfo {author} {\bibfnamefont {L.~P.}\ \bibnamefont {Gaffney}},
  \bibinfo {author} {\bibfnamefont {P.}~\bibnamefont {Spagnoletti}}, \bibinfo
  {author} {\bibfnamefont {K.}~\bibnamefont {Abrahams}}, \bibinfo {author}
  {\bibfnamefont {M.}~\bibnamefont {Bowry}}, \bibinfo {author} {\bibfnamefont
  {J.}~\bibnamefont {Cederk\"all}}, \bibinfo {author} {\bibfnamefont
  {G.}~\bibnamefont {de~Angelis}}, \bibinfo {author} {\bibfnamefont
  {H.}~\bibnamefont {De~Witte}}, \bibinfo {author} {\bibfnamefont {P.~E.}\
  \bibnamefont {Garrett}}, \bibinfo {author} {\bibfnamefont {A.}~\bibnamefont
  {Goldkuhle}}, \bibinfo {author} {\bibfnamefont {C.}~\bibnamefont {Henrich}},
  \bibinfo {author} {\bibfnamefont {A.}~\bibnamefont {Illana}}, \bibinfo
  {author} {\bibfnamefont {K.}~\bibnamefont {Johnston}}, \bibinfo {author}
  {\bibfnamefont {D.~T.}\ \bibnamefont {Joss}}, \bibinfo {author}
  {\bibfnamefont {J.~M.}\ \bibnamefont {Keatings}}, \bibinfo {author}
  {\bibfnamefont {N.~A.}\ \bibnamefont {Kelly}}, \bibinfo {author}
  {\bibfnamefont {M.}~\bibnamefont {Komorowska}}, \bibinfo {author}
  {\bibfnamefont {J.}~\bibnamefont {Konki}}, \bibinfo {author} {\bibfnamefont
  {T.}~\bibnamefont {Kr\"oll}}, \bibinfo {author} {\bibfnamefont
  {M.}~\bibnamefont {Lozano}}, \bibinfo {author} {\bibfnamefont {B.~S.}\
  \bibnamefont {Nara~Singh}}, \bibinfo {author} {\bibfnamefont
  {D.}~\bibnamefont {O'Donnell}}, \bibinfo {author} {\bibfnamefont
  {J.}~\bibnamefont {Ojala}}, \bibinfo {author} {\bibfnamefont {R.~D.}\
  \bibnamefont {Page}}, \bibinfo {author} {\bibfnamefont {L.~G.}\ \bibnamefont
  {Pedersen}}, \bibinfo {author} {\bibfnamefont {C.}~\bibnamefont {Raison}},
  \bibinfo {author} {\bibfnamefont {P.}~\bibnamefont {Reiter}}, \bibinfo
  {author} {\bibfnamefont {J.~A.}\ \bibnamefont {Rodriguez}}, \bibinfo {author}
  {\bibfnamefont {D.}~\bibnamefont {Rosiak}}, \bibinfo {author} {\bibfnamefont
  {S.}~\bibnamefont {Rothe}}, \bibinfo {author} {\bibfnamefont
  {M.}~\bibnamefont {Scheck}}, \bibinfo {author} {\bibfnamefont
  {M.}~\bibnamefont {Seidlitz}}, \bibinfo {author} {\bibfnamefont {T.~M.}\
  \bibnamefont {Shneidman}}, \bibinfo {author} {\bibfnamefont {B.}~\bibnamefont
  {Siebeck}}, \bibinfo {author} {\bibfnamefont {J.}~\bibnamefont {Sinclair}},
  \bibinfo {author} {\bibfnamefont {J.~F.}\ \bibnamefont {Smith}}, \bibinfo
  {author} {\bibfnamefont {M.}~\bibnamefont {Stryjczyk}}, \bibinfo {author}
  {\bibfnamefont {P.}~\bibnamefont {Van~Duppen}}, \bibinfo {author}
  {\bibfnamefont {S.}~\bibnamefont {Vinals}}, \bibinfo {author} {\bibfnamefont
  {V.}~\bibnamefont {Virtanen}}, \bibinfo {author} {\bibfnamefont
  {N.}~\bibnamefont {Warr}}, \bibinfo {author} {\bibfnamefont {K.}~\bibnamefont
  {Wrzosek-Lipska}},\ and\ \bibinfo {author} {\bibfnamefont {M.}~\bibnamefont
  {Zieli\ifmmode~\acute{n}\else \'{n}\fi{}ska}},\ }\href
  {https://doi.org/10.1103/PhysRevLett.124.042503} {\bibfield  {journal}
  {\bibinfo  {journal} {Phys. Rev. Lett.}\ }\textbf {\bibinfo {volume} {124}},\
  \bibinfo {pages} {042503} (\bibinfo {year} {2020})}\BibitemShut {NoStop}%
\bibitem [{\citenamefont {Chishti}\ \emph {et~al.}(2020)\citenamefont
  {Chishti}, \citenamefont {O'Donnell}, \citenamefont {Battaglia},
  \citenamefont {Bowry}, \citenamefont {Jaroszynski}, \citenamefont {Singh},
  \citenamefont {Scheck}, \citenamefont {Spagnoletti},\ and\ \citenamefont
  {Smith}}]{chishti2020}%
  \BibitemOpen
  \bibfield  {author} {\bibinfo {author} {\bibfnamefont {M.~M.~R.}\
  \bibnamefont {Chishti}}, \bibinfo {author} {\bibfnamefont {D.}~\bibnamefont
  {O'Donnell}}, \bibinfo {author} {\bibfnamefont {G.}~\bibnamefont
  {Battaglia}}, \bibinfo {author} {\bibfnamefont {M.}~\bibnamefont {Bowry}},
  \bibinfo {author} {\bibfnamefont {D.~A.}\ \bibnamefont {Jaroszynski}},
  \bibinfo {author} {\bibfnamefont {B.~S.~N.}\ \bibnamefont {Singh}}, \bibinfo
  {author} {\bibfnamefont {M.}~\bibnamefont {Scheck}}, \bibinfo {author}
  {\bibfnamefont {P.}~\bibnamefont {Spagnoletti}},\ and\ \bibinfo {author}
  {\bibfnamefont {J.~F.}\ \bibnamefont {Smith}},\ }\href
  {https://doi.org/10.1038/s41567-020-0899-4} {\bibfield  {journal} {\bibinfo
  {journal} {Nat. Phys.}\ }\textbf {\bibinfo {volume} {16}},\ \bibinfo {pages}
  {853} (\bibinfo {year} {2020})}\BibitemShut {NoStop}%
\bibitem [{\citenamefont {Bucher}\ \emph {et~al.}(2016)\citenamefont {Bucher},
  \citenamefont {Zhu}, \citenamefont {Wu}, \citenamefont {Janssens},
  \citenamefont {Cline}, \citenamefont {Hayes}, \citenamefont {Albers},
  \citenamefont {Ayangeakaa}, \citenamefont {Butler}, \citenamefont {Campbell},
  \citenamefont {Carpenter}, \citenamefont {Chiara}, \citenamefont {Clark},
  \citenamefont {Crawford}, \citenamefont {Cromaz}, \citenamefont {David},
  \citenamefont {Dickerson}, \citenamefont {Gregor}, \citenamefont {Harker},
  \citenamefont {Hoffman}, \citenamefont {Kay}, \citenamefont {Kondev},
  \citenamefont {Korichi}, \citenamefont {Lauritsen}, \citenamefont
  {Macchiavelli}, \citenamefont {Pardo}, \citenamefont {Richard}, \citenamefont
  {Riley}, \citenamefont {Savard}, \citenamefont {Scheck}, \citenamefont
  {Seweryniak}, \citenamefont {Smith}, \citenamefont {Vondrasek},\ and\
  \citenamefont {Wiens}}]{bucher2016}%
  \BibitemOpen
  \bibfield  {author} {\bibinfo {author} {\bibfnamefont {B.}~\bibnamefont
  {Bucher}}, \bibinfo {author} {\bibfnamefont {S.}~\bibnamefont {Zhu}},
  \bibinfo {author} {\bibfnamefont {C.~Y.}\ \bibnamefont {Wu}}, \bibinfo
  {author} {\bibfnamefont {R.~V.~F.}\ \bibnamefont {Janssens}}, \bibinfo
  {author} {\bibfnamefont {D.}~\bibnamefont {Cline}}, \bibinfo {author}
  {\bibfnamefont {A.~B.}\ \bibnamefont {Hayes}}, \bibinfo {author}
  {\bibfnamefont {M.}~\bibnamefont {Albers}}, \bibinfo {author} {\bibfnamefont
  {A.~D.}\ \bibnamefont {Ayangeakaa}}, \bibinfo {author} {\bibfnamefont
  {P.~A.}\ \bibnamefont {Butler}}, \bibinfo {author} {\bibfnamefont {C.~M.}\
  \bibnamefont {Campbell}}, \bibinfo {author} {\bibfnamefont {M.~P.}\
  \bibnamefont {Carpenter}}, \bibinfo {author} {\bibfnamefont {C.~J.}\
  \bibnamefont {Chiara}}, \bibinfo {author} {\bibfnamefont {J.~A.}\
  \bibnamefont {Clark}}, \bibinfo {author} {\bibfnamefont {H.~L.}\ \bibnamefont
  {Crawford}}, \bibinfo {author} {\bibfnamefont {M.}~\bibnamefont {Cromaz}},
  \bibinfo {author} {\bibfnamefont {H.~M.}\ \bibnamefont {David}}, \bibinfo
  {author} {\bibfnamefont {C.}~\bibnamefont {Dickerson}}, \bibinfo {author}
  {\bibfnamefont {E.~T.}\ \bibnamefont {Gregor}}, \bibinfo {author}
  {\bibfnamefont {J.}~\bibnamefont {Harker}}, \bibinfo {author} {\bibfnamefont
  {C.~R.}\ \bibnamefont {Hoffman}}, \bibinfo {author} {\bibfnamefont {B.~P.}\
  \bibnamefont {Kay}}, \bibinfo {author} {\bibfnamefont {F.~G.}\ \bibnamefont
  {Kondev}}, \bibinfo {author} {\bibfnamefont {A.}~\bibnamefont {Korichi}},
  \bibinfo {author} {\bibfnamefont {T.}~\bibnamefont {Lauritsen}}, \bibinfo
  {author} {\bibfnamefont {A.~O.}\ \bibnamefont {Macchiavelli}}, \bibinfo
  {author} {\bibfnamefont {R.~C.}\ \bibnamefont {Pardo}}, \bibinfo {author}
  {\bibfnamefont {A.}~\bibnamefont {Richard}}, \bibinfo {author} {\bibfnamefont
  {M.~A.}\ \bibnamefont {Riley}}, \bibinfo {author} {\bibfnamefont
  {G.}~\bibnamefont {Savard}}, \bibinfo {author} {\bibfnamefont
  {M.}~\bibnamefont {Scheck}}, \bibinfo {author} {\bibfnamefont
  {D.}~\bibnamefont {Seweryniak}}, \bibinfo {author} {\bibfnamefont {M.~K.}\
  \bibnamefont {Smith}}, \bibinfo {author} {\bibfnamefont {R.}~\bibnamefont
  {Vondrasek}},\ and\ \bibinfo {author} {\bibfnamefont {A.}~\bibnamefont
  {Wiens}},\ }\href {https://doi.org/10.1103/PhysRevLett.116.112503} {\bibfield
   {journal} {\bibinfo  {journal} {Phys. Rev. Lett.}\ }\textbf {\bibinfo
  {volume} {116}},\ \bibinfo {pages} {112503} (\bibinfo {year}
  {2016})}\BibitemShut {NoStop}%
\bibitem [{\citenamefont {Bucher}\ \emph {et~al.}(2017)\citenamefont {Bucher},
  \citenamefont {Zhu}, \citenamefont {Wu}, \citenamefont {Janssens},
  \citenamefont {Bernard}, \citenamefont {Robledo}, \citenamefont
  {Rodr\'{\i}guez}, \citenamefont {Cline}, \citenamefont {Hayes}, \citenamefont
  {Ayangeakaa}, \citenamefont {Buckner}, \citenamefont {Campbell},
  \citenamefont {Carpenter}, \citenamefont {Clark}, \citenamefont {Crawford},
  \citenamefont {David}, \citenamefont {Dickerson}, \citenamefont {Harker},
  \citenamefont {Hoffman}, \citenamefont {Kay}, \citenamefont {Kondev},
  \citenamefont {Lauritsen}, \citenamefont {Macchiavelli}, \citenamefont
  {Pardo}, \citenamefont {Savard}, \citenamefont {Seweryniak},\ and\
  \citenamefont {Vondrasek}}]{bucher2017}%
  \BibitemOpen
  \bibfield  {author} {\bibinfo {author} {\bibfnamefont {B.}~\bibnamefont
  {Bucher}}, \bibinfo {author} {\bibfnamefont {S.}~\bibnamefont {Zhu}},
  \bibinfo {author} {\bibfnamefont {C.~Y.}\ \bibnamefont {Wu}}, \bibinfo
  {author} {\bibfnamefont {R.~V.~F.}\ \bibnamefont {Janssens}}, \bibinfo
  {author} {\bibfnamefont {R.~N.}\ \bibnamefont {Bernard}}, \bibinfo {author}
  {\bibfnamefont {L.~M.}\ \bibnamefont {Robledo}}, \bibinfo {author}
  {\bibfnamefont {T.~R.}\ \bibnamefont {Rodr\'{\i}guez}}, \bibinfo {author}
  {\bibfnamefont {D.}~\bibnamefont {Cline}}, \bibinfo {author} {\bibfnamefont
  {A.~B.}\ \bibnamefont {Hayes}}, \bibinfo {author} {\bibfnamefont {A.~D.}\
  \bibnamefont {Ayangeakaa}}, \bibinfo {author} {\bibfnamefont {M.~Q.}\
  \bibnamefont {Buckner}}, \bibinfo {author} {\bibfnamefont {C.~M.}\
  \bibnamefont {Campbell}}, \bibinfo {author} {\bibfnamefont {M.~P.}\
  \bibnamefont {Carpenter}}, \bibinfo {author} {\bibfnamefont {J.~A.}\
  \bibnamefont {Clark}}, \bibinfo {author} {\bibfnamefont {H.~L.}\ \bibnamefont
  {Crawford}}, \bibinfo {author} {\bibfnamefont {H.~M.}\ \bibnamefont {David}},
  \bibinfo {author} {\bibfnamefont {C.}~\bibnamefont {Dickerson}}, \bibinfo
  {author} {\bibfnamefont {J.}~\bibnamefont {Harker}}, \bibinfo {author}
  {\bibfnamefont {C.~R.}\ \bibnamefont {Hoffman}}, \bibinfo {author}
  {\bibfnamefont {B.~P.}\ \bibnamefont {Kay}}, \bibinfo {author} {\bibfnamefont
  {F.~G.}\ \bibnamefont {Kondev}}, \bibinfo {author} {\bibfnamefont
  {T.}~\bibnamefont {Lauritsen}}, \bibinfo {author} {\bibfnamefont {A.~O.}\
  \bibnamefont {Macchiavelli}}, \bibinfo {author} {\bibfnamefont {R.~C.}\
  \bibnamefont {Pardo}}, \bibinfo {author} {\bibfnamefont {G.}~\bibnamefont
  {Savard}}, \bibinfo {author} {\bibfnamefont {D.}~\bibnamefont {Seweryniak}},\
  and\ \bibinfo {author} {\bibfnamefont {R.}~\bibnamefont {Vondrasek}},\ }\href
  {https://doi.org/10.1103/PhysRevLett.118.152504} {\bibfield  {journal}
  {\bibinfo  {journal} {Phys. Rev. Lett.}\ }\textbf {\bibinfo {volume} {118}},\
  \bibinfo {pages} {152504} (\bibinfo {year} {2017})}\BibitemShut {NoStop}%
\bibitem [{\citenamefont {Rugari}\ \emph {et~al.}(1993)\citenamefont {Rugari},
  \citenamefont {France}, \citenamefont {Lund}, \citenamefont {Zhao},
  \citenamefont {Gai}, \citenamefont {Butler}, \citenamefont {Holliday},
  \citenamefont {James}, \citenamefont {Jones}, \citenamefont {Poynter},
  \citenamefont {Tanner}, \citenamefont {Ying},\ and\ \citenamefont
  {Simpson}}]{rugari1993}%
  \BibitemOpen
  \bibfield  {author} {\bibinfo {author} {\bibfnamefont {S.~L.}\ \bibnamefont
  {Rugari}}, \bibinfo {author} {\bibfnamefont {R.~H.}\ \bibnamefont {France}},
  \bibinfo {author} {\bibfnamefont {B.~J.}\ \bibnamefont {Lund}}, \bibinfo
  {author} {\bibfnamefont {Z.}~\bibnamefont {Zhao}}, \bibinfo {author}
  {\bibfnamefont {M.}~\bibnamefont {Gai}}, \bibinfo {author} {\bibfnamefont
  {P.~A.}\ \bibnamefont {Butler}}, \bibinfo {author} {\bibfnamefont {V.~A.}\
  \bibnamefont {Holliday}}, \bibinfo {author} {\bibfnamefont {A.~N.}\
  \bibnamefont {James}}, \bibinfo {author} {\bibfnamefont {G.~D.}\ \bibnamefont
  {Jones}}, \bibinfo {author} {\bibfnamefont {R.~J.}\ \bibnamefont {Poynter}},
  \bibinfo {author} {\bibfnamefont {R.~J.}\ \bibnamefont {Tanner}}, \bibinfo
  {author} {\bibfnamefont {K.~L.}\ \bibnamefont {Ying}},\ and\ \bibinfo
  {author} {\bibfnamefont {J.}~\bibnamefont {Simpson}},\ }\href
  {https://doi.org/10.1103/PhysRevC.48.2078} {\bibfield  {journal} {\bibinfo
  {journal} {Phys. Rev. C}\ }\textbf {\bibinfo {volume} {48}},\ \bibinfo
  {pages} {2078} (\bibinfo {year} {1993})}\BibitemShut {NoStop}%
\bibitem [{\citenamefont {Fahlander}\ \emph {et~al.}(1994)\citenamefont
  {Fahlander}, \citenamefont {Seweryniak}, \citenamefont {Nyberg},
  \citenamefont {Dombrádi}, \citenamefont {Perez}, \citenamefont {Józsa},
  \citenamefont {Nyakó}, \citenamefont {Atac}, \citenamefont {Cederwall},
  \citenamefont {Johnson}, \citenamefont {Kerek}, \citenamefont {Kownacki},
  \citenamefont {Norlin}, \citenamefont {Wyss}, \citenamefont {Adamides},
  \citenamefont {Ideguchi}, \citenamefont {Julin}, \citenamefont {Juutinen},
  \citenamefont {Karczmarczyk}, \citenamefont {Mitarai}, \citenamefont
  {Piiparinen}, \citenamefont {Schubart}, \citenamefont {Sletten},
  \citenamefont {Törmänen},\ and\ \citenamefont {Virtanen}}]{FAHLANDER1994}%
  \BibitemOpen
  \bibfield  {author} {\bibinfo {author} {\bibfnamefont {C.}~\bibnamefont
  {Fahlander}}, \bibinfo {author} {\bibfnamefont {D.}~\bibnamefont
  {Seweryniak}}, \bibinfo {author} {\bibfnamefont {J.}~\bibnamefont {Nyberg}},
  \bibinfo {author} {\bibfnamefont {Z.}~\bibnamefont {Dombrádi}}, \bibinfo
  {author} {\bibfnamefont {G.}~\bibnamefont {Perez}}, \bibinfo {author}
  {\bibfnamefont {M.}~\bibnamefont {Józsa}}, \bibinfo {author} {\bibfnamefont
  {B.}~\bibnamefont {Nyakó}}, \bibinfo {author} {\bibfnamefont
  {A.}~\bibnamefont {Atac}}, \bibinfo {author} {\bibfnamefont {B.}~\bibnamefont
  {Cederwall}}, \bibinfo {author} {\bibfnamefont {A.}~\bibnamefont {Johnson}},
  \bibinfo {author} {\bibfnamefont {A.}~\bibnamefont {Kerek}}, \bibinfo
  {author} {\bibfnamefont {J.}~\bibnamefont {Kownacki}}, \bibinfo {author}
  {\bibfnamefont {L.-O.}\ \bibnamefont {Norlin}}, \bibinfo {author}
  {\bibfnamefont {R.}~\bibnamefont {Wyss}}, \bibinfo {author} {\bibfnamefont
  {E.}~\bibnamefont {Adamides}}, \bibinfo {author} {\bibfnamefont
  {E.}~\bibnamefont {Ideguchi}}, \bibinfo {author} {\bibfnamefont
  {R.}~\bibnamefont {Julin}}, \bibinfo {author} {\bibfnamefont
  {S.}~\bibnamefont {Juutinen}}, \bibinfo {author} {\bibfnamefont
  {W.}~\bibnamefont {Karczmarczyk}}, \bibinfo {author} {\bibfnamefont
  {S.}~\bibnamefont {Mitarai}}, \bibinfo {author} {\bibfnamefont
  {M.}~\bibnamefont {Piiparinen}}, \bibinfo {author} {\bibfnamefont
  {R.}~\bibnamefont {Schubart}}, \bibinfo {author} {\bibfnamefont
  {G.}~\bibnamefont {Sletten}}, \bibinfo {author} {\bibfnamefont
  {S.}~\bibnamefont {Törmänen}},\ and\ \bibinfo {author} {\bibfnamefont
  {A.}~\bibnamefont {Virtanen}},\ }\href
  {https://doi.org/https://doi.org/10.1016/0375-9474(94)90944-X} {\bibfield
  {journal} {\bibinfo  {journal} {Nucl. Phys. A}\ }\textbf {\bibinfo {volume}
  {577}},\ \bibinfo {pages} {773 } (\bibinfo {year} {1994})}\BibitemShut
  {NoStop}%
\bibitem [{\citenamefont {Smith}\ \emph {et~al.}(1998)\citenamefont {Smith},
  \citenamefont {Chiara}, \citenamefont {Fossan}, \citenamefont {Lane},
  \citenamefont {Sears}, \citenamefont {Thorslund}, \citenamefont {Amro},
  \citenamefont {Davids}, \citenamefont {Janssens}, \citenamefont {Seweryniak},
  \citenamefont {Hibbert}, \citenamefont {Wadsworth}, \citenamefont {Lee},\
  and\ \citenamefont {Macchiavelli}}]{smith1998}%
  \BibitemOpen
  \bibfield  {author} {\bibinfo {author} {\bibfnamefont {J.~F.}\ \bibnamefont
  {Smith}}, \bibinfo {author} {\bibfnamefont {C.~J.}\ \bibnamefont {Chiara}},
  \bibinfo {author} {\bibfnamefont {D.~B.}\ \bibnamefont {Fossan}}, \bibinfo
  {author} {\bibfnamefont {G.~J.}\ \bibnamefont {Lane}}, \bibinfo {author}
  {\bibfnamefont {J.~M.}\ \bibnamefont {Sears}}, \bibinfo {author}
  {\bibfnamefont {I.}~\bibnamefont {Thorslund}}, \bibinfo {author}
  {\bibfnamefont {H.}~\bibnamefont {Amro}}, \bibinfo {author} {\bibfnamefont
  {C.~N.}\ \bibnamefont {Davids}}, \bibinfo {author} {\bibfnamefont {R.~V.~F.}\
  \bibnamefont {Janssens}}, \bibinfo {author} {\bibfnamefont {D.}~\bibnamefont
  {Seweryniak}}, \bibinfo {author} {\bibfnamefont {I.~M.}\ \bibnamefont
  {Hibbert}}, \bibinfo {author} {\bibfnamefont {R.}~\bibnamefont {Wadsworth}},
  \bibinfo {author} {\bibfnamefont {I.~Y.}\ \bibnamefont {Lee}},\ and\ \bibinfo
  {author} {\bibfnamefont {A.~O.}\ \bibnamefont {Macchiavelli}},\ }\href
  {https://doi.org/10.1103/PhysRevC.57.R1037} {\bibfield  {journal} {\bibinfo
  {journal} {Phys. Rev. C}\ }\textbf {\bibinfo {volume} {57}},\ \bibinfo
  {pages} {R1037} (\bibinfo {year} {1998})}\BibitemShut {NoStop}%
\bibitem [{\citenamefont {DeGraaf}\ \emph {et~al.}(1998)\citenamefont
  {DeGraaf}, \citenamefont {Cromaz}, \citenamefont {Drake}, \citenamefont
  {Janzen}, \citenamefont {Radford},\ and\ \citenamefont {Ward}}]{degraaf1998}%
  \BibitemOpen
  \bibfield  {author} {\bibinfo {author} {\bibfnamefont {J.}~\bibnamefont
  {DeGraaf}}, \bibinfo {author} {\bibfnamefont {M.}~\bibnamefont {Cromaz}},
  \bibinfo {author} {\bibfnamefont {T.~E.}\ \bibnamefont {Drake}}, \bibinfo
  {author} {\bibfnamefont {V.~P.}\ \bibnamefont {Janzen}}, \bibinfo {author}
  {\bibfnamefont {D.~C.}\ \bibnamefont {Radford}},\ and\ \bibinfo {author}
  {\bibfnamefont {D.}~\bibnamefont {Ward}},\ }\href
  {https://doi.org/10.1103/PhysRevC.58.164} {\bibfield  {journal} {\bibinfo
  {journal} {Phys. Rev. C}\ }\textbf {\bibinfo {volume} {58}},\ \bibinfo
  {pages} {164} (\bibinfo {year} {1998})}\BibitemShut {NoStop}%
\bibitem [{\citenamefont {Paul}\ \emph {et~al.}(2000)\citenamefont {Paul},
  \citenamefont {Scraggs}, \citenamefont {Boston}, \citenamefont {Dorvaux},
  \citenamefont {Greenlees}, \citenamefont {Helariutta}, \citenamefont {Jones},
  \citenamefont {Julin}, \citenamefont {Juutinen}, \citenamefont
  {Kankaanp\"a\"a}, \citenamefont {Kettunen}, \citenamefont {Muikku},
  \citenamefont {Nieminen}, \citenamefont {Rahkila},\ and\ \citenamefont
  {Stezowski}}]{paul2000}%
  \BibitemOpen
  \bibfield  {author} {\bibinfo {author} {\bibfnamefont {E.}~\bibnamefont
  {Paul}}, \bibinfo {author} {\bibfnamefont {H.}~\bibnamefont {Scraggs}},
  \bibinfo {author} {\bibfnamefont {A.}~\bibnamefont {Boston}}, \bibinfo
  {author} {\bibfnamefont {O.}~\bibnamefont {Dorvaux}}, \bibinfo {author}
  {\bibfnamefont {P.}~\bibnamefont {Greenlees}}, \bibinfo {author}
  {\bibfnamefont {K.}~\bibnamefont {Helariutta}}, \bibinfo {author}
  {\bibfnamefont {P.}~\bibnamefont {Jones}}, \bibinfo {author} {\bibfnamefont
  {R.}~\bibnamefont {Julin}}, \bibinfo {author} {\bibfnamefont
  {S.}~\bibnamefont {Juutinen}}, \bibinfo {author} {\bibfnamefont
  {H.}~\bibnamefont {Kankaanp\"a\"a}}, \bibinfo {author} {\bibfnamefont
  {H.}~\bibnamefont {Kettunen}}, \bibinfo {author} {\bibfnamefont
  {M.}~\bibnamefont {Muikku}}, \bibinfo {author} {\bibfnamefont
  {P.}~\bibnamefont {Nieminen}}, \bibinfo {author} {\bibfnamefont
  {P.}~\bibnamefont {Rahkila}},\ and\ \bibinfo {author} {\bibfnamefont
  {O.}~\bibnamefont {Stezowski}},\ }\href
  {https://doi.org/https://doi.org/10.1016/S0375-9474(00)00093-2} {\bibfield
  {journal} {\bibinfo  {journal} {Nucl. Phys. A}\ }\textbf {\bibinfo {volume}
  {673}},\ \bibinfo {pages} {31} (\bibinfo {year} {2000})}\BibitemShut
  {NoStop}%
\bibitem [{\citenamefont {Rzaca-Urban}\ \emph {et~al.}(2000)\citenamefont
  {Rzaca-Urban}, \citenamefont {Urban}, \citenamefont {Kaczor}, \citenamefont
  {Durell}, \citenamefont {Leddy}, \citenamefont {Jones}, \citenamefont
  {Phillips}, \citenamefont {Smith}, \citenamefont {Varley}, \citenamefont
  {Ahmad}, \citenamefont {Morss}, \citenamefont {Bentaleb}, \citenamefont
  {Lubkiewicz},\ and\ \citenamefont {Schulz}}]{rzacaurban2000}%
  \BibitemOpen
  \bibfield  {author} {\bibinfo {author} {\bibfnamefont {T.}~\bibnamefont
  {Rzaca-Urban}}, \bibinfo {author} {\bibfnamefont {W.}~\bibnamefont {Urban}},
  \bibinfo {author} {\bibfnamefont {A.}~\bibnamefont {Kaczor}}, \bibinfo
  {author} {\bibfnamefont {J.~L.}\ \bibnamefont {Durell}}, \bibinfo {author}
  {\bibfnamefont {M.~J.}\ \bibnamefont {Leddy}}, \bibinfo {author}
  {\bibfnamefont {M.~A.}\ \bibnamefont {Jones}}, \bibinfo {author}
  {\bibfnamefont {W.~R.}\ \bibnamefont {Phillips}}, \bibinfo {author}
  {\bibfnamefont {A.~G.}\ \bibnamefont {Smith}}, \bibinfo {author}
  {\bibfnamefont {B.~J.}\ \bibnamefont {Varley}}, \bibinfo {author}
  {\bibfnamefont {I.}~\bibnamefont {Ahmad}}, \bibinfo {author} {\bibfnamefont
  {L.~R.}\ \bibnamefont {Morss}}, \bibinfo {author} {\bibfnamefont
  {M.}~\bibnamefont {Bentaleb}}, \bibinfo {author} {\bibfnamefont
  {E.}~\bibnamefont {Lubkiewicz}},\ and\ \bibinfo {author} {\bibfnamefont
  {N.}~\bibnamefont {Schulz}},\ }\href {https://doi.org/10.1007/s100500070033}
  {\bibfield  {journal} {\bibinfo  {journal} {Eur. Phys. J. A}\ }\textbf
  {\bibinfo {volume} {9}},\ \bibinfo {pages} {165} (\bibinfo {year}
  {2000})}\BibitemShut {NoStop}%
\bibitem [{\citenamefont {Smith}\ \emph {et~al.}(2001)\citenamefont {Smith},
  \citenamefont {Chiara}, \citenamefont {Fossan}, \citenamefont {LaFosse},
  \citenamefont {Lane}, \citenamefont {Sears}, \citenamefont {Starosta},
  \citenamefont {Devlin}, \citenamefont {Lerma}, \citenamefont {Sarantites},
  \citenamefont {Freeman}, \citenamefont {Leddy}, \citenamefont {Durell},
  \citenamefont {Boston}, \citenamefont {Paul}, \citenamefont {Semple},
  \citenamefont {Lee}, \citenamefont {Macchiavelli},\ and\ \citenamefont
  {Heenen}}]{SMITH2001}%
  \BibitemOpen
  \bibfield  {author} {\bibinfo {author} {\bibfnamefont {J.}~\bibnamefont
  {Smith}}, \bibinfo {author} {\bibfnamefont {C.}~\bibnamefont {Chiara}},
  \bibinfo {author} {\bibfnamefont {D.}~\bibnamefont {Fossan}}, \bibinfo
  {author} {\bibfnamefont {D.}~\bibnamefont {LaFosse}}, \bibinfo {author}
  {\bibfnamefont {G.}~\bibnamefont {Lane}}, \bibinfo {author} {\bibfnamefont
  {J.}~\bibnamefont {Sears}}, \bibinfo {author} {\bibfnamefont
  {K.}~\bibnamefont {Starosta}}, \bibinfo {author} {\bibfnamefont
  {M.}~\bibnamefont {Devlin}}, \bibinfo {author} {\bibfnamefont
  {F.}~\bibnamefont {Lerma}}, \bibinfo {author} {\bibfnamefont
  {D.}~\bibnamefont {Sarantites}}, \bibinfo {author} {\bibfnamefont
  {S.}~\bibnamefont {Freeman}}, \bibinfo {author} {\bibfnamefont
  {M.}~\bibnamefont {Leddy}}, \bibinfo {author} {\bibfnamefont
  {J.}~\bibnamefont {Durell}}, \bibinfo {author} {\bibfnamefont
  {A.}~\bibnamefont {Boston}}, \bibinfo {author} {\bibfnamefont
  {E.}~\bibnamefont {Paul}}, \bibinfo {author} {\bibfnamefont {A.}~\bibnamefont
  {Semple}}, \bibinfo {author} {\bibfnamefont {I.}~\bibnamefont {Lee}},
  \bibinfo {author} {\bibfnamefont {A.}~\bibnamefont {Macchiavelli}},\ and\
  \bibinfo {author} {\bibfnamefont {P.}~\bibnamefont {Heenen}},\ }\href
  {https://doi.org/https://doi.org/10.1016/S0370-2693(01)01339-9} {\bibfield
  {journal} {\bibinfo  {journal} {Phys. Lett. B}\ }\textbf {\bibinfo {volume}
  {523}},\ \bibinfo {pages} {13 } (\bibinfo {year} {2001})}\BibitemShut
  {NoStop}%
\bibitem [{\citenamefont {{de Angelis}}\ \emph {et~al.}(2002)\citenamefont {{de
  Angelis}}, \citenamefont {Gadea}, \citenamefont {Farnea}, \citenamefont
  {Isocrate}, \citenamefont {Petkov}, \citenamefont {Marginean}, \citenamefont
  {Napoli}, \citenamefont {Dewald}, \citenamefont {Bellato}, \citenamefont
  {Bracco}, \citenamefont {Camera}, \citenamefont {Curien}, \citenamefont
  {De Poli}, \citenamefont {Fioretto}, \citenamefont {Fitzler}, \citenamefont
  {Kasemann}, \citenamefont {Kintz}, \citenamefont {Klug}, \citenamefont
  {Lenzi}, \citenamefont {Lunardi}, \citenamefont {Menegazzo}, \citenamefont
  {Pavan}, \citenamefont {Pedroza}, \citenamefont {Pucknell}, \citenamefont
  {Ring}, \citenamefont {Sampson},\ and\ \citenamefont {Wyss}}]{DEANGELIS2002}%
  \BibitemOpen
  \bibfield  {author} {\bibinfo {author} {\bibfnamefont {G.}~\bibnamefont {{de
  Angelis}}}, \bibinfo {author} {\bibfnamefont {A.}~\bibnamefont {Gadea}},
  \bibinfo {author} {\bibfnamefont {E.}~\bibnamefont {Farnea}}, \bibinfo
  {author} {\bibfnamefont {R.}~\bibnamefont {Isocrate}}, \bibinfo {author}
  {\bibfnamefont {P.}~\bibnamefont {Petkov}}, \bibinfo {author} {\bibfnamefont
  {N.}~\bibnamefont {Marginean}}, \bibinfo {author} {\bibfnamefont
  {D.}~\bibnamefont {Napoli}}, \bibinfo {author} {\bibfnamefont
  {A.}~\bibnamefont {Dewald}}, \bibinfo {author} {\bibfnamefont
  {M.}~\bibnamefont {Bellato}}, \bibinfo {author} {\bibfnamefont
  {A.}~\bibnamefont {Bracco}}, \bibinfo {author} {\bibfnamefont
  {F.}~\bibnamefont {Camera}}, \bibinfo {author} {\bibfnamefont
  {D.}~\bibnamefont {Curien}}, \bibinfo {author} {\bibfnamefont
  {M.}~\bibnamefont {De Poli}}, \bibinfo {author} {\bibfnamefont
  {E.}~\bibnamefont {Fioretto}}, \bibinfo {author} {\bibfnamefont
  {A.}~\bibnamefont {Fitzler}}, \bibinfo {author} {\bibfnamefont
  {S.}~\bibnamefont {Kasemann}}, \bibinfo {author} {\bibfnamefont
  {N.}~\bibnamefont {Kintz}}, \bibinfo {author} {\bibfnamefont
  {T.}~\bibnamefont {Klug}}, \bibinfo {author} {\bibfnamefont {S.}~\bibnamefont
  {Lenzi}}, \bibinfo {author} {\bibfnamefont {S.}~\bibnamefont {Lunardi}},
  \bibinfo {author} {\bibfnamefont {R.}~\bibnamefont {Menegazzo}}, \bibinfo
  {author} {\bibfnamefont {P.}~\bibnamefont {Pavan}}, \bibinfo {author}
  {\bibfnamefont {J.}~\bibnamefont {Pedroza}}, \bibinfo {author} {\bibfnamefont
  {V.}~\bibnamefont {Pucknell}}, \bibinfo {author} {\bibfnamefont
  {C.}~\bibnamefont {Ring}}, \bibinfo {author} {\bibfnamefont {J.}~\bibnamefont
  {Sampson}},\ and\ \bibinfo {author} {\bibfnamefont {R.}~\bibnamefont
  {Wyss}},\ }\href
  {https://doi.org/https://doi.org/10.1016/S0370-2693(02)01728-8} {\bibfield
  {journal} {\bibinfo  {journal} {Phys. Lett. B}\ }\textbf {\bibinfo {volume}
  {535}},\ \bibinfo {pages} {93} (\bibinfo {year} {2002})}\BibitemShut
  {NoStop}%
\bibitem [{\citenamefont {Capponi}\ \emph {et~al.}(2016)\citenamefont
  {Capponi}, \citenamefont {Smith}, \citenamefont {Ruotsalainen}, \citenamefont
  {Scholey}, \citenamefont {Rahkila}, \citenamefont {Auranen}, \citenamefont
  {Bianco}, \citenamefont {Boston}, \citenamefont {Boston}, \citenamefont
  {Cullen}, \citenamefont {Derkx}, \citenamefont {Drummond}, \citenamefont
  {Grahn}, \citenamefont {Greenlees}, \citenamefont {Grocutt}, \citenamefont
  {Hadinia}, \citenamefont {Jakobsson}, \citenamefont {Joss}, \citenamefont
  {Julin}, \citenamefont {Juutinen}, \citenamefont {Labiche}, \citenamefont
  {Leino}, \citenamefont {Leach}, \citenamefont {McPeake}, \citenamefont
  {Mulholland}, \citenamefont {Nieminen}, \citenamefont {O'Donnell},
  \citenamefont {Paul}, \citenamefont {Peura}, \citenamefont {Sandzelius},
  \citenamefont {Sar\'en}, \citenamefont {Saygi}, \citenamefont {Sorri},
  \citenamefont {Stolze}, \citenamefont {Thornthwaite}, \citenamefont
  {Taylor},\ and\ \citenamefont {Uusitalo}}]{capponi2016}%
  \BibitemOpen
  \bibfield  {author} {\bibinfo {author} {\bibfnamefont {L.}~\bibnamefont
  {Capponi}}, \bibinfo {author} {\bibfnamefont {J.~F.}\ \bibnamefont {Smith}},
  \bibinfo {author} {\bibfnamefont {P.}~\bibnamefont {Ruotsalainen}}, \bibinfo
  {author} {\bibfnamefont {C.}~\bibnamefont {Scholey}}, \bibinfo {author}
  {\bibfnamefont {P.}~\bibnamefont {Rahkila}}, \bibinfo {author} {\bibfnamefont
  {K.}~\bibnamefont {Auranen}}, \bibinfo {author} {\bibfnamefont
  {L.}~\bibnamefont {Bianco}}, \bibinfo {author} {\bibfnamefont {A.~J.}\
  \bibnamefont {Boston}}, \bibinfo {author} {\bibfnamefont {H.~C.}\
  \bibnamefont {Boston}}, \bibinfo {author} {\bibfnamefont {D.~M.}\
  \bibnamefont {Cullen}}, \bibinfo {author} {\bibfnamefont {X.}~\bibnamefont
  {Derkx}}, \bibinfo {author} {\bibfnamefont {M.~C.}\ \bibnamefont {Drummond}},
  \bibinfo {author} {\bibfnamefont {T.}~\bibnamefont {Grahn}}, \bibinfo
  {author} {\bibfnamefont {P.~T.}\ \bibnamefont {Greenlees}}, \bibinfo {author}
  {\bibfnamefont {L.}~\bibnamefont {Grocutt}}, \bibinfo {author} {\bibfnamefont
  {B.}~\bibnamefont {Hadinia}}, \bibinfo {author} {\bibfnamefont
  {U.}~\bibnamefont {Jakobsson}}, \bibinfo {author} {\bibfnamefont {D.~T.}\
  \bibnamefont {Joss}}, \bibinfo {author} {\bibfnamefont {R.}~\bibnamefont
  {Julin}}, \bibinfo {author} {\bibfnamefont {S.}~\bibnamefont {Juutinen}},
  \bibinfo {author} {\bibfnamefont {M.}~\bibnamefont {Labiche}}, \bibinfo
  {author} {\bibfnamefont {M.}~\bibnamefont {Leino}}, \bibinfo {author}
  {\bibfnamefont {K.~G.}\ \bibnamefont {Leach}}, \bibinfo {author}
  {\bibfnamefont {C.}~\bibnamefont {McPeake}}, \bibinfo {author} {\bibfnamefont
  {K.~F.}\ \bibnamefont {Mulholland}}, \bibinfo {author} {\bibfnamefont
  {P.}~\bibnamefont {Nieminen}}, \bibinfo {author} {\bibfnamefont
  {D.}~\bibnamefont {O'Donnell}}, \bibinfo {author} {\bibfnamefont {E.~S.}\
  \bibnamefont {Paul}}, \bibinfo {author} {\bibfnamefont {P.}~\bibnamefont
  {Peura}}, \bibinfo {author} {\bibfnamefont {M.}~\bibnamefont {Sandzelius}},
  \bibinfo {author} {\bibfnamefont {J.}~\bibnamefont {Sar\'en}}, \bibinfo
  {author} {\bibfnamefont {B.}~\bibnamefont {Saygi}}, \bibinfo {author}
  {\bibfnamefont {J.}~\bibnamefont {Sorri}}, \bibinfo {author} {\bibfnamefont
  {S.}~\bibnamefont {Stolze}}, \bibinfo {author} {\bibfnamefont
  {A.}~\bibnamefont {Thornthwaite}}, \bibinfo {author} {\bibfnamefont {M.~J.}\
  \bibnamefont {Taylor}},\ and\ \bibinfo {author} {\bibfnamefont
  {J.}~\bibnamefont {Uusitalo}},\ }\href
  {https://doi.org/10.1103/PhysRevC.94.024314} {\bibfield  {journal} {\bibinfo
  {journal} {Phys. Rev. C}\ }\textbf {\bibinfo {volume} {94}},\ \bibinfo
  {pages} {024314} (\bibinfo {year} {2016})}\BibitemShut {NoStop}%
\bibitem [{\citenamefont {Gregor}\ \emph {et~al.}(2017)\citenamefont {Gregor},
  \citenamefont {Scheck}, \citenamefont {Chapman}, \citenamefont {Gaffney},
  \citenamefont {Keatings}, \citenamefont {Mashtakov}, \citenamefont
  {O'Donnell}, \citenamefont {Smith}, \citenamefont {Spagnoletti},
  \citenamefont {Th{\"u}rauf}, \citenamefont {Werner},\ and\ \citenamefont
  {Wiseman}}]{gregor2017}%
  \BibitemOpen
  \bibfield  {author} {\bibinfo {author} {\bibfnamefont {E.~T.}\ \bibnamefont
  {Gregor}}, \bibinfo {author} {\bibfnamefont {M.}~\bibnamefont {Scheck}},
  \bibinfo {author} {\bibfnamefont {R.}~\bibnamefont {Chapman}}, \bibinfo
  {author} {\bibfnamefont {L.~P.}\ \bibnamefont {Gaffney}}, \bibinfo {author}
  {\bibfnamefont {J.}~\bibnamefont {Keatings}}, \bibinfo {author}
  {\bibfnamefont {K.~R.}\ \bibnamefont {Mashtakov}}, \bibinfo {author}
  {\bibfnamefont {D.}~\bibnamefont {O'Donnell}}, \bibinfo {author}
  {\bibfnamefont {J.~F.}\ \bibnamefont {Smith}}, \bibinfo {author}
  {\bibfnamefont {P.}~\bibnamefont {Spagnoletti}}, \bibinfo {author}
  {\bibfnamefont {M.}~\bibnamefont {Th{\"u}rauf}}, \bibinfo {author}
  {\bibfnamefont {V.}~\bibnamefont {Werner}},\ and\ \bibinfo {author}
  {\bibfnamefont {C.}~\bibnamefont {Wiseman}},\ }\href
  {https://doi.org/10.1140/epja/i2017-12224-7} {\bibfield  {journal} {\bibinfo
  {journal} {Eur. Phys. J. A}\ }\textbf {\bibinfo {volume} {53}},\ \bibinfo
  {pages} {50} (\bibinfo {year} {2017})}\BibitemShut {NoStop}%
\bibitem [{\citenamefont {Nazarewicz}\ \emph {et~al.}(1984)\citenamefont
  {Nazarewicz}, \citenamefont {Olanders}, \citenamefont {Ragnarsson},
  \citenamefont {Dudek}, \citenamefont {Leander}, \citenamefont {M\"oller},\
  and\ \citenamefont {Ruchowsa}}]{naza1984b}%
  \BibitemOpen
  \bibfield  {author} {\bibinfo {author} {\bibfnamefont {W.}~\bibnamefont
  {Nazarewicz}}, \bibinfo {author} {\bibfnamefont {P.}~\bibnamefont
  {Olanders}}, \bibinfo {author} {\bibfnamefont {I.}~\bibnamefont
  {Ragnarsson}}, \bibinfo {author} {\bibfnamefont {J.}~\bibnamefont {Dudek}},
  \bibinfo {author} {\bibfnamefont {G.~A.}\ \bibnamefont {Leander}}, \bibinfo
  {author} {\bibfnamefont {P.}~\bibnamefont {M\"oller}},\ and\ \bibinfo
  {author} {\bibfnamefont {E.}~\bibnamefont {Ruchowsa}},\ }\href
  {https://doi.org/10.1016/0375-9474(84)90208-2} {\bibfield  {journal}
  {\bibinfo  {journal} {Nucl. Phys. A}\ }\textbf {\bibinfo {volume} {429}},\
  \bibinfo {pages} {269 } (\bibinfo {year} {1984})}\BibitemShut {NoStop}%
\bibitem [{\citenamefont {Leander}\ \emph {et~al.}(1985)\citenamefont
  {Leander}, \citenamefont {Nazarewicz}, \citenamefont {Olanders},
  \citenamefont {Ragnarsson},\ and\ \citenamefont {Dudek}}]{leander1985}%
  \BibitemOpen
  \bibfield  {author} {\bibinfo {author} {\bibfnamefont {G.}~\bibnamefont
  {Leander}}, \bibinfo {author} {\bibfnamefont {W.}~\bibnamefont {Nazarewicz}},
  \bibinfo {author} {\bibfnamefont {P.}~\bibnamefont {Olanders}}, \bibinfo
  {author} {\bibfnamefont {I.}~\bibnamefont {Ragnarsson}},\ and\ \bibinfo
  {author} {\bibfnamefont {J.}~\bibnamefont {Dudek}},\ }\href
  {https://doi.org/https://doi.org/10.1016/0370-2693(85)90496-4} {\bibfield
  {journal} {\bibinfo  {journal} {Phys. Lett. B}\ }\textbf {\bibinfo {volume}
  {152}},\ \bibinfo {pages} {284 } (\bibinfo {year} {1985})}\BibitemShut
  {NoStop}%
\bibitem [{\citenamefont {Möller}\ \emph {et~al.}(2008)\citenamefont
  {Möller}, \citenamefont {Bengtsson}, \citenamefont {Carlsson}, \citenamefont
  {Olivius}, \citenamefont {Ichikawa}, \citenamefont {Sagawa},\ and\
  \citenamefont {Iwamoto}}]{moeller2008}%
  \BibitemOpen
  \bibfield  {author} {\bibinfo {author} {\bibfnamefont {P.}~\bibnamefont
  {Möller}}, \bibinfo {author} {\bibfnamefont {R.}~\bibnamefont {Bengtsson}},
  \bibinfo {author} {\bibfnamefont {B.}~\bibnamefont {Carlsson}}, \bibinfo
  {author} {\bibfnamefont {P.}~\bibnamefont {Olivius}}, \bibinfo {author}
  {\bibfnamefont {T.}~\bibnamefont {Ichikawa}}, \bibinfo {author}
  {\bibfnamefont {H.}~\bibnamefont {Sagawa}},\ and\ \bibinfo {author}
  {\bibfnamefont {A.}~\bibnamefont {Iwamoto}},\ }\href
  {https://doi.org/https://doi.org/10.1016/j.adt.2008.05.002} {\bibfield
  {journal} {\bibinfo  {journal} {At. Dat. Nucl. Dat. Tab.}\ }\textbf {\bibinfo
  {volume} {94}},\ \bibinfo {pages} {758 } (\bibinfo {year}
  {2008})}\BibitemShut {NoStop}%
\bibitem [{\citenamefont {Marcos}\ \emph {et~al.}(1983)\citenamefont {Marcos},
  \citenamefont {Flocard},\ and\ \citenamefont {Heenen}}]{MARCOS1983}%
  \BibitemOpen
  \bibfield  {author} {\bibinfo {author} {\bibfnamefont {S.}~\bibnamefont
  {Marcos}}, \bibinfo {author} {\bibfnamefont {H.}~\bibnamefont {Flocard}},\
  and\ \bibinfo {author} {\bibfnamefont {P.}~\bibnamefont {Heenen}},\ }\href
  {https://doi.org/https://doi.org/10.1016/0375-9474(83)90405-0} {\bibfield
  {journal} {\bibinfo  {journal} {Nucl. Phys. A}\ }\textbf {\bibinfo {volume}
  {410}},\ \bibinfo {pages} {125 } (\bibinfo {year} {1983})}\BibitemShut
  {NoStop}%
\bibitem [{\citenamefont {Bonche}\ \emph {et~al.}(1986)\citenamefont {Bonche},
  \citenamefont {Heenen}, \citenamefont {Flocard},\ and\ \citenamefont
  {Vautherin}}]{BONCHE1986}%
  \BibitemOpen
  \bibfield  {author} {\bibinfo {author} {\bibfnamefont {P.}~\bibnamefont
  {Bonche}}, \bibinfo {author} {\bibfnamefont {P.}~\bibnamefont {Heenen}},
  \bibinfo {author} {\bibfnamefont {H.}~\bibnamefont {Flocard}},\ and\ \bibinfo
  {author} {\bibfnamefont {D.}~\bibnamefont {Vautherin}},\ }\href
  {https://doi.org/https://doi.org/10.1016/0370-2693(86)90609-X} {\bibfield
  {journal} {\bibinfo  {journal} {Phys. Lett. B}\ }\textbf {\bibinfo {volume}
  {175}},\ \bibinfo {pages} {387 } (\bibinfo {year} {1986})}\BibitemShut
  {NoStop}%
\bibitem [{\citenamefont {Bonche}\ \emph {et~al.}(1991)\citenamefont {Bonche},
  \citenamefont {Krieger}, \citenamefont {Weiss}, \citenamefont {Dobaczewski},
  \citenamefont {Flocard},\ and\ \citenamefont {Heenen}}]{BONCHE1991}%
  \BibitemOpen
  \bibfield  {author} {\bibinfo {author} {\bibfnamefont {P.}~\bibnamefont
  {Bonche}}, \bibinfo {author} {\bibfnamefont {S.~J.}\ \bibnamefont {Krieger}},
  \bibinfo {author} {\bibfnamefont {M.~S.}\ \bibnamefont {Weiss}}, \bibinfo
  {author} {\bibfnamefont {J.}~\bibnamefont {Dobaczewski}}, \bibinfo {author}
  {\bibfnamefont {H.}~\bibnamefont {Flocard}},\ and\ \bibinfo {author}
  {\bibfnamefont {P.-H.}\ \bibnamefont {Heenen}},\ }\href
  {https://doi.org/10.1103/PhysRevLett.66.876} {\bibfield  {journal} {\bibinfo
  {journal} {Phys. Rev. Lett.}\ }\textbf {\bibinfo {volume} {66}},\ \bibinfo
  {pages} {876} (\bibinfo {year} {1991})}\BibitemShut {NoStop}%
\bibitem [{\citenamefont {Heenen}\ \emph {et~al.}(1994)\citenamefont {Heenen},
  \citenamefont {Skalski}, \citenamefont {Bonche},\ and\ \citenamefont
  {Flocard}}]{heenen1994}%
  \BibitemOpen
  \bibfield  {author} {\bibinfo {author} {\bibfnamefont {P.-H.}\ \bibnamefont
  {Heenen}}, \bibinfo {author} {\bibfnamefont {J.}~\bibnamefont {Skalski}},
  \bibinfo {author} {\bibfnamefont {P.}~\bibnamefont {Bonche}},\ and\ \bibinfo
  {author} {\bibfnamefont {H.}~\bibnamefont {Flocard}},\ }\href
  {https://doi.org/10.1103/PhysRevC.50.802} {\bibfield  {journal} {\bibinfo
  {journal} {Phys. Rev. C}\ }\textbf {\bibinfo {volume} {50}},\ \bibinfo
  {pages} {802} (\bibinfo {year} {1994})}\BibitemShut {NoStop}%
\bibitem [{\citenamefont {Robledo}\ \emph {et~al.}(1987)\citenamefont
  {Robledo}, \citenamefont {Egido}, \citenamefont {Berger},\ and\ \citenamefont
  {Girod}}]{ROBLEDO1987}%
  \BibitemOpen
  \bibfield  {author} {\bibinfo {author} {\bibfnamefont {L.~M.}\ \bibnamefont
  {Robledo}}, \bibinfo {author} {\bibfnamefont {J.~L.}\ \bibnamefont {Egido}},
  \bibinfo {author} {\bibfnamefont {J.}~\bibnamefont {Berger}},\ and\ \bibinfo
  {author} {\bibfnamefont {M.}~\bibnamefont {Girod}},\ }\href
  {https://doi.org/https://doi.org/10.1016/0370-2693(87)91085-9} {\bibfield
  {journal} {\bibinfo  {journal} {Phys. Lett. B}\ }\textbf {\bibinfo {volume}
  {187}},\ \bibinfo {pages} {223 } (\bibinfo {year} {1987})}\BibitemShut
  {NoStop}%
\bibitem [{\citenamefont {Robledo}\ \emph {et~al.}(1988)\citenamefont
  {Robledo}, \citenamefont {Egido}, \citenamefont {Nerlo-Pomorska},\ and\
  \citenamefont {Pomorski}}]{ROBLEDO1988}%
  \BibitemOpen
  \bibfield  {author} {\bibinfo {author} {\bibfnamefont {L.~M.}\ \bibnamefont
  {Robledo}}, \bibinfo {author} {\bibfnamefont {J.~L.}\ \bibnamefont {Egido}},
  \bibinfo {author} {\bibfnamefont {B.}~\bibnamefont {Nerlo-Pomorska}},\ and\
  \bibinfo {author} {\bibfnamefont {K.}~\bibnamefont {Pomorski}},\ }\href
  {https://doi.org/https://doi.org/10.1016/0370-2693(88)90592-8} {\bibfield
  {journal} {\bibinfo  {journal} {Phys. Lett. B}\ }\textbf {\bibinfo {volume}
  {201}},\ \bibinfo {pages} {409 } (\bibinfo {year} {1988})}\BibitemShut
  {NoStop}%
\bibitem [{\citenamefont {Egido}\ and\ \citenamefont
  {Robledo}(1990)}]{EGIDO1990}%
  \BibitemOpen
  \bibfield  {author} {\bibinfo {author} {\bibfnamefont {J.~L.}\ \bibnamefont
  {Egido}}\ and\ \bibinfo {author} {\bibfnamefont {L.~M.}\ \bibnamefont
  {Robledo}},\ }\href
  {https://doi.org/https://doi.org/10.1016/0375-9474(90)90141-8} {\bibfield
  {journal} {\bibinfo  {journal} {Nucl. Phys. A}\ }\textbf {\bibinfo {volume}
  {518}},\ \bibinfo {pages} {475 } (\bibinfo {year} {1990})}\BibitemShut
  {NoStop}%
\bibitem [{\citenamefont {Egido}\ and\ \citenamefont
  {Robledo}(1991)}]{EGIDO1991}%
  \BibitemOpen
  \bibfield  {author} {\bibinfo {author} {\bibfnamefont {J.~L.}\ \bibnamefont
  {Egido}}\ and\ \bibinfo {author} {\bibfnamefont {L.~M.}\ \bibnamefont
  {Robledo}},\ }\href
  {https://doi.org/https://doi.org/10.1016/0375-9474(91)90016-Y} {\bibfield
  {journal} {\bibinfo  {journal} {Nucl. Phys. A}\ }\textbf {\bibinfo {volume}
  {524}},\ \bibinfo {pages} {65 } (\bibinfo {year} {1991})}\BibitemShut
  {NoStop}%
\bibitem [{\citenamefont {Egido}\ and\ \citenamefont
  {Robledo}(1992)}]{EGIDO1992}%
  \BibitemOpen
  \bibfield  {author} {\bibinfo {author} {\bibfnamefont {J.~L.}\ \bibnamefont
  {Egido}}\ and\ \bibinfo {author} {\bibfnamefont {L.~M.}\ \bibnamefont
  {Robledo}},\ }\href
  {https://doi.org/https://doi.org/10.1016/0375-9474(92)90294-T} {\bibfield
  {journal} {\bibinfo  {journal} {Nucl. Phys. A}\ }\textbf {\bibinfo {volume}
  {545}},\ \bibinfo {pages} {589 } (\bibinfo {year} {1992})}\BibitemShut
  {NoStop}%
\bibitem [{\citenamefont {Garrote}\ \emph {et~al.}(1998)\citenamefont
  {Garrote}, \citenamefont {Egido},\ and\ \citenamefont
  {Robledo}}]{GARROTE1998}%
  \BibitemOpen
  \bibfield  {author} {\bibinfo {author} {\bibfnamefont {E.}~\bibnamefont
  {Garrote}}, \bibinfo {author} {\bibfnamefont {J.~L.}\ \bibnamefont {Egido}},\
  and\ \bibinfo {author} {\bibfnamefont {L.~M.}\ \bibnamefont {Robledo}},\
  }\href {https://doi.org/10.1103/PhysRevLett.80.4398} {\bibfield  {journal}
  {\bibinfo  {journal} {Phys. Rev. Lett.}\ }\textbf {\bibinfo {volume} {80}},\
  \bibinfo {pages} {4398} (\bibinfo {year} {1998})}\BibitemShut {NoStop}%
\bibitem [{\citenamefont {Garrote}\ \emph {et~al.}(1999)\citenamefont
  {Garrote}, \citenamefont {Egido},\ and\ \citenamefont
  {Robledo}}]{GARROTE1999}%
  \BibitemOpen
  \bibfield  {author} {\bibinfo {author} {\bibfnamefont {E.}~\bibnamefont
  {Garrote}}, \bibinfo {author} {\bibfnamefont {J.~L.}\ \bibnamefont {Egido}},\
  and\ \bibinfo {author} {\bibfnamefont {L.~M.}\ \bibnamefont {Robledo}},\
  }\href {https://doi.org/https://doi.org/10.1016/S0375-9474(00)88535-8}
  {\bibfield  {journal} {\bibinfo  {journal} {Nucl. Phys. A}\ }\textbf
  {\bibinfo {volume} {654}},\ \bibinfo {pages} {723c } (\bibinfo {year}
  {1999})}\BibitemShut {NoStop}%
\bibitem [{\citenamefont {Long}\ \emph {et~al.}(2004)\citenamefont {Long},
  \citenamefont {Meng}, \citenamefont {Giai},\ and\ \citenamefont
  {Zhou}}]{long2004}%
  \BibitemOpen
  \bibfield  {author} {\bibinfo {author} {\bibfnamefont {W.}~\bibnamefont
  {Long}}, \bibinfo {author} {\bibfnamefont {J.}~\bibnamefont {Meng}}, \bibinfo
  {author} {\bibfnamefont {N.~V.}\ \bibnamefont {Giai}},\ and\ \bibinfo
  {author} {\bibfnamefont {S.-G.}\ \bibnamefont {Zhou}},\ }\href
  {https://doi.org/10.1103/PhysRevC.69.034319} {\bibfield  {journal} {\bibinfo
  {journal} {Phys. Rev. C}\ }\textbf {\bibinfo {volume} {69}},\ \bibinfo
  {pages} {034319} (\bibinfo {year} {2004})}\BibitemShut {NoStop}%
\bibitem [{\citenamefont {Robledo}\ \emph {et~al.}(2010)\citenamefont
  {Robledo}, \citenamefont {Baldo}, \citenamefont {Schuck},\ and\ \citenamefont
  {Vi\~nas}}]{robledo2010}%
  \BibitemOpen
  \bibfield  {author} {\bibinfo {author} {\bibfnamefont {L.~M.}\ \bibnamefont
  {Robledo}}, \bibinfo {author} {\bibfnamefont {M.}~\bibnamefont {Baldo}},
  \bibinfo {author} {\bibfnamefont {P.}~\bibnamefont {Schuck}},\ and\ \bibinfo
  {author} {\bibfnamefont {X.}~\bibnamefont {Vi\~nas}},\ }\href
  {https://doi.org/10.1103/PhysRevC.81.034315} {\bibfield  {journal} {\bibinfo
  {journal} {Phys. Rev. C}\ }\textbf {\bibinfo {volume} {81}},\ \bibinfo
  {pages} {034315} (\bibinfo {year} {2010})}\BibitemShut {NoStop}%
\bibitem [{\citenamefont {Robledo}\ and\ \citenamefont
  {Bertsch}(2011)}]{robledo2011}%
  \BibitemOpen
  \bibfield  {author} {\bibinfo {author} {\bibfnamefont {L.~M.}\ \bibnamefont
  {Robledo}}\ and\ \bibinfo {author} {\bibfnamefont {G.~F.}\ \bibnamefont
  {Bertsch}},\ }\href {https://doi.org/10.1103/PhysRevC.84.054302} {\bibfield
  {journal} {\bibinfo  {journal} {Phys. Rev. C}\ }\textbf {\bibinfo {volume}
  {84}},\ \bibinfo {pages} {054302} (\bibinfo {year} {2011})}\BibitemShut
  {NoStop}%
\bibitem [{\citenamefont {Erler}\ \emph {et~al.}(2012)\citenamefont {Erler},
  \citenamefont {Langanke}, \citenamefont {Loens}, \citenamefont
  {Mart\'{\i}nez-Pinedo},\ and\ \citenamefont {Reinhard}}]{erler2012}%
  \BibitemOpen
  \bibfield  {author} {\bibinfo {author} {\bibfnamefont {J.}~\bibnamefont
  {Erler}}, \bibinfo {author} {\bibfnamefont {K.}~\bibnamefont {Langanke}},
  \bibinfo {author} {\bibfnamefont {H.~P.}\ \bibnamefont {Loens}}, \bibinfo
  {author} {\bibfnamefont {G.}~\bibnamefont {Mart\'{\i}nez-Pinedo}},\ and\
  \bibinfo {author} {\bibfnamefont {P.-G.}\ \bibnamefont {Reinhard}},\ }\href
  {https://doi.org/10.1103/PhysRevC.85.025802} {\bibfield  {journal} {\bibinfo
  {journal} {Phys. Rev. C}\ }\textbf {\bibinfo {volume} {85}},\ \bibinfo
  {pages} {025802} (\bibinfo {year} {2012})}\BibitemShut {NoStop}%
\bibitem [{\citenamefont {Robledo}\ and\ \citenamefont
  {Rodríguez-Guzm\'an}(2012)}]{robledo2012}%
  \BibitemOpen
  \bibfield  {author} {\bibinfo {author} {\bibfnamefont {L.~M.}\ \bibnamefont
  {Robledo}}\ and\ \bibinfo {author} {\bibfnamefont {R.~R.}\ \bibnamefont
  {Rodríguez-Guzm\'an}},\ }\href
  {http://stacks.iop.org/0954-3899/39/i=10/a=105103} {\bibfield  {journal}
  {\bibinfo  {journal} {J. Phys. G: Nucl. Part. Phys.}\ }\textbf {\bibinfo
  {volume} {39}},\ \bibinfo {pages} {105103} (\bibinfo {year}
  {2012})}\BibitemShut {NoStop}%
\bibitem [{\citenamefont {Rodr\'iguez-Guzm\'an}\ \emph
  {et~al.}(2012)\citenamefont {Rodr\'iguez-Guzm\'an}, \citenamefont {Robledo},\
  and\ \citenamefont {Sarriguren}}]{rayner2012}%
  \BibitemOpen
  \bibfield  {author} {\bibinfo {author} {\bibfnamefont {R.}~\bibnamefont
  {Rodr\'iguez-Guzm\'an}}, \bibinfo {author} {\bibfnamefont {L.~M.}\
  \bibnamefont {Robledo}},\ and\ \bibinfo {author} {\bibfnamefont
  {P.}~\bibnamefont {Sarriguren}},\ }\href
  {https://doi.org/10.1103/PhysRevC.86.034336} {\bibfield  {journal} {\bibinfo
  {journal} {Phys. Rev. C}\ }\textbf {\bibinfo {volume} {86}},\ \bibinfo
  {pages} {034336} (\bibinfo {year} {2012})}\BibitemShut {NoStop}%
\bibitem [{\citenamefont {Robledo}\ and\ \citenamefont
  {Butler}(2013)}]{robledo2013}%
  \BibitemOpen
  \bibfield  {author} {\bibinfo {author} {\bibfnamefont {L.~M.}\ \bibnamefont
  {Robledo}}\ and\ \bibinfo {author} {\bibfnamefont {P.~A.}\ \bibnamefont
  {Butler}},\ }\href {https://doi.org/10.1103/PhysRevC.88.051302} {\bibfield
  {journal} {\bibinfo  {journal} {Phys. Rev. C}\ }\textbf {\bibinfo {volume}
  {88}},\ \bibinfo {pages} {051302} (\bibinfo {year} {2013})}\BibitemShut
  {NoStop}%
\bibitem [{\citenamefont {Robledo}(2015)}]{robledo2015}%
  \BibitemOpen
  \bibfield  {author} {\bibinfo {author} {\bibfnamefont {L.~M.}\ \bibnamefont
  {Robledo}},\ }\href@noop {} {\bibfield  {journal} {\bibinfo  {journal} {J.
  Phys. G: Nucl. Part. Phys.}\ }\textbf {\bibinfo {volume} {42}},\ \bibinfo
  {pages} {055109} (\bibinfo {year} {2015})}\BibitemShut {NoStop}%
\bibitem [{\citenamefont {Bernard}\ \emph {et~al.}(2016)\citenamefont
  {Bernard}, \citenamefont {Robledo},\ and\ \citenamefont
  {Rodr\'{\i}guez}}]{bernard2016}%
  \BibitemOpen
  \bibfield  {author} {\bibinfo {author} {\bibfnamefont {R.~N.}\ \bibnamefont
  {Bernard}}, \bibinfo {author} {\bibfnamefont {L.~M.}\ \bibnamefont
  {Robledo}},\ and\ \bibinfo {author} {\bibfnamefont {T.~R.}\ \bibnamefont
  {Rodr\'{\i}guez}},\ }\href {https://doi.org/10.1103/PhysRevC.93.061302}
  {\bibfield  {journal} {\bibinfo  {journal} {Phys. Rev. C}\ }\textbf {\bibinfo
  {volume} {93}},\ \bibinfo {pages} {061302} (\bibinfo {year}
  {2016})}\BibitemShut {NoStop}%
\bibitem [{\citenamefont {Agbemava}\ \emph {et~al.}(2016)\citenamefont
  {Agbemava}, \citenamefont {Afanasjev},\ and\ \citenamefont
  {Ring}}]{agbemava2016}%
  \BibitemOpen
  \bibfield  {author} {\bibinfo {author} {\bibfnamefont {S.~E.}\ \bibnamefont
  {Agbemava}}, \bibinfo {author} {\bibfnamefont {A.~V.}\ \bibnamefont
  {Afanasjev}},\ and\ \bibinfo {author} {\bibfnamefont {P.}~\bibnamefont
  {Ring}},\ }\href {https://doi.org/10.1103/PhysRevC.93.044304} {\bibfield
  {journal} {\bibinfo  {journal} {Phys. Rev. C}\ }\textbf {\bibinfo {volume}
  {93}},\ \bibinfo {pages} {044304} (\bibinfo {year} {2016})}\BibitemShut
  {NoStop}%
\bibitem [{\citenamefont {Agbemava}\ and\ \citenamefont
  {Afanasjev}(2017)}]{agbemava2017}%
  \BibitemOpen
  \bibfield  {author} {\bibinfo {author} {\bibfnamefont {S.~E.}\ \bibnamefont
  {Agbemava}}\ and\ \bibinfo {author} {\bibfnamefont {A.~V.}\ \bibnamefont
  {Afanasjev}},\ }\href {https://doi.org/10.1103/PhysRevC.96.024301} {\bibfield
   {journal} {\bibinfo  {journal} {Phys. Rev. C}\ }\textbf {\bibinfo {volume}
  {96}},\ \bibinfo {pages} {024301} (\bibinfo {year} {2017})}\BibitemShut
  {NoStop}%
\bibitem [{\citenamefont {Xu}\ and\ \citenamefont {Li}(2017)}]{xu2017}%
  \BibitemOpen
  \bibfield  {author} {\bibinfo {author} {\bibfnamefont {Z.}~\bibnamefont
  {Xu}}\ and\ \bibinfo {author} {\bibfnamefont {Z.-P.}\ \bibnamefont {Li}},\
  }\href {https://doi.org/10.1088/1674-1137/41/12/124107} {\bibfield  {journal}
  {\bibinfo  {journal} {Chin. Phys. C}\ }\textbf {\bibinfo {volume} {41}},\
  \bibinfo {pages} {124107} (\bibinfo {year} {2017})}\BibitemShut {NoStop}%
\bibitem [{\citenamefont {Xia}\ \emph {et~al.}(2017)\citenamefont {Xia},
  \citenamefont {Tao}, \citenamefont {Lu}, \citenamefont {Li}, \citenamefont
  {Nik\ifmmode \check{s}\else \v{s}\fi{}i\ifmmode~\acute{c}\else \'{c}\fi{}},\
  and\ \citenamefont {Vretenar}}]{xia2017}%
  \BibitemOpen
  \bibfield  {author} {\bibinfo {author} {\bibfnamefont {S.~Y.}\ \bibnamefont
  {Xia}}, \bibinfo {author} {\bibfnamefont {H.}~\bibnamefont {Tao}}, \bibinfo
  {author} {\bibfnamefont {Y.}~\bibnamefont {Lu}}, \bibinfo {author}
  {\bibfnamefont {Z.~P.}\ \bibnamefont {Li}}, \bibinfo {author} {\bibfnamefont
  {T.}~\bibnamefont {Nik\ifmmode \check{s}\else
  \v{s}\fi{}i\ifmmode~\acute{c}\else \'{c}\fi{}}},\ and\ \bibinfo {author}
  {\bibfnamefont {D.}~\bibnamefont {Vretenar}},\ }\href
  {https://doi.org/10.1103/PhysRevC.96.054303} {\bibfield  {journal} {\bibinfo
  {journal} {Phys. Rev. C}\ }\textbf {\bibinfo {volume} {96}},\ \bibinfo
  {pages} {054303} (\bibinfo {year} {2017})}\BibitemShut {NoStop}%
\bibitem [{\citenamefont {Ebata}\ and\ \citenamefont
  {Nakatsukasa}(2017)}]{ebata2017}%
  \BibitemOpen
  \bibfield  {author} {\bibinfo {author} {\bibfnamefont {S.}~\bibnamefont
  {Ebata}}\ and\ \bibinfo {author} {\bibfnamefont {T.}~\bibnamefont
  {Nakatsukasa}},\ }\href {https://doi.org/10.1088/1402-4896/aa6c4c} {\bibfield
   {journal} {\bibinfo  {journal} {Physica Scripta}\ }\textbf {\bibinfo
  {volume} {92}},\ \bibinfo {pages} {064005} (\bibinfo {year}
  {2017})}\BibitemShut {NoStop}%
\bibitem [{\citenamefont {Rodr\'{\i}guez-Guzm\'an}\ \emph
  {et~al.}(2020)\citenamefont {Rodr\'{\i}guez-Guzm\'an}, \citenamefont
  {Humadi},\ and\ \citenamefont {Robledo}}]{rayner2020}%
  \BibitemOpen
  \bibfield  {author} {\bibinfo {author} {\bibfnamefont {R.}~\bibnamefont
  {Rodr\'{\i}guez-Guzm\'an}}, \bibinfo {author} {\bibfnamefont {Y.~M.}\
  \bibnamefont {Humadi}},\ and\ \bibinfo {author} {\bibfnamefont {L.~M.}\
  \bibnamefont {Robledo}},\ }\href
  {https://doi.org/10.1140/epja/s10050-020-00051-w} {\bibfield  {journal}
  {\bibinfo  {journal} {Eur. Phys. J. A}\ }\textbf {\bibinfo {volume} {56}},\
  \bibinfo {pages} {43} (\bibinfo {year} {2020})}\BibitemShut {NoStop}%
\bibitem [{\citenamefont {Cao}\ \emph {et~al.}(2020)\citenamefont {Cao},
  \citenamefont {Agbemava}, \citenamefont {Afanasjev}, \citenamefont
  {Nazarewicz},\ and\ \citenamefont {Olsen}}]{cao2020}%
  \BibitemOpen
  \bibfield  {author} {\bibinfo {author} {\bibfnamefont {Y.}~\bibnamefont
  {Cao}}, \bibinfo {author} {\bibfnamefont {S.~E.}\ \bibnamefont {Agbemava}},
  \bibinfo {author} {\bibfnamefont {A.~V.}\ \bibnamefont {Afanasjev}}, \bibinfo
  {author} {\bibfnamefont {W.}~\bibnamefont {Nazarewicz}},\ and\ \bibinfo
  {author} {\bibfnamefont {E.}~\bibnamefont {Olsen}},\ }\href
  {https://doi.org/10.1103/PhysRevC.102.024311} {\bibfield  {journal} {\bibinfo
   {journal} {Phys. Rev. C}\ }\textbf {\bibinfo {volume} {102}},\ \bibinfo
  {pages} {024311} (\bibinfo {year} {2020})}\BibitemShut {NoStop}%
\bibitem [{\citenamefont {Rodr{\'{\i}}guez-Guzm{\'{a}}n}\ \emph
  {et~al.}(2020)\citenamefont {Rodr{\'{\i}}guez-Guzm{\'{a}}n}, \citenamefont
  {Humadi},\ and\ \citenamefont {Robledo}}]{rayner2020oct}%
  \BibitemOpen
  \bibfield  {author} {\bibinfo {author} {\bibfnamefont {R.}~\bibnamefont
  {Rodr{\'{\i}}guez-Guzm{\'{a}}n}}, \bibinfo {author} {\bibfnamefont {Y.~M.}\
  \bibnamefont {Humadi}},\ and\ \bibinfo {author} {\bibfnamefont {L.~M.}\
  \bibnamefont {Robledo}},\ }\href {https://doi.org/10.1088/1361-6471/abb000}
  {\bibfield  {journal} {\bibinfo  {journal} {J. Phys. G: Nucl. Part. Phys.}\
  }\textbf {\bibinfo {volume} {48}},\ \bibinfo {pages} {015103} (\bibinfo
  {year} {2020})}\BibitemShut {NoStop}%
\bibitem [{\citenamefont {Rodr\'{\i}guez-Guzm\'an}\ and\ \citenamefont
  {Robledo}(2021)}]{rayner2021}%
  \BibitemOpen
  \bibfield  {author} {\bibinfo {author} {\bibfnamefont {R.}~\bibnamefont
  {Rodr\'{\i}guez-Guzm\'an}}\ and\ \bibinfo {author} {\bibfnamefont {L.~M.}\
  \bibnamefont {Robledo}},\ }\href
  {https://doi.org/10.1103/PhysRevC.103.044301} {\bibfield  {journal} {\bibinfo
   {journal} {Phys. Rev. C}\ }\textbf {\bibinfo {volume} {103}},\ \bibinfo
  {pages} {044301} (\bibinfo {year} {2021})}\BibitemShut {NoStop}%
\bibitem [{\citenamefont {Nomura}\ \emph
  {et~al.}(2021{\natexlab{a}})\citenamefont {Nomura}, \citenamefont {Lotina},
  \citenamefont {Nik\ifmmode \check{s}\else \v{s}\fi{}i\ifmmode~\acute{c}\else
  \'{c}\fi{}},\ and\ \citenamefont {Vretenar}}]{nomura2021qoch}%
  \BibitemOpen
  \bibfield  {author} {\bibinfo {author} {\bibfnamefont {K.}~\bibnamefont
  {Nomura}}, \bibinfo {author} {\bibfnamefont {L.}~\bibnamefont {Lotina}},
  \bibinfo {author} {\bibfnamefont {T.}~\bibnamefont {Nik\ifmmode
  \check{s}\else \v{s}\fi{}i\ifmmode~\acute{c}\else \'{c}\fi{}}},\ and\
  \bibinfo {author} {\bibfnamefont {D.}~\bibnamefont {Vretenar}},\ }\href
  {https://doi.org/10.1103/PhysRevC.103.054301} {\bibfield  {journal} {\bibinfo
   {journal} {Phys. Rev. C}\ }\textbf {\bibinfo {volume} {103}},\ \bibinfo
  {pages} {054301} (\bibinfo {year} {2021}{\natexlab{a}})}\BibitemShut
  {NoStop}%
\bibitem [{\citenamefont {Engel}\ and\ \citenamefont
  {Iachello}(1985)}]{engel1985}%
  \BibitemOpen
  \bibfield  {author} {\bibinfo {author} {\bibfnamefont {J.}~\bibnamefont
  {Engel}}\ and\ \bibinfo {author} {\bibfnamefont {F.}~\bibnamefont
  {Iachello}},\ }\href {https://doi.org/10.1103/PhysRevLett.54.1126} {\bibfield
   {journal} {\bibinfo  {journal} {Phys. Rev. Lett.}\ }\textbf {\bibinfo
  {volume} {54}},\ \bibinfo {pages} {1126} (\bibinfo {year}
  {1985})}\BibitemShut {NoStop}%
\bibitem [{\citenamefont {Engel}\ and\ \citenamefont
  {Iachello}(1987)}]{engel1987}%
  \BibitemOpen
  \bibfield  {author} {\bibinfo {author} {\bibfnamefont {J.}~\bibnamefont
  {Engel}}\ and\ \bibinfo {author} {\bibfnamefont {F.}~\bibnamefont
  {Iachello}},\ }\href {https://doi.org/10.1016/0375-9474(87)90220-X}
  {\bibfield  {journal} {\bibinfo  {journal} {Nucl. Phys. A}\ }\textbf
  {\bibinfo {volume} {472}},\ \bibinfo {pages} {61 } (\bibinfo {year}
  {1987})}\BibitemShut {NoStop}%
\bibitem [{\citenamefont {Kusnezov}\ and\ \citenamefont
  {Iachello}(1988)}]{kusnezov1988}%
  \BibitemOpen
  \bibfield  {author} {\bibinfo {author} {\bibfnamefont {D.}~\bibnamefont
  {Kusnezov}}\ and\ \bibinfo {author} {\bibfnamefont {F.}~\bibnamefont
  {Iachello}},\ }\href
  {https://doi.org/https://doi.org/10.1016/0370-2693(88)91166-5} {\bibfield
  {journal} {\bibinfo  {journal} {Phys. Lett. B}\ }\textbf {\bibinfo {volume}
  {209}},\ \bibinfo {pages} {420 } (\bibinfo {year} {1988})}\BibitemShut
  {NoStop}%
\bibitem [{\citenamefont {Yoshinaga}\ \emph {et~al.}(1993)\citenamefont
  {Yoshinaga}, \citenamefont {Mizusaki},\ and\ \citenamefont
  {Otsuka}}]{yoshinaga1993}%
  \BibitemOpen
  \bibfield  {author} {\bibinfo {author} {\bibfnamefont {N.}~\bibnamefont
  {Yoshinaga}}, \bibinfo {author} {\bibfnamefont {T.}~\bibnamefont
  {Mizusaki}},\ and\ \bibinfo {author} {\bibfnamefont {T.}~\bibnamefont
  {Otsuka}},\ }\href
  {https://doi.org/https://doi.org/10.1016/0375-9474(93)90186-2} {\bibfield
  {journal} {\bibinfo  {journal} {Nucl. Phys. A}\ }\textbf {\bibinfo {volume}
  {559}},\ \bibinfo {pages} {193} (\bibinfo {year} {1993})}\BibitemShut
  {NoStop}%
\bibitem [{\citenamefont {Zamfir}\ and\ \citenamefont
  {Kusnezov}(2001)}]{zamfir2001}%
  \BibitemOpen
  \bibfield  {author} {\bibinfo {author} {\bibfnamefont {N.~V.}\ \bibnamefont
  {Zamfir}}\ and\ \bibinfo {author} {\bibfnamefont {D.}~\bibnamefont
  {Kusnezov}},\ }\href {https://doi.org/10.1103/PhysRevC.63.054306} {\bibfield
  {journal} {\bibinfo  {journal} {Phys. Rev. C}\ }\textbf {\bibinfo {volume}
  {63}},\ \bibinfo {pages} {054306} (\bibinfo {year} {2001})}\BibitemShut
  {NoStop}%
\bibitem [{\citenamefont {Zamfir}\ and\ \citenamefont
  {Kusnezov}(2003)}]{zamfir2003}%
  \BibitemOpen
  \bibfield  {author} {\bibinfo {author} {\bibfnamefont {N.~V.}\ \bibnamefont
  {Zamfir}}\ and\ \bibinfo {author} {\bibfnamefont {D.}~\bibnamefont
  {Kusnezov}},\ }\href {https://doi.org/10.1103/PhysRevC.67.014305} {\bibfield
  {journal} {\bibinfo  {journal} {Phys. Rev. C}\ }\textbf {\bibinfo {volume}
  {67}},\ \bibinfo {pages} {014305} (\bibinfo {year} {2003})}\BibitemShut
  {NoStop}%
\bibitem [{\citenamefont {Nomura}\ \emph {et~al.}(2013)\citenamefont {Nomura},
  \citenamefont {Vretenar},\ and\ \citenamefont {Lu}}]{nomura2013oct}%
  \BibitemOpen
  \bibfield  {author} {\bibinfo {author} {\bibfnamefont {K.}~\bibnamefont
  {Nomura}}, \bibinfo {author} {\bibfnamefont {D.}~\bibnamefont {Vretenar}},\
  and\ \bibinfo {author} {\bibfnamefont {B.-N.}\ \bibnamefont {Lu}},\ }\href
  {https://doi.org/10.1103/PhysRevC.88.021303} {\bibfield  {journal} {\bibinfo
  {journal} {Phys. Rev. C}\ }\textbf {\bibinfo {volume} {88}},\ \bibinfo
  {pages} {021303} (\bibinfo {year} {2013})}\BibitemShut {NoStop}%
\bibitem [{\citenamefont {Nomura}\ \emph {et~al.}(2014)\citenamefont {Nomura},
  \citenamefont {Vretenar}, \citenamefont {Nik\ifmmode \check{s}\else
  \v{s}\fi{}i\ifmmode~\acute{c}\else \'{c}\fi{}},\ and\ \citenamefont
  {Lu}}]{nomura2014}%
  \BibitemOpen
  \bibfield  {author} {\bibinfo {author} {\bibfnamefont {K.}~\bibnamefont
  {Nomura}}, \bibinfo {author} {\bibfnamefont {D.}~\bibnamefont {Vretenar}},
  \bibinfo {author} {\bibfnamefont {T.}~\bibnamefont {Nik\ifmmode
  \check{s}\else \v{s}\fi{}i\ifmmode~\acute{c}\else \'{c}\fi{}}},\ and\
  \bibinfo {author} {\bibfnamefont {B.-N.}\ \bibnamefont {Lu}},\ }\href
  {https://doi.org/10.1103/PhysRevC.89.024312} {\bibfield  {journal} {\bibinfo
  {journal} {Phys. Rev. C}\ }\textbf {\bibinfo {volume} {89}},\ \bibinfo
  {pages} {024312} (\bibinfo {year} {2014})}\BibitemShut {NoStop}%
\bibitem [{\citenamefont {Nomura}\ \emph {et~al.}(2015)\citenamefont {Nomura},
  \citenamefont {Rodr\'{\i}guez-Guzm\'an},\ and\ \citenamefont
  {Robledo}}]{nomura2015}%
  \BibitemOpen
  \bibfield  {author} {\bibinfo {author} {\bibfnamefont {K.}~\bibnamefont
  {Nomura}}, \bibinfo {author} {\bibfnamefont {R.}~\bibnamefont
  {Rodr\'{\i}guez-Guzm\'an}},\ and\ \bibinfo {author} {\bibfnamefont {L.~M.}\
  \bibnamefont {Robledo}},\ }\href {https://doi.org/10.1103/PhysRevC.92.014312}
  {\bibfield  {journal} {\bibinfo  {journal} {Phys. Rev. C}\ }\textbf {\bibinfo
  {volume} {92}},\ \bibinfo {pages} {014312} (\bibinfo {year}
  {2015})}\BibitemShut {NoStop}%
\bibitem [{\citenamefont {Nomura}\ \emph {et~al.}(2020)\citenamefont {Nomura},
  \citenamefont {Rodr\'{\i}guez-Guzm\'an}, \citenamefont {Humadi},
  \citenamefont {Robledo},\ and\ \citenamefont
  {Garc\'{\i}a-Ramos}}]{nomura2020oct}%
  \BibitemOpen
  \bibfield  {author} {\bibinfo {author} {\bibfnamefont {K.}~\bibnamefont
  {Nomura}}, \bibinfo {author} {\bibfnamefont {R.}~\bibnamefont
  {Rodr\'{\i}guez-Guzm\'an}}, \bibinfo {author} {\bibfnamefont {Y.~M.}\
  \bibnamefont {Humadi}}, \bibinfo {author} {\bibfnamefont {L.~M.}\
  \bibnamefont {Robledo}},\ and\ \bibinfo {author} {\bibfnamefont {J.~E.}\
  \bibnamefont {Garc\'{\i}a-Ramos}},\ }\href
  {https://doi.org/10.1103/PhysRevC.102.064326} {\bibfield  {journal} {\bibinfo
   {journal} {Phys. Rev. C}\ }\textbf {\bibinfo {volume} {102}},\ \bibinfo
  {pages} {064326} (\bibinfo {year} {2020})}\BibitemShut {NoStop}%
\bibitem [{\citenamefont {Vallejos}\ and\ \citenamefont
  {Barea}(2021)}]{vallejos2021}%
  \BibitemOpen
  \bibfield  {author} {\bibinfo {author} {\bibfnamefont {O.}~\bibnamefont
  {Vallejos}}\ and\ \bibinfo {author} {\bibfnamefont {J.}~\bibnamefont
  {Barea}},\ }\href {https://doi.org/10.1103/PhysRevC.104.014308} {\bibfield
  {journal} {\bibinfo  {journal} {Phys. Rev. C}\ }\textbf {\bibinfo {volume}
  {104}},\ \bibinfo {pages} {014308} (\bibinfo {year} {2021})}\BibitemShut
  {NoStop}%
\bibitem [{\citenamefont {Nomura}\ \emph
  {et~al.}(2021{\natexlab{b}})\citenamefont {Nomura}, \citenamefont
  {Rodr\'{\i}guez-Guzm\'an}, \citenamefont {Robledo},\ and\ \citenamefont
  {Garc\'{\i}a-Ramos}}]{nomura2021oct-u}%
  \BibitemOpen
  \bibfield  {author} {\bibinfo {author} {\bibfnamefont {K.}~\bibnamefont
  {Nomura}}, \bibinfo {author} {\bibfnamefont {R.}~\bibnamefont
  {Rodr\'{\i}guez-Guzm\'an}}, \bibinfo {author} {\bibfnamefont
  {L.}~\bibnamefont {Robledo}},\ and\ \bibinfo {author} {\bibfnamefont
  {J.}~\bibnamefont {Garc\'{\i}a-Ramos}},\ }\href
  {https://doi.org/10.1103/PhysRevC.103.044311} {\bibfield  {journal} {\bibinfo
   {journal} {Phys. Rev. C}\ }\textbf {\bibinfo {volume} {103}},\ \bibinfo
  {pages} {044311} (\bibinfo {year} {2021}{\natexlab{b}})}\BibitemShut
  {NoStop}%
\bibitem [{\citenamefont {Nomura}\ \emph
  {et~al.}(2021{\natexlab{c}})\citenamefont {Nomura}, \citenamefont
  {Rodr\'{\i}guez-Guzm\'an}, \citenamefont {Robledo}, \citenamefont
  {Garc\'{\i}a-Ramos},\ and\ \citenamefont {Hern\'andez}}]{nomura2021oct-ba}%
  \BibitemOpen
  \bibfield  {author} {\bibinfo {author} {\bibfnamefont {K.}~\bibnamefont
  {Nomura}}, \bibinfo {author} {\bibfnamefont {R.}~\bibnamefont
  {Rodr\'{\i}guez-Guzm\'an}}, \bibinfo {author} {\bibfnamefont {L.~M.}\
  \bibnamefont {Robledo}}, \bibinfo {author} {\bibfnamefont {J.~E.}\
  \bibnamefont {Garc\'{\i}a-Ramos}},\ and\ \bibinfo {author} {\bibfnamefont
  {N.~C.}\ \bibnamefont {Hern\'andez}},\ }\href
  {https://doi.org/10.1103/PhysRevC.104.044324} {\bibfield  {journal} {\bibinfo
   {journal} {Phys. Rev. C}\ }\textbf {\bibinfo {volume} {104}},\ \bibinfo
  {pages} {044324} (\bibinfo {year} {2021}{\natexlab{c}})}\BibitemShut
  {NoStop}%
\bibitem [{\citenamefont {Bonatsos}\ \emph {et~al.}(2005)\citenamefont
  {Bonatsos}, \citenamefont {Lenis}, \citenamefont {Minkov}, \citenamefont
  {Petrellis},\ and\ \citenamefont {Yotov}}]{bonatsos2005}%
  \BibitemOpen
  \bibfield  {author} {\bibinfo {author} {\bibfnamefont {D.}~\bibnamefont
  {Bonatsos}}, \bibinfo {author} {\bibfnamefont {D.}~\bibnamefont {Lenis}},
  \bibinfo {author} {\bibfnamefont {N.}~\bibnamefont {Minkov}}, \bibinfo
  {author} {\bibfnamefont {D.}~\bibnamefont {Petrellis}},\ and\ \bibinfo
  {author} {\bibfnamefont {P.}~\bibnamefont {Yotov}},\ }\href
  {https://doi.org/10.1103/PhysRevC.71.064309} {\bibfield  {journal} {\bibinfo
  {journal} {Phys. Rev. C}\ }\textbf {\bibinfo {volume} {71}},\ \bibinfo
  {pages} {064309} (\bibinfo {year} {2005})}\BibitemShut {NoStop}%
\bibitem [{\citenamefont {Lenis}\ and\ \citenamefont
  {Bonatsos}(2006)}]{lenis2006}%
  \BibitemOpen
  \bibfield  {author} {\bibinfo {author} {\bibfnamefont {D.}~\bibnamefont
  {Lenis}}\ and\ \bibinfo {author} {\bibfnamefont {D.}~\bibnamefont
  {Bonatsos}},\ }\href {https://doi.org/10.1016/j.physletb.2005.12.016}
  {\bibfield  {journal} {\bibinfo  {journal} {Phys. Lett. B}\ }\textbf
  {\bibinfo {volume} {633}},\ \bibinfo {pages} {474} (\bibinfo {year}
  {2006})}\BibitemShut {NoStop}%
\bibitem [{\citenamefont {Bizzeti}\ and\ \citenamefont
  {Bizzeti-Sona}(2013)}]{bizzeti2013}%
  \BibitemOpen
  \bibfield  {author} {\bibinfo {author} {\bibfnamefont {P.~G.}\ \bibnamefont
  {Bizzeti}}\ and\ \bibinfo {author} {\bibfnamefont {A.~M.}\ \bibnamefont
  {Bizzeti-Sona}},\ }\href {https://doi.org/10.1103/PhysRevC.88.011305}
  {\bibfield  {journal} {\bibinfo  {journal} {Phys. Rev. C}\ }\textbf {\bibinfo
  {volume} {88}},\ \bibinfo {pages} {011305} (\bibinfo {year}
  {2013})}\BibitemShut {NoStop}%
\bibitem [{\citenamefont {Shneidman}\ \emph {et~al.}(2002)\citenamefont
  {Shneidman}, \citenamefont {Adamian}, \citenamefont {Antonenko},
  \citenamefont {Jolos},\ and\ \citenamefont {Scheid}}]{shneidman2002}%
  \BibitemOpen
  \bibfield  {author} {\bibinfo {author} {\bibfnamefont {T.~M.}\ \bibnamefont
  {Shneidman}}, \bibinfo {author} {\bibfnamefont {G.~G.}\ \bibnamefont
  {Adamian}}, \bibinfo {author} {\bibfnamefont {N.~V.}\ \bibnamefont
  {Antonenko}}, \bibinfo {author} {\bibfnamefont {R.~V.}\ \bibnamefont
  {Jolos}},\ and\ \bibinfo {author} {\bibfnamefont {W.}~\bibnamefont
  {Scheid}},\ }\href
  {https://doi.org/http://dx.doi.org/10.1016/S0370-2693(01)01512-X} {\bibfield
  {journal} {\bibinfo  {journal} {Phys. Lett. B}\ }\textbf {\bibinfo {volume}
  {526}},\ \bibinfo {pages} {322 } (\bibinfo {year} {2002})}\BibitemShut
  {NoStop}%
\bibitem [{\citenamefont {Shneidman}\ \emph {et~al.}(2003)\citenamefont
  {Shneidman}, \citenamefont {Adamian}, \citenamefont {Antonenko},
  \citenamefont {Jolos},\ and\ \citenamefont {Scheid}}]{shneidman2003}%
  \BibitemOpen
  \bibfield  {author} {\bibinfo {author} {\bibfnamefont {T.~M.}\ \bibnamefont
  {Shneidman}}, \bibinfo {author} {\bibfnamefont {G.~G.}\ \bibnamefont
  {Adamian}}, \bibinfo {author} {\bibfnamefont {N.~V.}\ \bibnamefont
  {Antonenko}}, \bibinfo {author} {\bibfnamefont {R.~V.}\ \bibnamefont
  {Jolos}},\ and\ \bibinfo {author} {\bibfnamefont {W.}~\bibnamefont
  {Scheid}},\ }\href {https://doi.org/10.1103/PhysRevC.67.014313} {\bibfield
  {journal} {\bibinfo  {journal} {Phys. Rev. C}\ }\textbf {\bibinfo {volume}
  {67}},\ \bibinfo {pages} {014313} (\bibinfo {year} {2003})}\BibitemShut
  {NoStop}%
\bibitem [{\citenamefont {Ring}\ and\ \citenamefont {Schuck}(1980)}]{RS}%
  \BibitemOpen
  \bibfield  {author} {\bibinfo {author} {\bibfnamefont {P.}~\bibnamefont
  {Ring}}\ and\ \bibinfo {author} {\bibfnamefont {P.}~\bibnamefont {Schuck}},\
  }\href@noop {} {\emph {\bibinfo {title} {The nuclear many-body problem}}}\
  (\bibinfo  {publisher} {Berlin: Springer-Verlag},\ \bibinfo {year}
  {1980})\BibitemShut {NoStop}%
\bibitem [{\citenamefont {Robledo}\ \emph {et~al.}(2019)\citenamefont
  {Robledo}, \citenamefont {Rodríguez},\ and\ \citenamefont
  {Rodríguez-Guzmán}}]{robledo2019}%
  \BibitemOpen
  \bibfield  {author} {\bibinfo {author} {\bibfnamefont {L.~M.}\ \bibnamefont
  {Robledo}}, \bibinfo {author} {\bibfnamefont {T.~R.}\ \bibnamefont
  {Rodríguez}},\ and\ \bibinfo {author} {\bibfnamefont {R.~R.}\ \bibnamefont
  {Rodríguez-Guzmán}},\ }\href
  {http://stacks.iop.org/0954-3899/46/i=1/a=013001} {\bibfield  {journal}
  {\bibinfo  {journal} {J. Phys. G: Nucl. Part. Phys.}\ }\textbf {\bibinfo
  {volume} {46}},\ \bibinfo {pages} {013001} (\bibinfo {year}
  {2019})}\BibitemShut {NoStop}%
\bibitem [{\citenamefont {Skalski}(1990)}]{SKALSKI1990}%
  \BibitemOpen
  \bibfield  {author} {\bibinfo {author} {\bibfnamefont {J.}~\bibnamefont
  {Skalski}},\ }\href
  {https://doi.org/https://doi.org/10.1016/0370-2693(90)92090-6} {\bibfield
  {journal} {\bibinfo  {journal} {Phys. Lett. B}\ }\textbf {\bibinfo {volume}
  {238}},\ \bibinfo {pages} {6 } (\bibinfo {year} {1990})}\BibitemShut
  {NoStop}%
\bibitem [{\citenamefont {Nomura}\ \emph {et~al.}(2008)\citenamefont {Nomura},
  \citenamefont {Shimizu},\ and\ \citenamefont {Otsuka}}]{nomura2008}%
  \BibitemOpen
  \bibfield  {author} {\bibinfo {author} {\bibfnamefont {K.}~\bibnamefont
  {Nomura}}, \bibinfo {author} {\bibfnamefont {N.}~\bibnamefont {Shimizu}},\
  and\ \bibinfo {author} {\bibfnamefont {T.}~\bibnamefont {Otsuka}},\ }\href
  {https://doi.org/10.1103/PhysRevLett.101.142501} {\bibfield  {journal}
  {\bibinfo  {journal} {Phys. Rev. Lett.}\ }\textbf {\bibinfo {volume} {101}},\
  \bibinfo {pages} {142501} (\bibinfo {year} {2008})}\BibitemShut {NoStop}%
\bibitem [{\citenamefont {Goriely}\ \emph {et~al.}(2009)\citenamefont
  {Goriely}, \citenamefont {Hilaire}, \citenamefont {Girod},\ and\
  \citenamefont {P\'eru}}]{D1M}%
  \BibitemOpen
  \bibfield  {author} {\bibinfo {author} {\bibfnamefont {S.}~\bibnamefont
  {Goriely}}, \bibinfo {author} {\bibfnamefont {S.}~\bibnamefont {Hilaire}},
  \bibinfo {author} {\bibfnamefont {M.}~\bibnamefont {Girod}},\ and\ \bibinfo
  {author} {\bibfnamefont {S.}~\bibnamefont {P\'eru}},\ }\href
  {https://doi.org/10.1103/PhysRevLett.102.242501} {\bibfield  {journal}
  {\bibinfo  {journal} {Phys. Rev. Lett.}\ }\textbf {\bibinfo {volume} {102}},\
  \bibinfo {pages} {242501} (\bibinfo {year} {2009})}\BibitemShut {NoStop}%
\bibitem [{\citenamefont {Nik\ifmmode \check{s}\else
  \v{s}\fi{}i\ifmmode~\acute{c}\else \'{c}\fi{}}\ \emph
  {et~al.}(2008)\citenamefont {Nik\ifmmode \check{s}\else
  \v{s}\fi{}i\ifmmode~\acute{c}\else \'{c}\fi{}}, \citenamefont {Vretenar},\
  and\ \citenamefont {Ring}}]{DDPC1}%
  \BibitemOpen
  \bibfield  {author} {\bibinfo {author} {\bibfnamefont {T.}~\bibnamefont
  {Nik\ifmmode \check{s}\else \v{s}\fi{}i\ifmmode~\acute{c}\else \'{c}\fi{}}},
  \bibinfo {author} {\bibfnamefont {D.}~\bibnamefont {Vretenar}},\ and\
  \bibinfo {author} {\bibfnamefont {P.}~\bibnamefont {Ring}},\ }\href
  {https://doi.org/10.1103/PhysRevC.78.034318} {\bibfield  {journal} {\bibinfo
  {journal} {Phys. Rev. C}\ }\textbf {\bibinfo {volume} {78}},\ \bibinfo
  {pages} {034318} (\bibinfo {year} {2008})}\BibitemShut {NoStop}%
\bibitem [{\citenamefont {Otsuka}\ \emph
  {et~al.}(1978{\natexlab{a}})\citenamefont {Otsuka}, \citenamefont {Arima},
  \citenamefont {Iachello},\ and\ \citenamefont {Talmi}}]{OAIT}%
  \BibitemOpen
  \bibfield  {author} {\bibinfo {author} {\bibfnamefont {T.}~\bibnamefont
  {Otsuka}}, \bibinfo {author} {\bibfnamefont {A.}~\bibnamefont {Arima}},
  \bibinfo {author} {\bibfnamefont {F.}~\bibnamefont {Iachello}},\ and\
  \bibinfo {author} {\bibfnamefont {I.}~\bibnamefont {Talmi}},\ }\href
  {https://doi.org/10.1016/0370-2693(78)90260-5} {\bibfield  {journal}
  {\bibinfo  {journal} {Phys. Lett. B}\ }\textbf {\bibinfo {volume} {76}},\
  \bibinfo {pages} {139 } (\bibinfo {year} {1978}{\natexlab{a}})}\BibitemShut
  {NoStop}%
\bibitem [{\citenamefont {Otsuka}\ \emph
  {et~al.}(1978{\natexlab{b}})\citenamefont {Otsuka}, \citenamefont {Arima},\
  and\ \citenamefont {Iachello}}]{OAI}%
  \BibitemOpen
  \bibfield  {author} {\bibinfo {author} {\bibfnamefont {T.}~\bibnamefont
  {Otsuka}}, \bibinfo {author} {\bibfnamefont {A.}~\bibnamefont {Arima}},\ and\
  \bibinfo {author} {\bibfnamefont {F.}~\bibnamefont {Iachello}},\ }\href
  {https://doi.org/10.1016/0375-9474(78)90532-8} {\bibfield  {journal}
  {\bibinfo  {journal} {Nucl. Phys. A}\ }\textbf {\bibinfo {volume} {309}},\
  \bibinfo {pages} {1} (\bibinfo {year} {1978}{\natexlab{b}})}\BibitemShut
  {NoStop}%
\bibitem [{\citenamefont {Iachello}\ and\ \citenamefont {Arima}(1987)}]{IBM}%
  \BibitemOpen
  \bibfield  {author} {\bibinfo {author} {\bibfnamefont {F.}~\bibnamefont
  {Iachello}}\ and\ \bibinfo {author} {\bibfnamefont {A.}~\bibnamefont
  {Arima}},\ }\href@noop {} {\emph {\bibinfo {title} {The interacting boson
  model}}}\ (\bibinfo  {publisher} {Cambridge University Press, Cambridge},\
  \bibinfo {year} {1987})\BibitemShut {NoStop}%
\bibitem [{\citenamefont {Elliot}\ and\ \citenamefont
  {White}(1980)}]{ELLIOT1980}%
  \BibitemOpen
  \bibfield  {author} {\bibinfo {author} {\bibfnamefont {J.}~\bibnamefont
  {Elliot}}\ and\ \bibinfo {author} {\bibfnamefont {A.}~\bibnamefont {White}},\
  }\href {https://doi.org/https://doi.org/10.1016/0370-2693(80)90573-0}
  {\bibfield  {journal} {\bibinfo  {journal} {Phys. Lett. B}\ }\textbf
  {\bibinfo {volume} {97}},\ \bibinfo {pages} {169} (\bibinfo {year}
  {1980})}\BibitemShut {NoStop}%
\bibitem [{\citenamefont {Elliot}\ and\ \citenamefont
  {Evans}(1981)}]{ELLIOT1981}%
  \BibitemOpen
  \bibfield  {author} {\bibinfo {author} {\bibfnamefont {J.}~\bibnamefont
  {Elliot}}\ and\ \bibinfo {author} {\bibfnamefont {J.}~\bibnamefont {Evans}},\
  }\href {https://doi.org/https://doi.org/10.1016/0370-2693(81)90297-5}
  {\bibfield  {journal} {\bibinfo  {journal} {Phys. Lett. B}\ }\textbf
  {\bibinfo {volume} {101}},\ \bibinfo {pages} {216} (\bibinfo {year}
  {1981})}\BibitemShut {NoStop}%
\bibitem [{\citenamefont {Ginocchio}\ and\ \citenamefont
  {Kirson}(1980)}]{ginocchio1980}%
  \BibitemOpen
  \bibfield  {author} {\bibinfo {author} {\bibfnamefont {J.~N.}\ \bibnamefont
  {Ginocchio}}\ and\ \bibinfo {author} {\bibfnamefont {M.~W.}\ \bibnamefont
  {Kirson}},\ }\href {https://doi.org/10.1016/0375-9474(80)90387-5} {\bibfield
  {journal} {\bibinfo  {journal} {Nucl. Phys. A}\ }\textbf {\bibinfo {volume}
  {350}},\ \bibinfo {pages} {31} (\bibinfo {year} {1980})}\BibitemShut
  {NoStop}%
\bibitem [{\citenamefont {Nomura}\ \emph {et~al.}(2011)\citenamefont {Nomura},
  \citenamefont {Otsuka}, \citenamefont {Shimizu},\ and\ \citenamefont
  {Guo}}]{nomura2011rot}%
  \BibitemOpen
  \bibfield  {author} {\bibinfo {author} {\bibfnamefont {K.}~\bibnamefont
  {Nomura}}, \bibinfo {author} {\bibfnamefont {T.}~\bibnamefont {Otsuka}},
  \bibinfo {author} {\bibfnamefont {N.}~\bibnamefont {Shimizu}},\ and\ \bibinfo
  {author} {\bibfnamefont {L.}~\bibnamefont {Guo}},\ }\href
  {https://doi.org/10.1103/PhysRevC.83.041302} {\bibfield  {journal} {\bibinfo
  {journal} {Phys. Rev. C}\ }\textbf {\bibinfo {volume} {83}},\ \bibinfo
  {pages} {041302} (\bibinfo {year} {2011})}\BibitemShut {NoStop}%
\bibitem [{\citenamefont {Schaaser}\ and\ \citenamefont
  {Brink}(1986)}]{schaaser1986}%
  \BibitemOpen
  \bibfield  {author} {\bibinfo {author} {\bibfnamefont {H.}~\bibnamefont
  {Schaaser}}\ and\ \bibinfo {author} {\bibfnamefont {D.~M.}\ \bibnamefont
  {Brink}},\ }\href@noop {} {\bibfield  {journal} {\bibinfo  {journal} {Nucl.
  Phys. A}\ }\textbf {\bibinfo {volume} {452}},\ \bibinfo {pages} {1 }
  (\bibinfo {year} {1986})}\BibitemShut {NoStop}%
\bibitem [{\citenamefont {Thouless}\ and\ \citenamefont {Valatin}(1962)}]{TV}%
  \BibitemOpen
  \bibfield  {author} {\bibinfo {author} {\bibfnamefont {D.~J.}\ \bibnamefont
  {Thouless}}\ and\ \bibinfo {author} {\bibfnamefont {J.~G.}\ \bibnamefont
  {Valatin}},\ }\href {https://doi.org/10.1016/0029-5582(62)90741-1} {\bibfield
   {journal} {\bibinfo  {journal} {Nucl. Phys.}\ }\textbf {\bibinfo {volume}
  {31}},\ \bibinfo {pages} {211 } (\bibinfo {year} {1962})}\BibitemShut
  {NoStop}%
\bibitem [{\citenamefont {{S. Heinze}}(2008)}]{arbmodel}%
  \BibitemOpen
  \bibfield  {author} {\bibinfo {author} {\bibnamefont {{S. Heinze}}}}
  (\bibinfo {year} {2008}),\ \bibinfo {note} {computer program ARBMODEL
  (University of Cologne)}\BibitemShut {NoStop}%
\bibitem [{\citenamefont {{Brookhaven National Nuclear Data Center}}()}]{data}%
  \BibitemOpen
  \bibfield  {author} {\bibinfo {author} {\bibnamefont {{Brookhaven National
  Nuclear Data Center}}},\ }\href@noop {} {}\bibinfo {howpublished}
  {{http://www.nndc.bnl.gov}}\BibitemShut {NoStop}%
\bibitem [{\citenamefont {M\"uller-Gatermann}\ \emph
  {et~al.}(2020)\citenamefont {M\"uller-Gatermann}, \citenamefont {Dewald},
  \citenamefont {Fransen}, \citenamefont {Beckers}, \citenamefont {Blazhev},
  \citenamefont {Braunroth}, \citenamefont {Goldkuhle}, \citenamefont {Jolie},
  \citenamefont {Kornwebel}, \citenamefont {Reviol}, \citenamefont {von Spee},\
  and\ \citenamefont {Zell}}]{muellergatermann2020}%
  \BibitemOpen
  \bibfield  {author} {\bibinfo {author} {\bibfnamefont {C.}~\bibnamefont
  {M\"uller-Gatermann}}, \bibinfo {author} {\bibfnamefont {A.}~\bibnamefont
  {Dewald}}, \bibinfo {author} {\bibfnamefont {C.}~\bibnamefont {Fransen}},
  \bibinfo {author} {\bibfnamefont {M.}~\bibnamefont {Beckers}}, \bibinfo
  {author} {\bibfnamefont {A.}~\bibnamefont {Blazhev}}, \bibinfo {author}
  {\bibfnamefont {T.}~\bibnamefont {Braunroth}}, \bibinfo {author}
  {\bibfnamefont {A.}~\bibnamefont {Goldkuhle}}, \bibinfo {author}
  {\bibfnamefont {J.}~\bibnamefont {Jolie}}, \bibinfo {author} {\bibfnamefont
  {L.}~\bibnamefont {Kornwebel}}, \bibinfo {author} {\bibfnamefont
  {W.}~\bibnamefont {Reviol}}, \bibinfo {author} {\bibfnamefont
  {F.}~\bibnamefont {von Spee}},\ and\ \bibinfo {author} {\bibfnamefont
  {K.~O.}\ \bibnamefont {Zell}},\ }\href
  {https://doi.org/10.1103/PhysRevC.102.064318} {\bibfield  {journal} {\bibinfo
   {journal} {Phys. Rev. C}\ }\textbf {\bibinfo {volume} {102}},\ \bibinfo
  {pages} {064318} (\bibinfo {year} {2020})}\BibitemShut {NoStop}%
\bibitem [{\citenamefont {Puddu}\ \emph {et~al.}(1980)\citenamefont {Puddu},
  \citenamefont {Scholten},\ and\ \citenamefont {Otsuka}}]{puddu1980}%
  \BibitemOpen
  \bibfield  {author} {\bibinfo {author} {\bibfnamefont {G.}~\bibnamefont
  {Puddu}}, \bibinfo {author} {\bibfnamefont {O.}~\bibnamefont {Scholten}},\
  and\ \bibinfo {author} {\bibfnamefont {T.}~\bibnamefont {Otsuka}},\ }\href
  {https://doi.org/https://doi.org/10.1016/0375-9474(80)90548-5} {\bibfield
  {journal} {\bibinfo  {journal} {Nucl. Phys. A}\ }\textbf {\bibinfo {volume}
  {348}},\ \bibinfo {pages} {109} (\bibinfo {year} {1980})}\BibitemShut
  {NoStop}%
\bibitem [{\citenamefont {Davydov}\ and\ \citenamefont
  {Filippov}(1958)}]{Davydov58}%
  \BibitemOpen
  \bibfield  {author} {\bibinfo {author} {\bibfnamefont {A.~S.}\ \bibnamefont
  {Davydov}}\ and\ \bibinfo {author} {\bibfnamefont {G.~F.}\ \bibnamefont
  {Filippov}},\ }\href {https://doi.org/10.1016/0029-5582(58)90153-6}
  {\bibfield  {journal} {\bibinfo  {journal} {Nucl. Phys.}\ }\textbf {\bibinfo
  {volume} {8}},\ \bibinfo {pages} {237 } (\bibinfo {year} {1958})}\BibitemShut
  {NoStop}%
\bibitem [{\citenamefont {Wilets}\ and\ \citenamefont {Jean}(1956)}]{gsoft}%
  \BibitemOpen
  \bibfield  {author} {\bibinfo {author} {\bibfnamefont {L.}~\bibnamefont
  {Wilets}}\ and\ \bibinfo {author} {\bibfnamefont {M.}~\bibnamefont {Jean}},\
  }\href {https://doi.org/10.1103/PhysRev.102.788} {\bibfield  {journal}
  {\bibinfo  {journal} {Phys. Rev.}\ }\textbf {\bibinfo {volume} {102}},\
  \bibinfo {pages} {788} (\bibinfo {year} {1956})}\BibitemShut {NoStop}%
\bibitem [{\citenamefont {Sears}\ \emph {et~al.}(1998)\citenamefont {Sears},
  \citenamefont {Fossan}, \citenamefont {Gluckman}, \citenamefont {Smith},
  \citenamefont {Thorslund}, \citenamefont {Paul}, \citenamefont {Hibbert},\
  and\ \citenamefont {Wadsworth}}]{sears1998}%
  \BibitemOpen
  \bibfield  {author} {\bibinfo {author} {\bibfnamefont {J.~M.}\ \bibnamefont
  {Sears}}, \bibinfo {author} {\bibfnamefont {D.~B.}\ \bibnamefont {Fossan}},
  \bibinfo {author} {\bibfnamefont {G.~R.}\ \bibnamefont {Gluckman}}, \bibinfo
  {author} {\bibfnamefont {J.~F.}\ \bibnamefont {Smith}}, \bibinfo {author}
  {\bibfnamefont {I.}~\bibnamefont {Thorslund}}, \bibinfo {author}
  {\bibfnamefont {E.~S.}\ \bibnamefont {Paul}}, \bibinfo {author}
  {\bibfnamefont {I.~M.}\ \bibnamefont {Hibbert}},\ and\ \bibinfo {author}
  {\bibfnamefont {R.}~\bibnamefont {Wadsworth}},\ }\href
  {https://doi.org/10.1103/PhysRevC.57.2991} {\bibfield  {journal} {\bibinfo
  {journal} {Phys. Rev. C}\ }\textbf {\bibinfo {volume} {57}},\ \bibinfo
  {pages} {2991} (\bibinfo {year} {1998})}\BibitemShut {NoStop}%
\bibitem [{\citenamefont {Cline}(1986)}]{cline1986}%
  \BibitemOpen
  \bibfield  {author} {\bibinfo {author} {\bibfnamefont {D.}~\bibnamefont
  {Cline}},\ }\href {https://doi.org/10.1146/annurev.ns.36.120186.003343}
  {\bibfield  {journal} {\bibinfo  {journal} {Ann. Rev. Nucl. Part. Sci.}\
  }\textbf {\bibinfo {volume} {36}},\ \bibinfo {pages} {683} (\bibinfo {year}
  {1986})}\BibitemShut {NoStop}%
\bibitem [{\citenamefont {Werner}\ \emph {et~al.}(2000)\citenamefont {Werner},
  \citenamefont {Pietralla}, \citenamefont {von Brentano}, \citenamefont
  {Casten},\ and\ \citenamefont {Jolos}}]{werner2000}%
  \BibitemOpen
  \bibfield  {author} {\bibinfo {author} {\bibfnamefont {V.}~\bibnamefont
  {Werner}}, \bibinfo {author} {\bibfnamefont {N.}~\bibnamefont {Pietralla}},
  \bibinfo {author} {\bibfnamefont {P.}~\bibnamefont {von Brentano}}, \bibinfo
  {author} {\bibfnamefont {R.~F.}\ \bibnamefont {Casten}},\ and\ \bibinfo
  {author} {\bibfnamefont {R.~V.}\ \bibnamefont {Jolos}},\ }\href@noop {}
  {\bibfield  {journal} {\bibinfo  {journal} {Phys. Rev. C}\ }\textbf {\bibinfo
  {volume} {61}},\ \bibinfo {pages} {021301} (\bibinfo {year}
  {2000})}\BibitemShut {NoStop}%
\end{thebibliography}%

\end{document}